\documentclass[]{preprintRSC}

\usepackage{amsmath,amssymb,amsfonts,mathrsfs}
\usepackage[latin1]{inputenc}
\usepackage{graphicx,subfigure}
\usepackage{stmaryrd}
\usepackage{setspace}
\usepackage{epstopdf}
\usepackage[pdftex,colorlinks,bookmarksopen,bookmarksnumbered,citecolor=red,urlcolor=red]{hyperref}

\newcommand{\bm}[1]{\text{\boldmath $#1$\unboldmath}}

\newcommand{\nsd}  {\ensuremath{\texttt{n}_{\texttt{sd}}}}
\newcommand{\bx}{\bm{x}}

\newcommand{\bv}{\bm{v}}

\newcommand{\nel}{\ensuremath{\texttt{n}_{\texttt{el}}}}
\newcommand{\nelB}{\nel^\texttt{B}}
\newcommand{\nno}{\ensuremath{\texttt{n}_{\texttt{no}}}}
\newcommand{\nnoB}{\nno^\texttt{B}}
\newcommand{\nq}{\ensuremath{\texttt{n}_{\texttt{q}}}}

\newcommand{\nk}{\ensuremath{\texttt{n}_{\texttt{k}}}}
\newcommand{\ncp}{\ensuremath{\texttt{n}_{\texttt{cp}}}}
\newcommand{\curve}{\bm{C}} 
\newcommand{\bB}{\bm{B}}

\newcommand{\grad}{\bm{\nabla}}

\newcommand{\Id}{\mathbf{I}}

\newcommand{\ttimes}{\!\!\times\!\!}

\newtheorem{Rk}{Remark}

\begin{document}

\begin{frontmatter}
		
\title{A machine learning approach to predict near-optimal meshes for turbulent compressible flow simulations}

\author{Sergi Sanchez-Gamero}
\author{Oubay Hassan}
\author{Ruben Sevilla\corref{mycorrespondingauthor}}
\ead{r.sevilla@swansea.ac.uk}
\address{Zienkiewicz Centre for Computational Engineering, Faculty of Science and Engineering \\ Swansea University, Swansea, SA1 8EN, Wales, UK}
		
\cortext[mycorrespondingauthor]{Corresponding author}
				
\begin{abstract}
This work presents a methodology to predict a near-optimal spacing function, which defines the element sizes, suitable to perform steady RANS turbulent viscous flow simulations. The strategy aims at utilising existing high fidelity simulations to compute a target spacing function and train an artificial neural network (ANN) to predict the spacing function for new simulations, either unseen operating conditions or unseen geometric configurations. Several challenges induced by the use of highly stretched elements are addressed. The final goal is to substantially reduce the time and human expertise that is nowadays required to produce suitable meshes for simulations. Numerical examples involving turbulent compressible flows in two dimensions are used to demonstrate the ability of the trained ANN to predict a suitable spacing function. The influence of the NN architecture and the size of the training dataset are discussed. Finally, the suitability of the predicted meshes to perform simulations is investigated.
\end{abstract}
		
		
\end{frontmatter}
	
\section{Introduction}

Generating a mesh that represents the geometry of a domain is a requirement for the majority of numerical methods used to solve partial differential equations. Fully automatic unstructured mesh generators are nowadays common and available in many commercial, research and open-source codes. However, generating tailored meshes for specific applications such as computational fluid dynamics (CFD) is still a time consuming task that requires substantial human intervention and expertise~\cite{dawes2001reducing,slotnick2014cfd}. This is mainly due to the need to generate tailored meshes for different geometric configurations and/or operating conditions during a design optimisation process. An obvious alternative to alleviate the need for excessive human interaction is to produce a highly refined mesh that is capable of capturing the solution features for all the configurations to be tested~\cite{MichalTetrahedron,lock2023meshing}. This removes the cost associated with human intervention, but leads to undesired large execution times. In addition, it leads to substantially larger carbon emissions associated to the simulation process, as discussed in~\cite{lock2023meshing}.

Automatic mesh adaptive algorithms have gained popularity in the last decades~\cite{meshECM}. The user only needs to generate an initial mesh and an iterative process automatically adapts the mesh based on an error estimation. Depending on the error estimator used, the initial mesh requires a level of pre-adaptation to be performed by a human, because if the initial mesh fails to capture some features of the solution, the error estimator driving the adaptive process will be unable to correctly identify the areas where refinement is necessary. Furthermore, error indicators that are based on a single entity will fail to capture many phenomena. Hence, to ensure sensing all flow features, multiple indicators that uses the primitive variables and derived variables are required to enable capturing features such as shock waves, contact discontinuity, flow separation and turbulence-related quantities. In addition, it has been shown that, for complex three dimensional simulations, the complete adaptation process can involve between 20 and 30 iterations~\cite{loseille2010fully}. Each iteration requires the computation of a solution, the estimation of the error, the generation of an adapted mesh and the interpolation of the solution from the original to the adapted mesh.

Machine learning (ML) techniques for simulating physical phenomena have gained significant popularity within the computational engineering community in the last years. However, there is a fundamental concern about the potential uncertainty of predictions which, for some applications, cannot be tolerated. Instead of predicting the physical phenomena directly, ML approaches are nowadays being increasingly applied for predicting near-optimal meshes for simulations. This is motivated by two main facts. First, mesh generation is still referred to as one of the main bottlenecks in the simulation pipeline and, second, the uncertainty associated to ML predictions is tolerated when the output is a mesh that can be further improved before obtaining a high fidelity simulation.

The earliest use of ML to aid the mesh generation stage is attributed to~\cite{chang1991self}. Self-organising maps were used to train an artificial neural network (ANN) capable of predicting the position of the interior mesh nodes, given the boundary discretisation and a desired mesh density function. During the 1990s many approaches followed this work by proposing several enhancements~\cite{dyck1992determining,lowther1993density,chedid1996automatic,manevitz1997finite}. Most of the work in this decade focused on two dimensional problems with simple geometries and in the context of electromagnetic simulations. Further work during the second half of the 1990s and the early 2000s introduced the use of let-it-grow or adaptive ANNs to solve some of the limitations of previous works based on self-organising maps~\cite{martinetz1994topology,fritzke1994growing,alfonzetti1996automatic,alfonzetti1998finite,triantafyllidis2002finite} and the first three dimensional results started to emerge~\cite{alfonzetti2003neural}.

In the last two decades the work in the area of ML for mesh generation has gained momentum, with numerous approaches to aid the mesh generation stage. In~\cite{zhang2020meshingnet,zhang2021meshingnet3d} the authors proposed employing ANNs to predict the spacing at a certain location for a given set of parameters associated to a partial differential equation, geometric parameters and boundary conditions. Another approach for predicting a suitable spacing for CFD applications was recently proposed in~\cite{huang2021machine}. This work produces highly accurate CFD data using adaptively refined meshes. The spacing from the adaptive computation is transferred to a Cartesian grid that covers the domain of interest and translated into a grey-scale image that is used to train an ANN. These approaches are related to the current work in the sense that they involve the prediction of a spacing function that is suitable for generating meshes suitable for unseen cases.

Other works have also presented the potential of using ANNs to assist mesh adaptivity or degree adaptivity strategies~\cite{manevitz2005neural,bohn2021recurrent,yang2021reinforcement,wallwork2022e2n,foucart2023deep,tlales2022machine}, to predict near-optimal spacing for complex three dimensional CFD applications~\cite{lock2023meshing,LockIMR2023} or to predict the mesh anisotropy~\cite{fidkowski2021metric}. A complete review, up to the first half of 2022, can be found in~\cite{lei2023s}. 

This work presents a new approach to predict the near-optimal spacing for steady RANS turbulent compressible flow simulations involving unseen geometric configurations or operating flow conditions. The proposed approach follows the rationale introduced in~\cite{lock2023meshing,LockIMR2023} and presents solutions to the challenges that appear when considering highly stretched elements, as required to capture boundary and shear layers, and when working with more than one key variable to determine the desired spacing. In addition, for problems involving variable geometric configurations, a strategy is presented that is able to provide a tight coupling with the computer aided design (CAD) model by considering the control points of the non-uniform rational B-splines (NURBS) curves as the inputs of the ANN.

The proposed methodology consists of the following stages. First, with a computed solution, a strategy to compute the so-called \textit{target spacing} is presented. The target spacing, defined as the spacing function that will induce a mesh capable of capturing all the features present in a solution, is obtained by using standard concepts of error analysis. The strategy is based on the computation of the eigenvalues of the Hessian matrix of a selected key variable. Second, the target spacing is transferred to a background coarse mesh. This is done to provide a discrete representation of the spacing with the same number of values for any solution, even if the meshes used to perform different simulations have different number of nodes and/or elements. When problems with variable geometry are considered, mesh morphing of the background mesh is performed using a simple elasticity analogy. The third step consists of training a feed-forward ANN to predict the spacing at the nodes of the background mesh for a given set of inputs, which could be operating conditions or geometric parameters. Finally, the trained ANN can be used to predict a near-optimal spacing function suitable for new simulations. 

The trained ANN can be used to produce a near-optimal mesh in the sense that is expected to be a very close representation of the mesh that an adaptive process would provide. As such, the predicted meshes can be integrated into an adaptive process and it is anticipated that the number of adaptive iterations will be significantly reduced when compared to starting the process with a very coarse mesh.
This approach has already proved to be beneficial when compared to the current industrial practice of generating a single refined mesh to perform a range of simulations for different flow conditions~\cite{lock2023meshing}. It has also the potential to substantially accelerate optimisation and inverse design processes because in such scenarios it is often required to evaluate the objective functions, i.e. to run a CFD simulation, hundreds or thousands of times~\cite{balla2022inverse}. In this context, manually generating a mesh for each configurations is clearly unfeasible and having a tool to predict a near-optimal mesh is particularly beneficial. It is worth remarking that the proposed approach also provides a path to utilise the vast amount of data that is available in industry to inform mesh generation.

The remainder of the paper is organised as follows. Section~\ref{sc:background} briefly recalls the concepts necessary to introduce the proposed methodology. These include the governing partial differential equations for a compressible turbulent viscous flow, NURBS curves, the definition of the spacing function using a background mesh, the use of an elastic analogy to morph an existing mesh and feed-forward ANNs. In Section~\ref{sc:nearOptimalSpacing} the main contribution of the work is presented. First, the computation of the target spacing for a given simulation is presented. Three important aspects discussed in detail are the computation of the Hessian of a key variable, the choice of the key variables for compressible turbulent viscous flows and the challenges introduced by the use of highly stretched elements. A smoothing procedure is proposed to avoid that small variations in the pressure within the inflation layer having a major effect on the computed target spacing. This section also presents the strategy utilised to transfer the spacing from a given mesh to a coarse background mesh. Particular attention is paid to the transfer of the spacing computed with two different key variables. In Section~\ref{sc:examples} two numerical examples are considered. The first example involves the prediction of near-optimal meshes for variable flow conditions, whereas in the second example the case of varying geometries is considered. The parameters for the second example are the position of the control points of the NURBS curves defining an aerofoil geometry and the example considers 23 geometric parameters. This Section discusses the effect of using different ANN architectures as well as the effect of using an increasing number of cases in the training dataset. The predicted meshes are compared to the target meshes and also utilised to perform simulations in order to assess the suitability of the predicted meshes. Finally, Section~\ref{sc:conclusions} summarises the conclusions of the work that has been presented.

\section{Background} \label{sc:background}

This Section summarises the fundamental concepts that are required to present the proposed approach to predict the near-optimal spacing suitable to generate meshes for unseen simulations. 

\subsection{Governing equations}

This work considers problems governed by the compressible Reynolds-averaged Navier-Stokes (RANS) equations coupled with the Spalart-Allmaras (SA) turbulence model~\cite{SA1992}. The non-dimensional conservative form of the RANS equations, without the turbulence equations, can be written as
\begin{equation} \label{eq:RANS}
\frac{\partial \bm{U}}{\partial t} + \grad \cdot \left( \bm{F} (\bm{U})- \bm{G} (\bm{U}, \grad \bm{U}) \right) = \bm{0},
\end{equation}
where $\bm{U}$ is the vector of conserved quantities, $\bm{F}$ is the advection flux tensor and $\bm{G}$ is the diffusion flux tensor, given by
\begin{equation}\label{eq:NSterms}
\bm{U} = 
\begin{Bmatrix} 
	\rho\\ 
	\rho\bv\\ 
	\rho E 
\end{Bmatrix}\,
\quad
\bm{F} (\bm{U}) = 
\begin{bmatrix} 
	\rho \bv^T\\
	\rho\bv \otimes\bv + p \Id\\
	(\rho E + p) \bv^T 
\end{bmatrix},
\quad
\bm{G} (\bm{U}, \grad \bm{U}) = 
\begin{bmatrix} 
	\bm{0}\\
	\bm{\sigma} \\
	(\bm{\sigma} \bv + \bm{q} )^T 
\end{bmatrix}.
\end{equation}
In the above expressions, $\rho$ denotes the density, $\bv$ the velocity vector, $E$ the total specific energy, $p$ the pressure and $\Id$ the identity matrix. The viscous stress tensor $\bm{\sigma}$ and the heat flux vector $\bm{q}$ are given by
\begin{equation}\label{eq:StressTensorHeatFlux}
\bm{\sigma} = \frac{\mu+\mu_t}{Re_\infty} \left(\grad \bv + \grad \bv^T - \frac{2}{3} (\grad \cdot \bv) \Id \right), 
\qquad 
\bm{q} =  \frac{1}{ Re_\infty} \left( \frac{\mu}{Pr_\infty} + \frac{\mu_t}{Pr_t}\right)\grad T,
\end{equation}
where $\mu$ is the viscosity, $T$ the temperature, $Re_\infty$ is the Reynolds number and $Pr_\infty$ and $Pr_t$ are the laminar and turbulent Prandtl numbers, respectively.

It is assumed that the viscosity varies with temperature according to the Sutherland's law, i.e. $\mu = \left(T/T_\infty\right)^{3/2} (T_\infty + S)/(T + S)$, where the free-stream temperature is $T_\infty = 1/\left((\gamma - 1) M_\infty^2\right)$ and the Sutherland constant is $S = S_0/\left((\gamma - 1) T_0 M_\infty^2\right)$, with $S_0 = 110K$ and $T_0 = 273K$. Assuming the fluid to be calorically perfect, the pressure is $p = (\gamma - 1) \rho \left( E -  \|\bv\|^2/2 \right)$, where $\gamma = c_p / c_v$ (equal to 1.4 for air in standard conditions) is the ratio of specific heats at constant pressure, $c_p$, and constant volume, $c_v$. 

The flow is typically characterised by the Prandtl, Reynolds and Mach numbers, given by
\begin{equation} \label{eq:non-dimensionalQuantities}
Pr_\infty = \frac{c_p \mu_\infty}{\kappa}, 
\qquad
Re_\infty = \frac{\rho_\infty v_\infty L}{\mu_\infty}, 
\qquad
M_\infty = \frac{v_\infty}{c_\infty}.
\end{equation}
where $c = \sqrt{\gamma p/\rho}$ is the speed of sound, $L$ a characteristic length and $\kappa$ the thermal conductivity.

In this work the eddy viscosity $\mu_t$ appearing in Equation~\eqref{eq:StressTensorHeatFlux} is calculated from the SA turbulence model~\cite{SA1992}, which is described in detail in many references and it is therefore not repeated here.

It is worth noting that only steady cases are considered in this work. Therefore, the time $t$ is actually an artificial pseudo-time that is simply utilised to facilitate convergence towards the steady state.

\subsection{Non-uniform rational B-splines} \label{sc:NURBS}

NURBS are commonly used in CAD to provide a boundary representation of the computational domain~\cite{piegl1996nurbs}. This work considers well known aerofoils such as the RAE2828 for problems with fixed geometry and NURBS-shaped aerofoils for problems with variable geometry. 

A NURBS curve of degree $q$ is a piecewise rational function defined in parametric form as
\begin{equation*} \label{eq:nurbsDefinition}
	\curve(\lambda) = \biggl(\sum_{i=0}^{\ncp} \nu_i\,\bB_i\,C_i^q(\lambda)\biggr)
	\biggm/
	\biggl(\sum_{i=0}^{\ncp} \nu_i\,          C_i^q(\lambda)\biggr)
	\qquad 0 \leq  \lambda \leq 1,
\end{equation*}
where the $\ncp+1$ control points are denoted by $\{\bB_i\}$ and the corresponding control weights are $\{\nu_i\}$. The normalised B-spline basis functions, $\{C_i^q(\lambda)\}$, are defined using the recursion
\begin{align*}
	C_i^0(\lambda) &= \begin{cases}
		1 & \text{if $\lambda\in    [\lambda_i,\lambda_{i+1}[$,}  \\
		0 & \text{elsewhere,}  \\
	\end{cases}\\
	C_i^k(\lambda) &= \frac{\lambda-\lambda_i}{\lambda_{i+k}-\lambda_i}
	C_i^{k-1}(\lambda)+
	\frac{\lambda_{i+k+1}-\lambda}{\lambda_{i+k+1}-\lambda_{i+1}}
	C_{i+1}^{k-1}(\lambda),
\end{align*}
for $k=1\hdots q$, where $\lambda_i$, for $i=0,\dotsc, \nk$, are the knots, where it is assumed that they are ordered $0\le\lambda_i\le\lambda_{i+1}\le 1$. The knots are commonly arranged into the so-called knot vector
\begin{equation*}
\mathcal{N} = \{ \underbrace{0,\hdots,0}_{q+1}, \lambda_{q+1},\dotsc,\lambda_{\nk-q-1}, \underbrace{1,\dotsc ,1}_{q+1} \},
\end{equation*}
which uniquely defines the B-spline basis functions.  The number of control points, $\ncp+1$, the number of knows, $\nk+1$, and the degree, $q$, are related using the expression  $\nk = \ncp + q + 1$.

\begin{Rk} \label{rk:nurbs}
NURBS curves possess a variety of attractive properties~\cite{piegl1996nurbs}, but for the purpose of this work only two of the properties are briefly recalled here. For a given NURBS curve with control points $\{\bB_i\}_{i=0,\ldots,\ncp}$ the end point interpolation property means that $\curve(0)=\bB_0$ and $\curve(1)=\bB_{\ncp}$. In addition the tangent to the initial (end) point is aligned with the two first (last) control points, namely $\curve'(0) = \mu \bv_{0,1}$ and $\curve'(1) = \kappa \bv_{\ncp-1,\ncp}$, where $\mu$ and $\kappa$ are real numbers and $\bv_{i,j}$ is the vector formed by connecting control points $\bB_i$ and $\bB_j$. In this work these two properties are relevant when building NURBS-shaped aerofoils as they provide a simple guideline to join NURBS curves whilst ensuring $\mathcal{G}^1$ continuity.
\end{Rk}

\subsection{Spacing function using a background mesh} \label{sc:backgroundMesh}

The generation of suitable meshes for turbulent compressible flow simulations usually relies on a combination of human expertise and techniques to define a user-controlled local refinement. Among all the available techniques, the use of a background mesh and/or mesh sources provide great flexibility. Previous works on machine learning for predicting near-optimal meshes for inviscid compressible flows~\cite{lock2023meshing,LockIMR2023}  have shown that, despite the greater flexibility of mesh sources, the use of a background mesh leads to a faster training of the ANN and, at the same time, a higher accuracy of the predictions. This is mainly due to the reduced number of outputs of the ANN. Therefore, this work only considers the use of a background mesh to define the mesh spacing.

A background mesh, denoted by $\mathcal{B}_h$, is defined as a collection of disjoint elements $\{B_e\}_{e=1,\ldots,\nelB}$, where $\nelB$ is the number of elements. A discrete nodal spacing field is then defined on the background mesh, namely $\{\delta_i^\texttt{B}\}_{i=1,\ldots,\nnoB}$, where $\nnoB$ is the number of mesh nodes. The background mesh covers the whole computational domain $\Omega$, namely $\Omega \subseteq \cup_{e=1}^{\nelB} B_e$, and it is used during the mesh generation stage to query the required spacing at a point $\bx \in \Omega$. This is done by identifying the element of the background mesh, $B_e$, that contains the point $\bx$ and interpolating the spacing at $\bx$ from the nodal spacing values of the background element $B_e$.

In addition, to capture the high gradient of the velocity field in the normal direction to a wall, an inflation layer of quadrilateral elements is typically employed. A common approach involves defining the non-dimensional height of the first layer of quadrilateral elements, $h_1$, based on the Reynolds number, for instance $h_1= C Re_\infty^{-3/4}$, where $C$ is a user-defined constant, and building a geometric progression to successively increase the height of the next layers. A growing factor $G$, which usually takes value between 1.1 and 1.5, is commonly employed to define the height of the next layer of quadrilateral elements, namely $h_i = Gh_{i-1}$ for $i=2,\ldots,\nq$, where $\nq$ is the maximum number of layers of quadrilateral elements. In the proposed implementation, $\nq$ can be selected to be an arbitrarily large number because the mesh generator employed is capable of creating different number of quadrilateral elements in the normal direction in different regions. The mesh generator stops creating a new layer automatically when the spacing of the layer to be generated is approximately half of the spacing given by the background mesh. If the mesh generator does not provide such capability, the approach presented here is still valid, but a structured layer with a fixed number of elements might need to be used.

In this work, the insertion of the inflation layer is considered to be independent upon the spacing of the background mesh. Although the background mesh does not need to contain an inflation layer, it is sometimes beneficial to include such a layer to ensure that a better resolution of the spacing is provided near the obstacle. In the numerical examples, both options are considered.

\subsection{Mesh morphing} \label{sc:meshMorphing}

When problems with varying geometric configurations are considered, a fixed fitted background mesh cannot be used. To ensure that a background mesh with the same topology is available for each geometric configuration, a mesh morphing approach based on linear elasticity is considered in this work. The strategy follows the rationale employed to generate high-order curved meshes in~\cite{xie2013generation}.

First, a background mesh is generated for a geometry with a pre-defined set of control points, namely $\{\bm{B}_i^{\texttt{B}}\}_{i=0,\ldots,\ncp}$ for each NURBS curve $\bm{C}^{\texttt{B}}$. 
The parametric coordinates of each node of the background mesh on the wall boundary, $\bx^w$, namely $\lambda_i$ such that $\bx^w_i= \bm{C}^{\texttt{B}}(\lambda_i)$, are computed using a standard NURBS inversion algorithm~\cite{piegl1996nurbs}. For a different geometric configuration with NURBS boundary parametrised by $\bm{C}$, with control points $\{\bm{B}_i\}_{i=0,\ldots,\ncp}$, the desired position of the background mesh nodes on the wall boundary is computed as $\hat{\bx}^w_i= \bm{C}(\lambda_i)$. This means that the parametric position of background mesh nodes on the wall boundary is kept constant and only the position of the control points is altered between different geometric configurations. Finally, using the initial and desired position of the background mesh nodes on the wall boundary, a boundary value problem governed by the linear elastic equation is defined. Homogeneous Dirichlet boundary conditions are imposed on the far field and Dirichlet boundary conditions, corresponding to $\hat{\bx}^w_i-\bx^w_i$, are imposed on each node on the wall. The solution of the elastic problem provides the desired position of all the nodes of the background mesh for a different geometric configuration.

It is worth noting that maintaining a good quality of the deformed background mesh is not a requirement as this mesh is only used to define the spacing function, not to perform a simulation. If required, multiple steps can be employed during the deformation, to ensure the validity of the deformed mesh~\cite{xie2013generation}.

The computational cost of deforming a background mesh is negligible due to the use of coarse background meshes. However, if desired, it is possible to deform the mesh layer-by-layer, starting with the elements in contact with a wall boundary~\cite{xie2013generation}. This strategy substantially reduces the computational cost of mesh deformation and also enables to stop deforming the mesh when the deformation computed in one layer is negligible.

\subsection{Feed-forward neural networks}

A feed-forward ANN is an arrangement of neurons organised in layers, where each neuron is only connected to all the neurons of the previous and next layers. A weight is associated to each connection between two neurons and a bias, which can be seen as an extra neuron with fixed value, is assigned to each layer. The first layer contains the so-called inputs of the network, whereas the neurons of the last layer are called outputs. It is common to number the remaining layers, named hidden layers, using an index $l = 1,\ldots,N_l$, where $N_l$ is the number of hidden layers~\cite{hagan1997neural}.

\begin{figure}[!tb]
	\centering
	\includegraphics[width=0.8\textwidth]{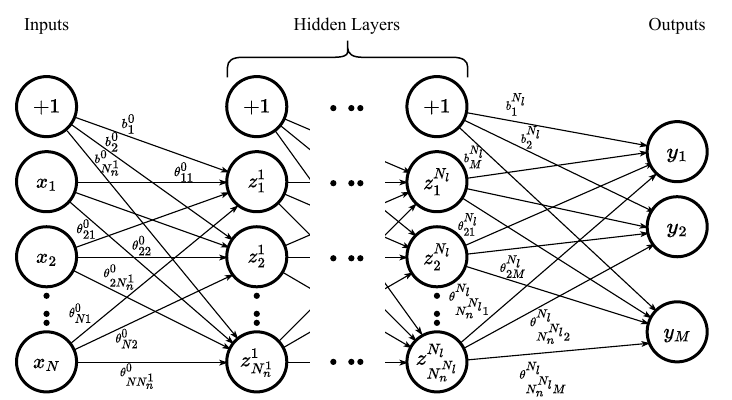}
	\caption{Sketch of a multi-layer feed-forward ANN.}
	\label{fig:NN}
\end{figure}

The forward propagation involves the calculation of the value associated to each neuron from the values of the neurons from the previous layer, the weights of the associated connections and an activation function. More precisely, the value of the $j$-th neuron of the $l+1$ layer is computed as
\begin{equation}\label{eq:neuronEq}
	z^{l+1}_{j} = F^{l+1} \left( \sum_{i=1}^{N_n^l}\theta^{l}_{ij}z^{l}_{i} + b^{l}_{j} \right),
\end{equation}
where $F^l$ is the activation function of the $l$-th hidden layer, $\theta^{l}_{ij}$ is the weight of the connection between the $i$-th neuron of layer $l$ and the $j$-th neuron of layer $l+1$, $b^l$ is the bias of the $l$-th layer and $N_n^l$ is the number of neurons of the $l$-th layer. Figure~\ref{fig:NN} shows a sketch of a feed-forward neural network.

For a given case, the inputs and outputs of the ANN are arranged in two vectors, $\mathbf{x} = \{x_{1},...,x_{N} \}^T$ and $\mathbf{y} = \{y_{1},...,y_{M} \}^T$, respectively, where $N$ is the number of inputs and $M$ is the number of outputs. The so-called cost or error function is defined as
\begin{equation} \label{eq:costFunction}
	E(\bm{\theta},\mathbf{b}) = \frac{1}{N_{tr}M} \sum_{k=1}^{N_{tr}} \sum_{i=1}^M [ y^k_i(\mathbf{x}) - h_i^k(\bm{\mathbf{x},\theta,\mathbf{b}}) ]^2,
\end{equation}
where $h_i^k(\mathbf{x},\bm{\theta})$ denotes the output predicted by the ANN during the forward propagation and $N_{tr}$ is the number of training cases. The training process consists of finding the weights $\bm{\theta}$ and biases $\mathbf{b}$ that minimise the error function. Several approaches are available to perform this optimisation. The current implementation, based on TensorFlow 2.7.0~\cite{abadi2016tensorflow}, utilises the ADAM optimiser~\cite{kingma2014adam}. The specific parameters of the optimiser considered here are the ones reported in~\cite{lock2023meshing,LockIMR2023}, where a similar approach was used to devise an ANN to predict near-optimal meshes for inviscid compressible flow simulations.

As usual in the context of ANNs, the hyperparameters, i.e. the number of neurons and hidden layers, can influence substantially the accuracy of the predictions. In the numerical examples considered, a simple grid search is performed in this work to select the hyperparameters. Further numerical experiments have shown that for the relative reduced datasets considered here, the rectified linear unit (ReLu) activation function produced more accurate results than other popular activation functions. Therefore, in all the numerical examples the ReLu activation function is used in all hidden layers and a linear function is employed in the output layer, namely
\begin{equation} \label{eq:activation}
F^l(x) = \begin{cases}
x & \text{if $l=N_l$,}  \\
\max\{0,x\} & \text{otherwise.}  \\
\end{cases}
\end{equation}

\section{Near-optimal spacing function prediction using ANNs} \label{sc:nearOptimalSpacing}

\begin{figure}[!tb]
	\centering
	\subfigure[$p$]{\includegraphics[width=0.32\textwidth]{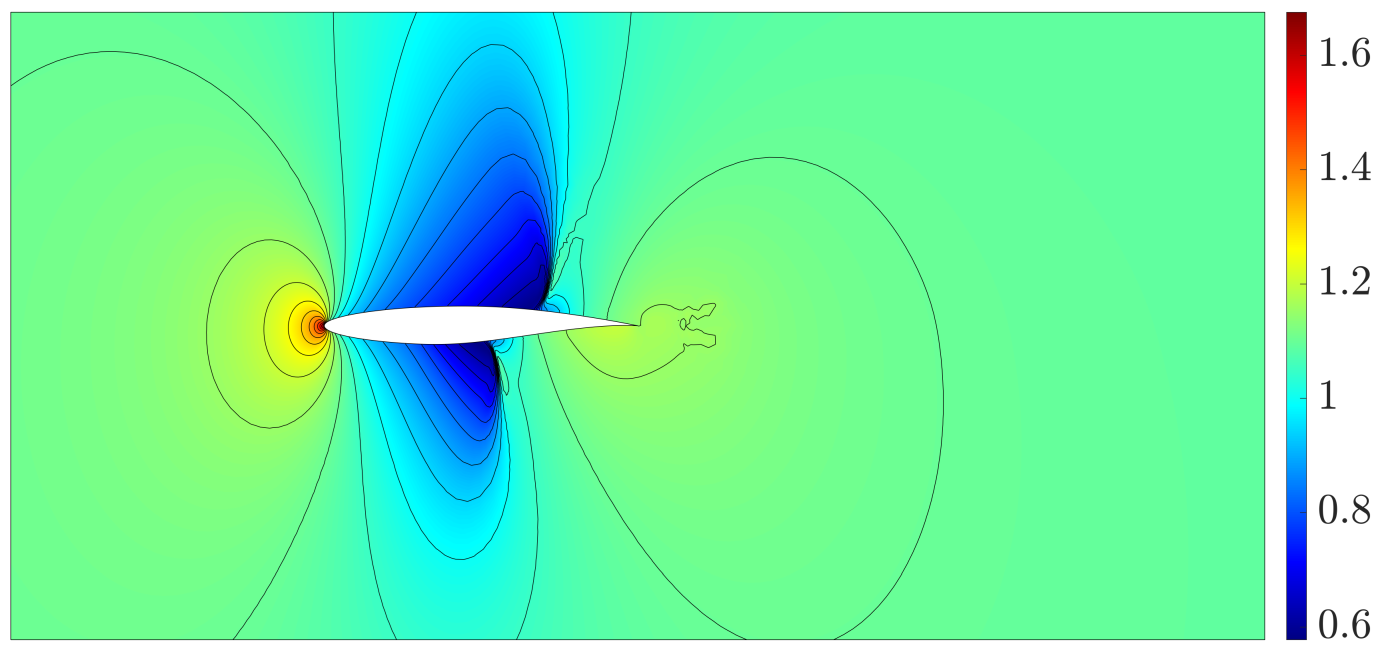}}
	\subfigure[$M$]{\includegraphics[width=0.32\textwidth]{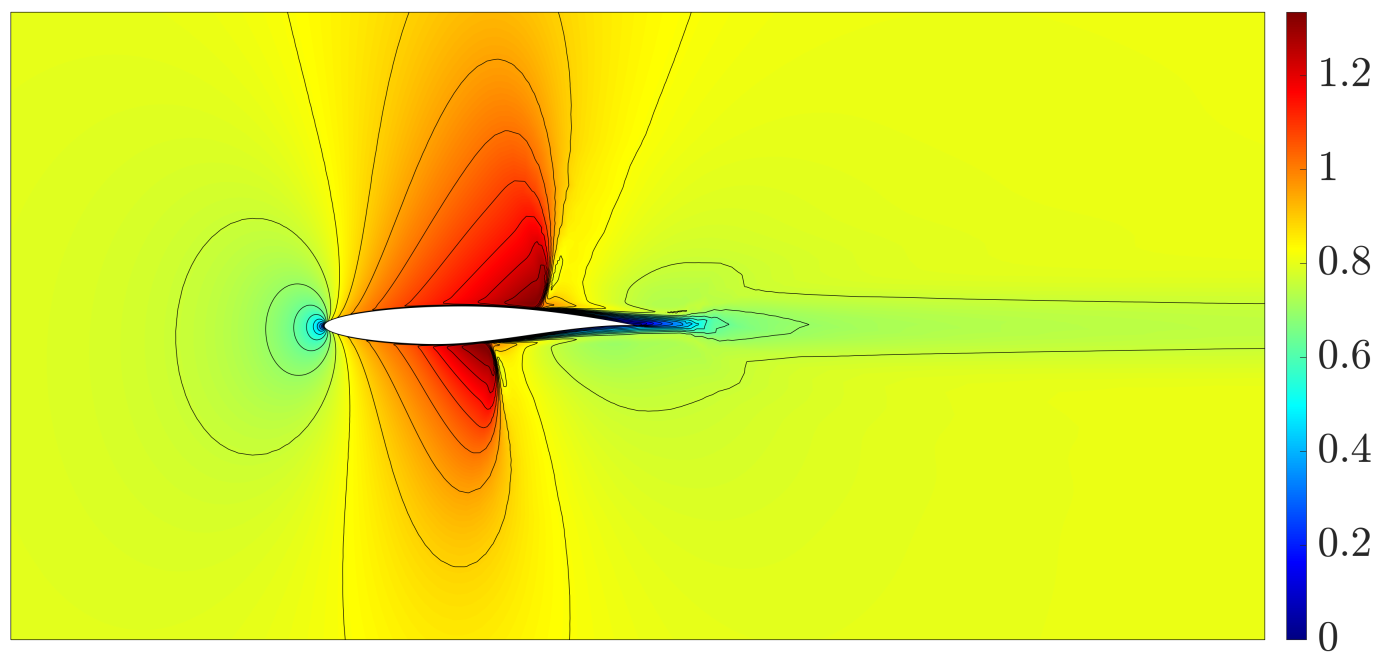}}
	\subfigure[$\rho$]{\includegraphics[width=0.32\textwidth]{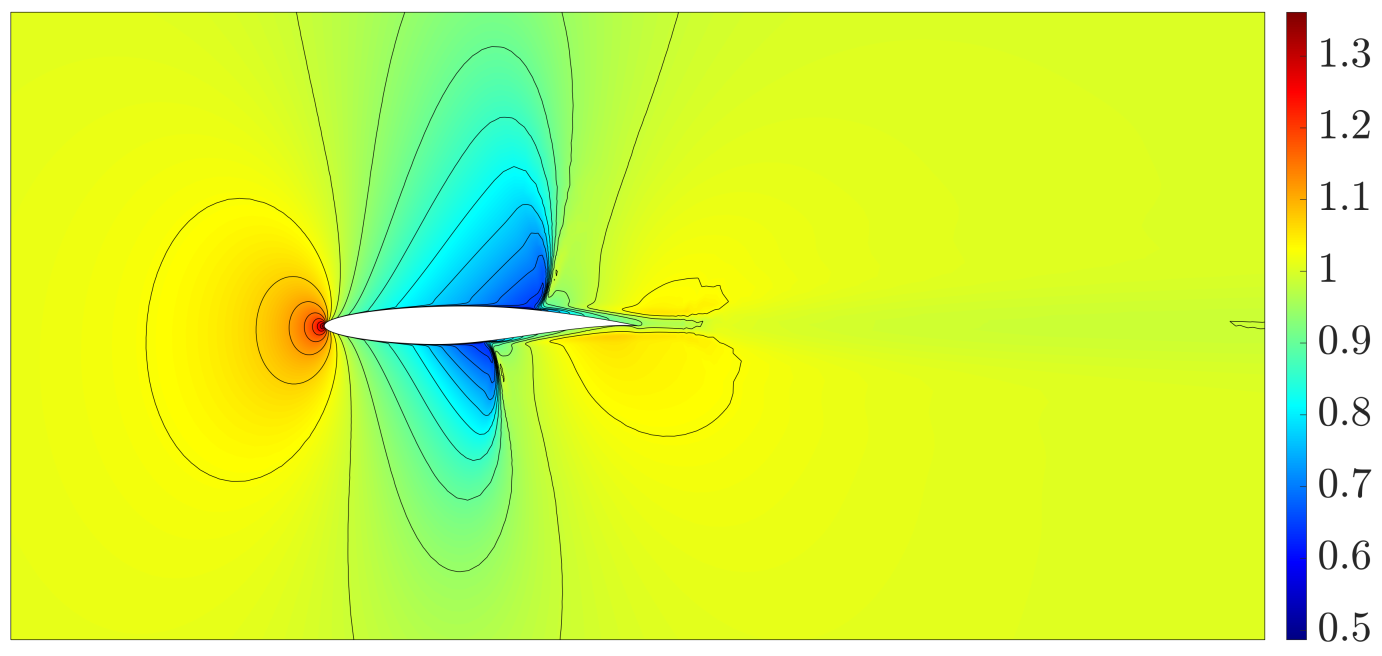}}
	\caption{Solution obtained on a fine mesh for $Re_\infty = 6.5 \times 10^{6}$, $M_{\infty} = 0.811$,  and $\alpha= 0.27^\circ$.}
	\label{fig:solution}
\end{figure}

This Section presents a novel strategy to predict a near-optimal spacing function to generate meshes suitable for performing turbulent compressible flow simulations. The strategy consists of training an ANN to predict a near-optimal discrete spacing in a background mesh. The training data is made of historical data available from previous high fidelity analysis. 
As an example, Figure~\ref{fig:solution} shows the pressure, Mach number and density for one simulation using a reference mesh. The geometry corresponds to the RAE2822 aerofoil and the flow conditions are $Re_\infty = 6.5 \times 10^{6}$, $M_{\infty} = 0.811$,  and $\alpha= 0.27^\circ$.
The objective is to predict the spacing for unseen operating flow conditions or geometric configurations. 

The proposed strategy consists of the following stages:
\begin{enumerate}
	\item For each available solution, a discrete target spacing function is computed on the same mesh that was used to perform the simulation. 
	\item The discrete spacing function is transferred from the computational mesh to a coarse background mesh. 
	\item A feed-forward ANN is trained to predict the discrete spacing function for the background mesh. The inputs of the ANN are flow conditions or geometric parameters, whereas the output of the ANN is a discrete spacing function, i.e. the spacing at the nodes of the background mesh.
\end{enumerate}

It is worth noting that, for different simulations, the computational mesh is usually different, whereas the same background mesh is considered for all cases. When geometric parameters are considered, the mesh morphing strategy described in Section~\ref{sc:meshMorphing} is applied to the background mesh to ensure that the topology of the background mesh is maintained.

\subsection{Computation of the target spacing in the computational mesh} \label{sc:targetSpacingComp}

Following principles traditionally employed in error analysis~\cite{peraire1987adaptive}, the procedure to compute a discrete spacing function from a given solution, requires the evaluation of the Hessian matrix of a chosen key variable at every point of the computational mesh. More precisely, the spacing at a node $\bx_i$ along a generic direction given by a unit vector $\bm{\beta}$ is related to the derivatives of the key variable as
\begin{equation} \label{eq:errorAnalysis}
	\delta_{i,\bm{\beta}}^2 \: \left ( \sum_{k,l=1}^{\nsd} (H_i)_{kl} \beta_k \beta_l \right ) = K,
\end{equation}
where $\nsd$ is the number of spatial dimensions, $K$ is a user-defined constant and 
\begin{equation} \label{eq:hessian}
	(H_i)_{kl} = \frac{ \partial^2 \sigma_i} {\partial x_k \partial x_l}
\end{equation}
are the components of the Hessian matrix at node $\bx_i$, $\mathbf{H}_i$, for the key variable $\sigma$. The evaluation of the derivatives of the selected key variable, $\sigma$, is described in detail in Section~\ref{sc:hessian}. After the derivatives are evaluated, the spacing at a node $\bx_i$ of the computational mesh is computed as
\begin{equation} \label{eq:spacingMin}
\delta_i  =
\begin{cases}
\delta_{min} & \text{if $\lambda_{i,max} > K/\delta_{min}^2$,} \\
\delta_{max} & \text{if $\lambda_{i,max} < K/\delta_{max}^2$,} \\
\sqrt{ K/\lambda_{i,max}} & \text{otherwise},
\end{cases}
\end{equation}
where $\{\lambda_{i,j}\}_{j=1,\ldots,\nsd}$ are the eigenvalues of $\mathbf{H}_i$ and $\lambda_{i,max} = \max_{j=1,\ldots, \nsd}\{\lambda_{i,j}\}$.

It can be seen that Equation~\eqref{eq:spacingMin} imposes that $\delta_i \in [\delta_{min}, \delta_{max}]$. The minimum value $\delta_{min}$ is used to avoid an excessive local refinement in regions with very steep gradients, for instance in the vicinity of strong shocks. Analogously, the maximum value $\delta_{max}$ is used to avoid an excessive large value in regions of undisturbed flow, where the key variable is smooth or nearly constant. To simplify the implementation, this work considers the values of $\delta_{min}$ and $\delta_{max}$ as the minimum and maximum spacing in the computational mesh, respectively.

\begin{figure}[!tb]
	\centering
	\subfigure[$S = 1$]{\includegraphics[width=0.32\textwidth]{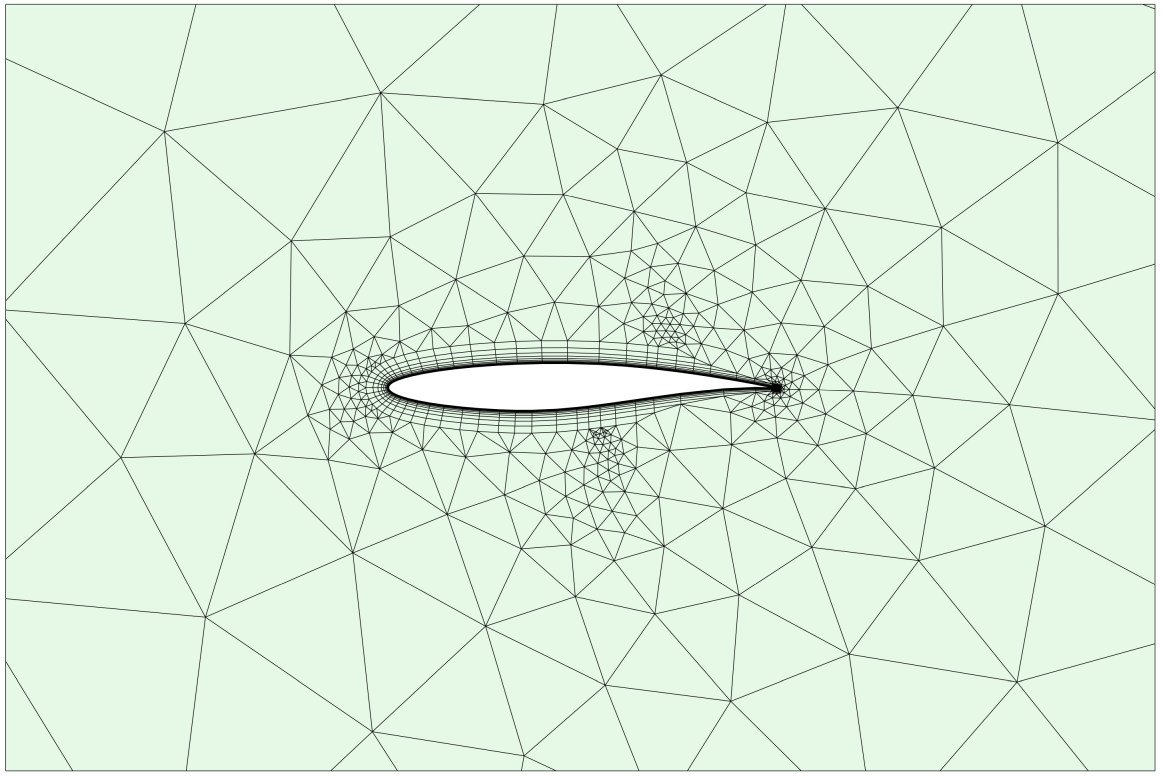}}
	\subfigure[$S = 0.5$]{\includegraphics[width=0.32\textwidth]{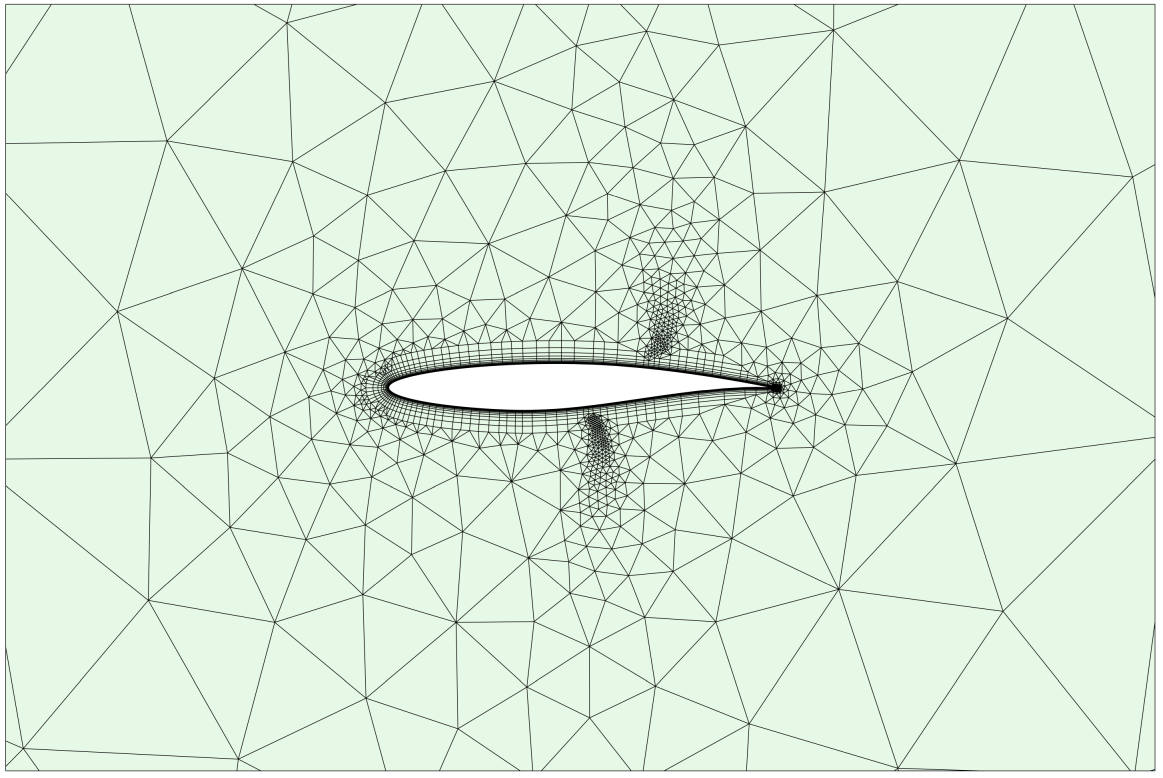}}
	\subfigure[$S = 0.2$]{\includegraphics[width=0.32\textwidth]{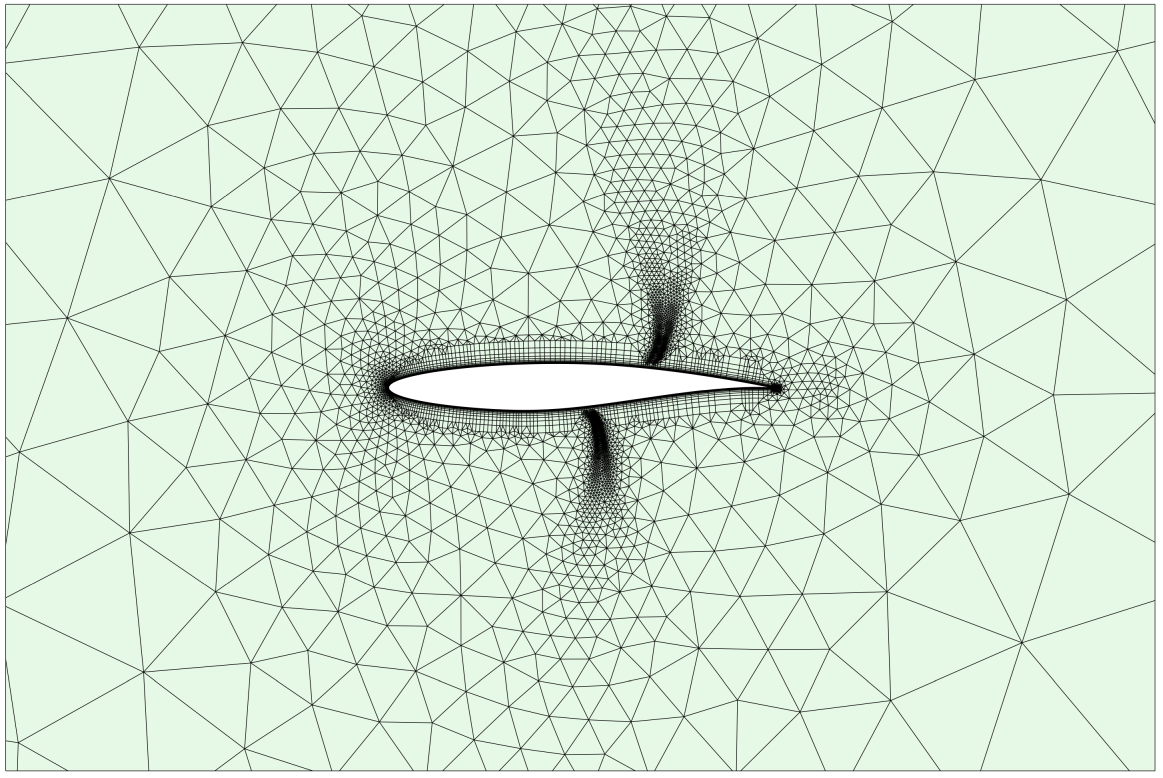}}
	\caption{Meshes obtained by varying the minimum spacing using the pressure as key variable for the solution of Figure~\ref{fig:solution}(a).}
	\label{fig:meshPScaling}
\end{figure}

In the current implementation, the value of $K$ is defined as 
\begin{equation} \label{eq:spacingK}
K = S^2 \delta_{min}^2 \lambda_{max},
\end{equation}
where $S \in (0,1]$ is a user-defined scaling factor and $\lambda_{max} = \max_{i=1,\ldots, \nno}\{\lambda_{i,max}\}$. For a value of $S=1$, the minimum spacing is only assigned to the node/s $\bx_i$ with a maximum eigenvalue of the Hessian matrix, $\lambda_{i,max}$, equal to the overall maximum $\lambda_{max}$. The lower the scaling factor is, the more nodes in the mesh will be assigned the minimum spacing. This is done in practice to ensure that the refinement is not just concentrated in one region with a very steep gradient of the key variable, missing most of the other features of the solution. 
Figure~\ref{fig:meshPScaling} illustrates the effect of the scaling factor. The spacing is computed using the pressure field of Figure~\ref{fig:solution}(a) as the key variable.
Unless otherwise stated, the value of the scaling factor used in this work is $S = 0.2$.

The spacing defined by Equation~\ref{eq:spacingMin} is a single scalar per node. Although it would be possible to compute an anisotropic metric per node, this is beyond the scope of the current work.

The next three Sections discuss the evaluation of the Hessian, the choice of the key variable, $\sigma$, and the proposed strategy to compute the spacing in the inflation layer, where the elements are quadrilaterals, with a significant stretching, designed to capture the large gradient of the velocity in the normal direction to the wall.

\subsubsection{Computation of the Hessian matrix} \label{sc:hessian}

The components of the Hessian are computed as the derivatives of the derivatives. Therefore, it is only necessary to detail how to compute a first order derivative.

Three different strategies to compute the first derivatives of the selected key variable on an unstructured mesh have been considered. The first strategy considers a vertex-centred FV approximation, i.e. using a dual mesh, of the derivatives at the nodes~\cite{sorensen2003b} given by
\begin{equation} \label{eq:FVderivative}
\grad\sigma_i \approx \frac{1}{|V_i|} \left[ \sum_{j \in \mathcal{N}_i} \frac{1}{2}(\sigma_i + \sigma_j) \bm{C}_{ij} +  \sum_{j \in \mathcal{N}_i^\partial} \sigma_i \bm{D}_{ij}  \right]
\end{equation}
where $V_i$ is the the control volume associated to node $\bx_i$, $\mathcal{N}_i$ is the set of nodes connected to $\bx_i$ by an interior edge (i.e., an edge not on the boundary) and $\mathcal{N}_i^\partial$ is the set of nodes connected to $\bx_i$ by a boundary edge. The geometric coefficient vectors $\bm{C}_{ij}$ and $\bm{D}_{ij}$  are defined as
\begin{equation} \label{eq:FVweights}
\bm{C}_{ij} = \sum_{\Gamma_i^k \in \mathcal{F}_{ij}} | \Gamma_i^k | \bm{n}_i^k,
\qquad 
\bm{D}_{ij} = \sum_{\Gamma_i^k \in \mathcal{F}_{ij}^\partial} | \Gamma_i^k | \bm{n}_i^k,
\end{equation}
where $\mathcal{F}_{ij}$ and $\mathcal{F}_{ij}^\partial$ are the sets of interior and boundary facets of control volume $V_i$, respectively, that intersect the edge $\Gamma_{ij}$ connecting nodes $\bx_i$ and $\bx_j$, and $\bm{n}_i^k$ is the unit outward normal to $\Gamma_i^k$ from the point of view of node $\bx_i$.

The second approach considers the standard recovery process employed in the finite element method~\cite{zienkiewicz1992superconvergent1,zienkiewicz1992superconvergent2}. In this case the derivative at a node $\bx_i$ is evaluated as a weighted average of the derivative computed at the barycentre of the elements that contain node $\bx_i$, namely
\begin{equation} \label{eq:FEMderivative}
\grad\sigma_i \approx \frac{\displaystyle \sum_{\Omega_e \in \mathcal{P}_i} |\Omega_e| \grad \sigma_e }{\displaystyle \sum_{\Omega_e \in \mathcal{P}_i} |\Omega_e|},
\end{equation}
where the gradient at the barycentre of the element, $\sigma_e$, is evaluated using the finite element shape functions of degree one and $\mathcal{P}_i$ is the patch of elements that contain the node $\bx_i$.

The third alternative considered here employs again the finite element recovery process, but after splitting the quadrilateral elements into two triangular elements.

\begin{figure}[!tb]
\centering
\subfigure[FV, $S=0.1$]				{\includegraphics[width=0.32\textwidth]{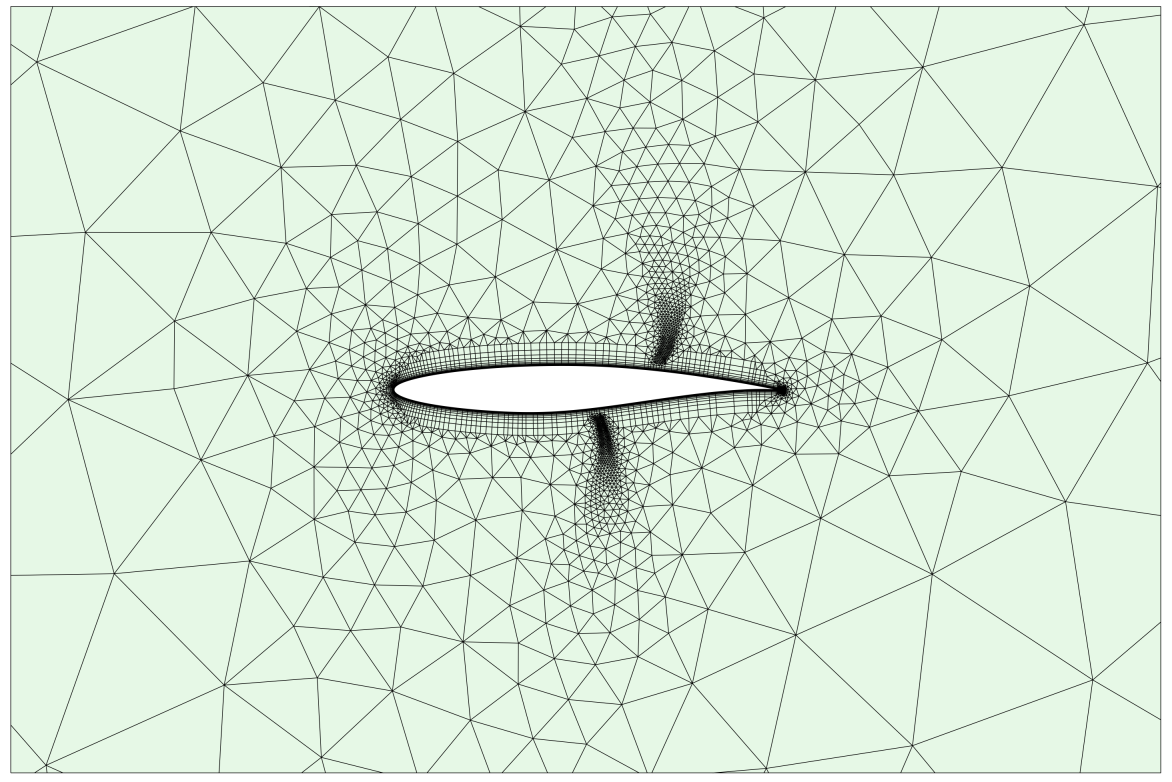}}
\subfigure[FEM hybrid, $S=0.2$]		{\includegraphics[width=0.32\textwidth]{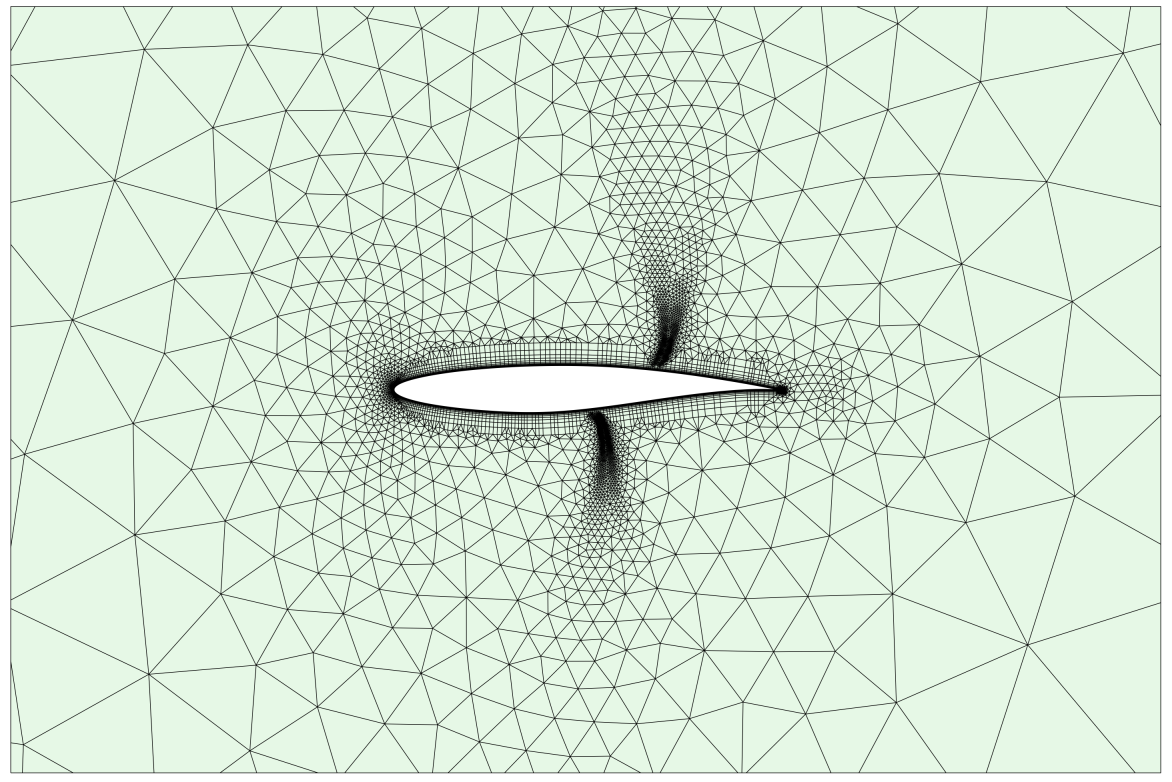}}	
\subfigure[FEM triangular, $S=0.2$]	{\includegraphics[width=0.32\textwidth]{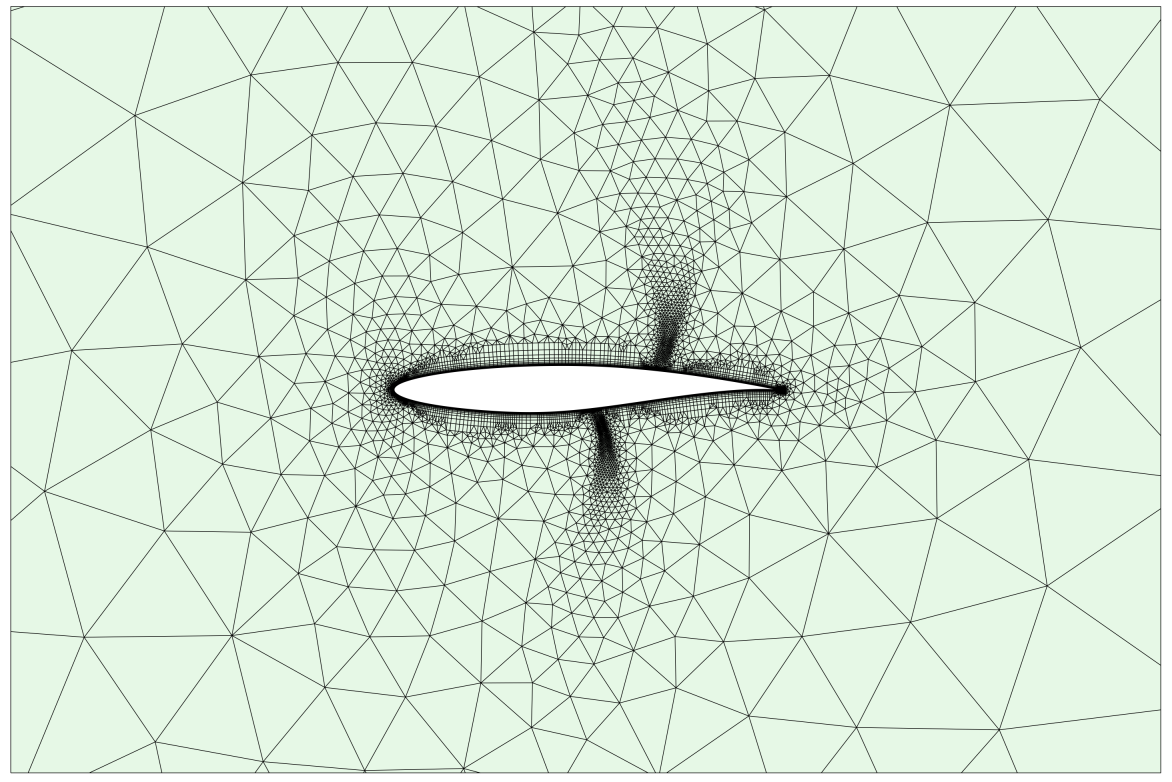}}	
\caption{Meshes obtained with the minimum spacing provided by pressure as key variable and with three different approaches to evaluate the Hessian for the solution of Figure~\ref{fig:solution}(a).}
\label{fig:meshPHessian}
\end{figure}

\begin{figure}[!tb]
\centering
\subfigure[FV, $S=0.1$]				{\includegraphics[width=0.32\textwidth]{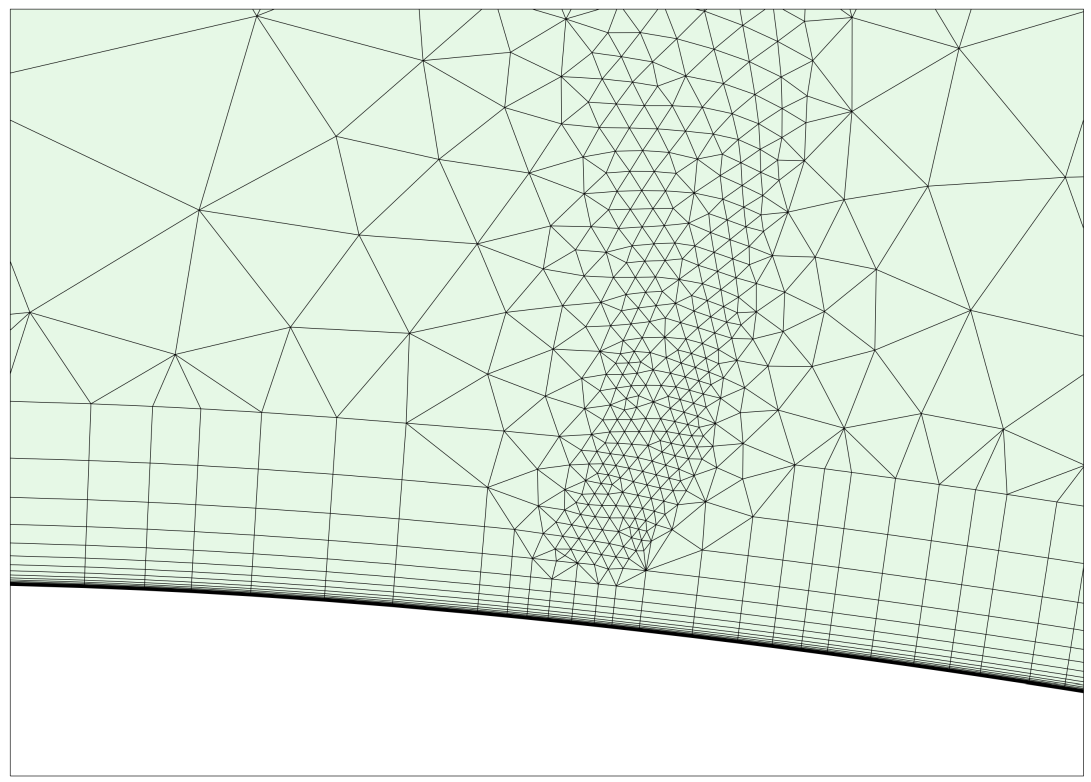}}
\subfigure[FEM hybrid, $S=0.2$]		{\includegraphics[width=0.32\textwidth]{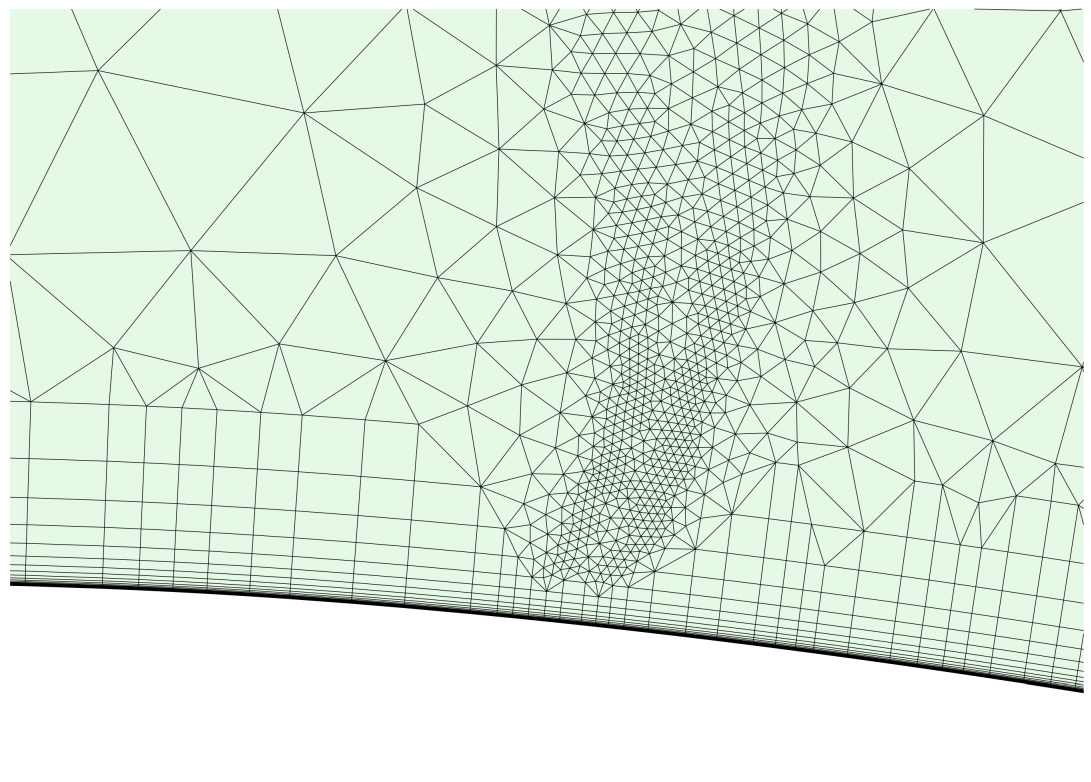}}	
\subfigure[FEM triangular, $S=0.2$]	{\includegraphics[width=0.32\textwidth]{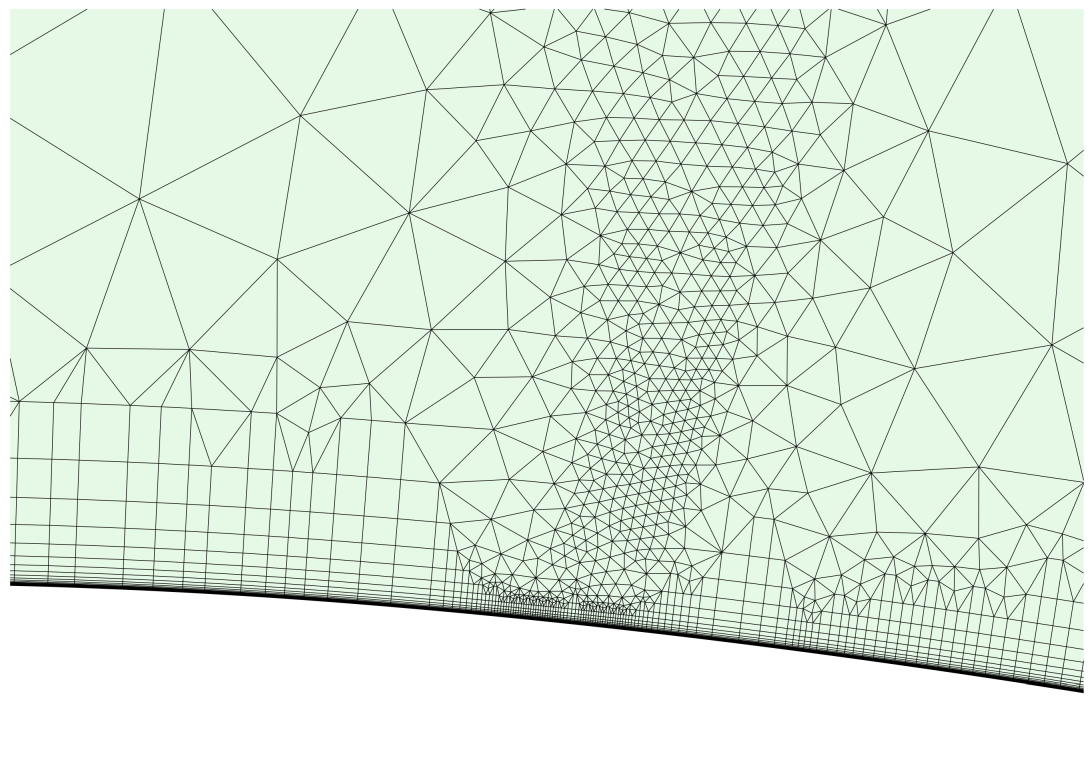}}	
\caption{Detail of the three meshes of Figure~\ref{fig:meshPHessian} near the region where the shock of the upper surface interacts with the boundary layer.}
\label{fig:meshPHessianZoom}
\end{figure}

Figure~\ref{fig:meshPHessian} shows the meshes obtained with the three approaches described above. The results show that the spacing computed with the finite element recovery process to evaluate the Hessian matrix is slightly smaller in the regions of interest, i.e. at the leading edge and near the shocks, compared to the spacing obtained by using a finite volume approximation of the derivatives in the Hessian matrix.
The finite volume approach requires a smaller scaling in order to produce a similar refinement as the one obtained with a finite element approximation of the derivatives. In this example the spacing computed after evaluating the derivatives with the finite volume formulae utilised a scaling of $S=0.1$, whereas the finite element recovery approaches employed $S=0.2$. To illustrate the differences between the three strategies, Figure~\ref{fig:meshPHessianZoom} shows a detailed view of the meshes in the vicinity of the region where the shock on the upper surface interacts with the boundary layer.
As it can be observed, the finite element recovery process performed after splitting the quadrilateral elements into triangles produces a very localised refinement in the region where the shock on the upper surface interacts with the boundary layer. This refinement is not achieved by the finite volume approximation of the derivatives or the finite element recovery process on hybrid meshes. For this reason, in all the remaining examples, the finite element recovery process after splitting the quadrilateral elements into triangles is considered.

As seen in Figure~\ref{fig:meshPHessianZoom}, when the required element size near the obstacle is below the element size of the inflation layer, triangular elements are generated close to the obstacle. With the mesh generator employed, this does not pose any issue as the number of quadrilateral elements in the normal direction can vary in different regions.

\subsubsection{Choice of the key variables} \label{sc:keyVariable}

\begin{figure}[!tb]
\centering
\subfigure[Key variable: $p$]			{\includegraphics[width=0.32\textwidth]{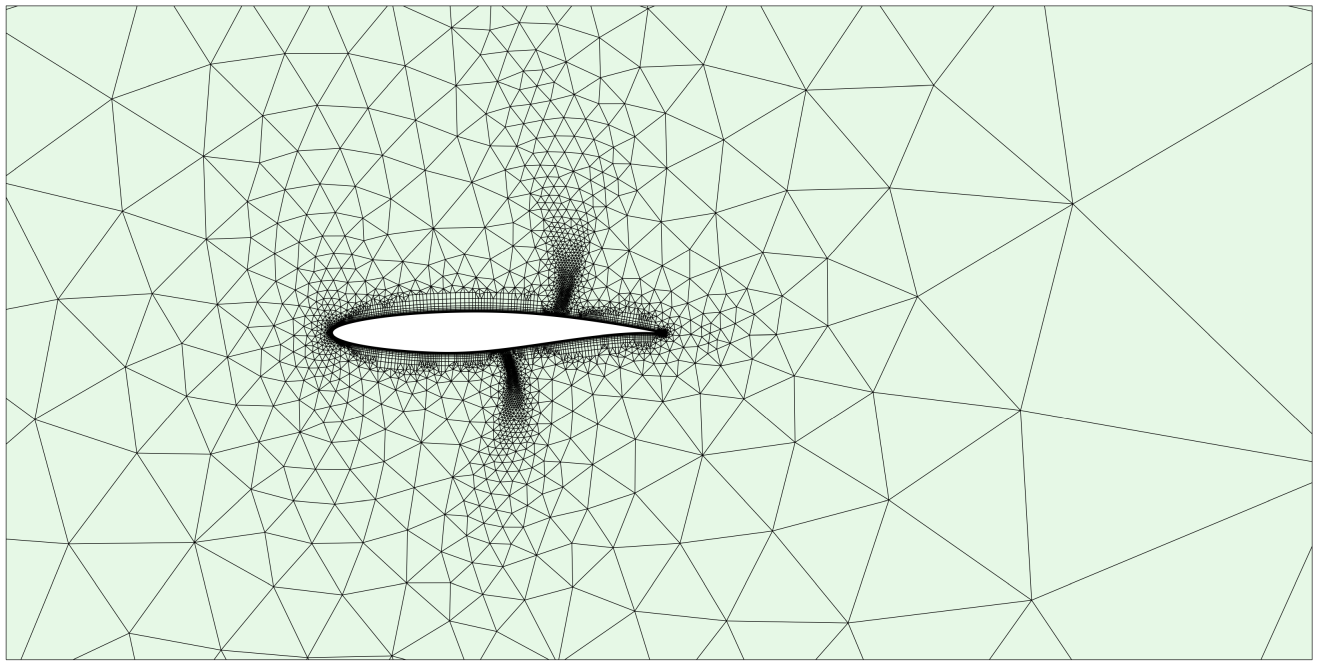}}
\subfigure[Key variables: $p$ and $\rho$]{\includegraphics[width=0.32\textwidth]{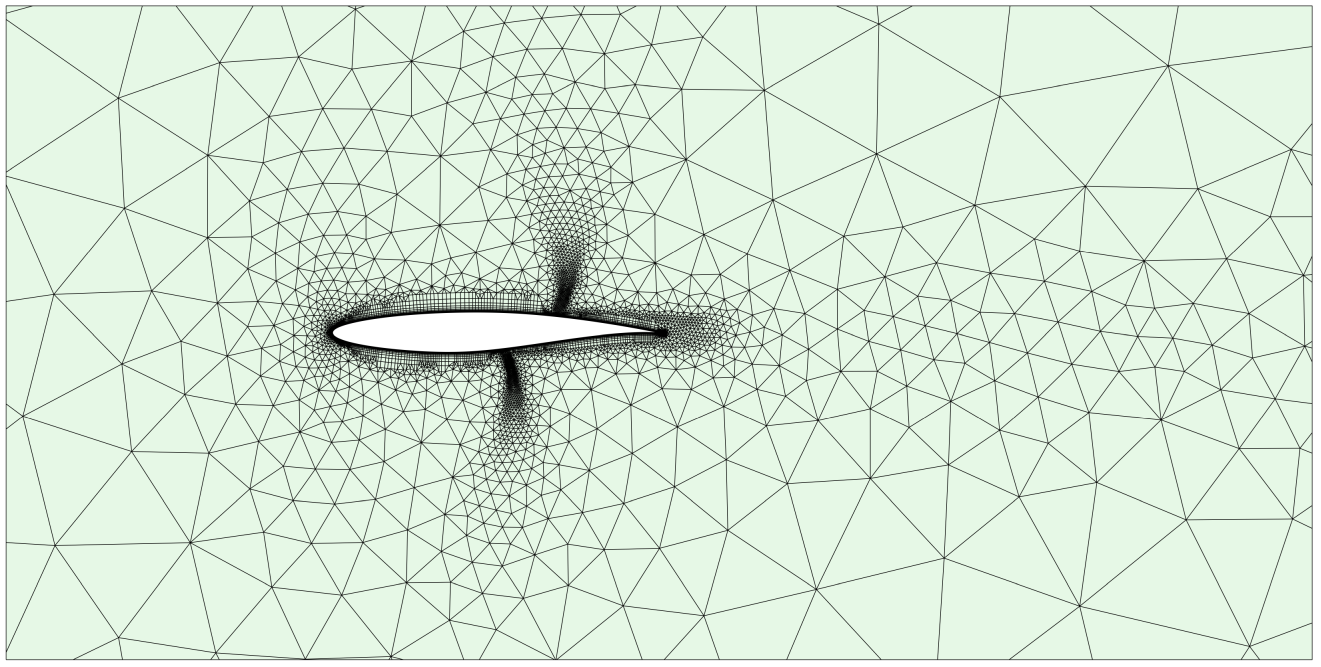}}	
\subfigure[Key variables: $p$ and $M$]	{\includegraphics[width=0.32\textwidth]{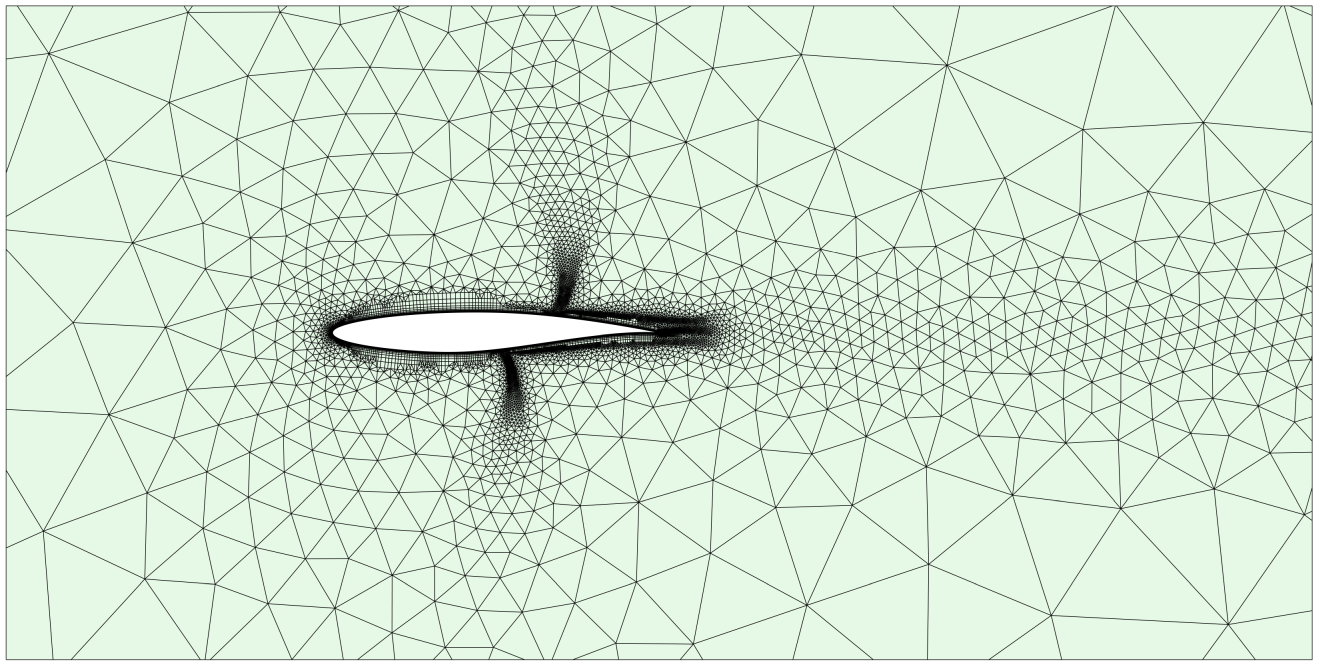}}	
\caption{Meshes obtained with the minimum spacing provided by different sets of key variables for the solution of Figure~\ref{fig:solution}.}
\label{fig:meshKeyVariables}
\end{figure}

For inviscid compressible flow simulations, previous works~\cite{lock2023meshing,LockIMR2023} have shown the effectiveness of using the pressure as the only key variable to capture flow features in the vicinity of an aerodynamic object. However, for viscous compressible flows, the second derivatives of the pressure in the region of the wake are not strong enough to provide relevant information. For this reason, a second key variable is considered to provide the required information that enables a local refinement of the wake. Two alternatives have been considered, namely the density and the Mach number. Figure~\ref{fig:meshKeyVariables} shows the meshes obtained by considering only the pressure and the pressure combined with the density and the Mach number as key variables.

As it can be observed, using the Mach number as the second key variable produces an extra refinement in the region of the wake. Further numerical experiments, not reported here for brevity confirm this observation and, therefore, for the numerical examples in this work, the pressure and the Mach number are the key variables considered. Obviously, when two key variables are employed, namely $\sigma^1$ and $\sigma^2$, Equation~\eqref{eq:spacingMin} is used to compute the spacing at each node for both variables, namely $\delta_i^1$ and $\delta_i^2$, and the minimum of the two spacings computed at each node is kept, $\delta_i = \min\{\delta_i^1, \delta_i^2\}$.

It is worth noting that in the example of Figure~\ref{fig:meshKeyVariables}, the spacing was computed using Equation~\eqref{eq:spacingMin}, but only for the nodes belonging to a triangular element. The next section details the computation of the spacing on the quadrilateral elements within the inflation layer. 

\subsubsection{Computation of the target spacing in the inflation layer} \label{sc:BL}

There are a number of difficulties that arise when computing the spacing in the inflation layer, caused by a significant variation of some quantities in the normal direction to the wall and the large stretching of the quadrilateral elements near the wall.

First, the Mach number presents a huge variation in the normal direction to the wall, due to the no-slip boundary condition. If Equation~\eqref{eq:spacingMin} is applied, the minimum spacing is completely dominated by the large values of the second derivatives of the velocity or Mach number in the normal direction to the wall. However, as described in Section~\ref{sc:backgroundMesh}, there are effective and easy-to-implement strategies to define the spacing in the normal direction to the wall that are widely adopted and do not require any human intervention. Therefore, this work assumes that there is no need to devise a strategy to predict the spacing in the normal direction to a wall within the inflation layer. To this end, the use of the Mach number as a key variable is restricted to nodes that belong to a triangular element.

Ignoring the computation of the spacing with the Mach number as key variable in the quadrilateral elements of the inflation layer might lead to a large discrepancy between the spacing computed by the pressure and the Mach number in the interface between the unstructured mesh and inflation layer. This ultimately can lead to an unfeasible discrete spacing distribution that cannot be realised by a mesh generator. To avoid this incompatibility, the spacing function obtained by using the Mach number as key variable is extended into the inflation layer by considering a constant extension in the normal direction.

\begin{figure}[!tb]
\centering
\includegraphics[width=0.45\textwidth]{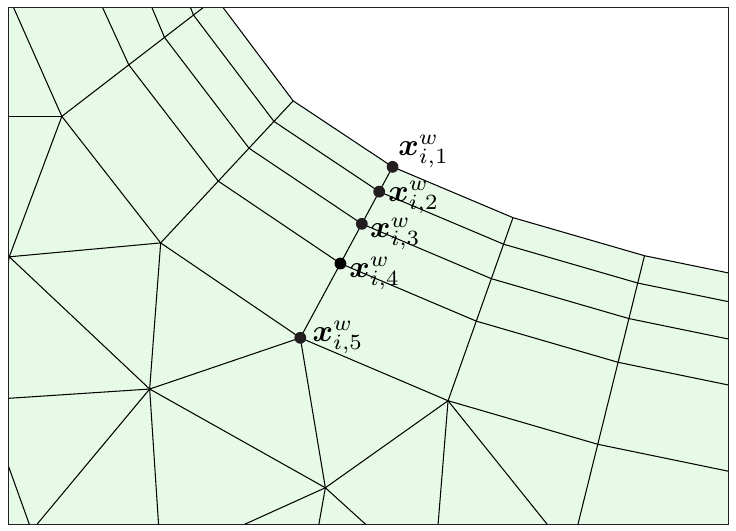}
\caption{Sketch of a hybrid mesh showing the numbering of the nodes in the inflation layer region.}
\label{fig:boundaryLayerMesh}
\end{figure}

Assuming that the nodes on the wall boundary are denoted by $\bx_{i,1}^w$ for $i=1,\ldots,\nno^w$, the nodes on the inflation layer that follow the normal direction to the wall from the $i$-th node $\bx_{i,1}^w$ are denoted by $\bx_{i,j}^w$ for $j=1,\ldots,\nno^{i,w}$, as illustrated in Figure~\ref{fig:boundaryLayerMesh}.

The spacing is computed at all nodes of the mesh that belong to a triangular element using Equation~\eqref{eq:spacingMin}, with the Mach number as key variable. Then the spacing associated to the rest of the nodes in the inflation layer is simply given by
\begin{equation} \label{eq:spacingMinflation}
\delta(\bx_{i,j}^w) = \delta(\bx_{i,\nno^{i,w}}^w),
\end{equation}
for $j=1,\ldots,\nno^{i,w}-1$.

\begin{figure}[!tb]
	\centering
	\subfigure[NASA] {\includegraphics[width=0.49\textwidth]{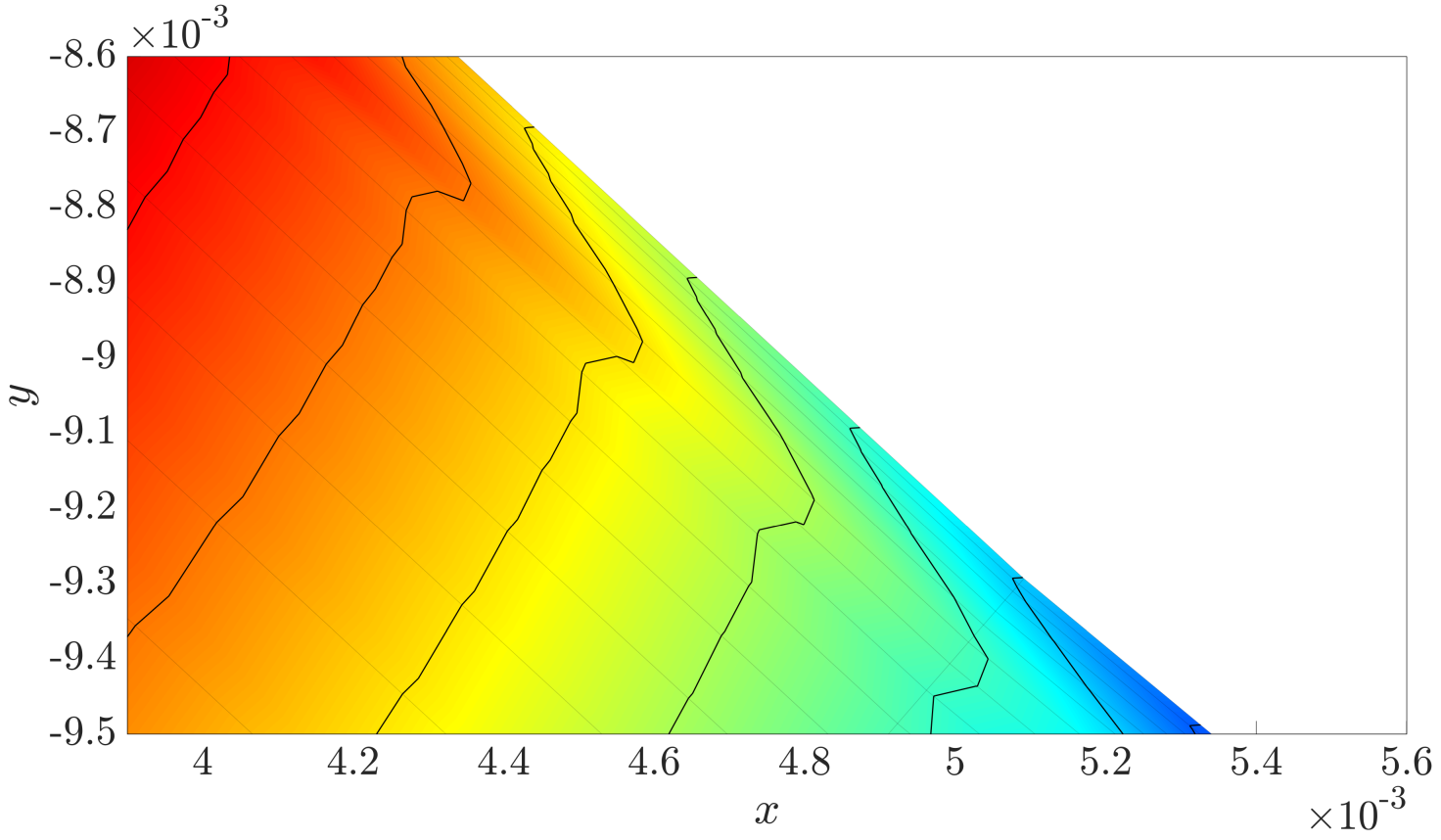}}
	\subfigure[FLITE]{\includegraphics[width=0.49\textwidth]{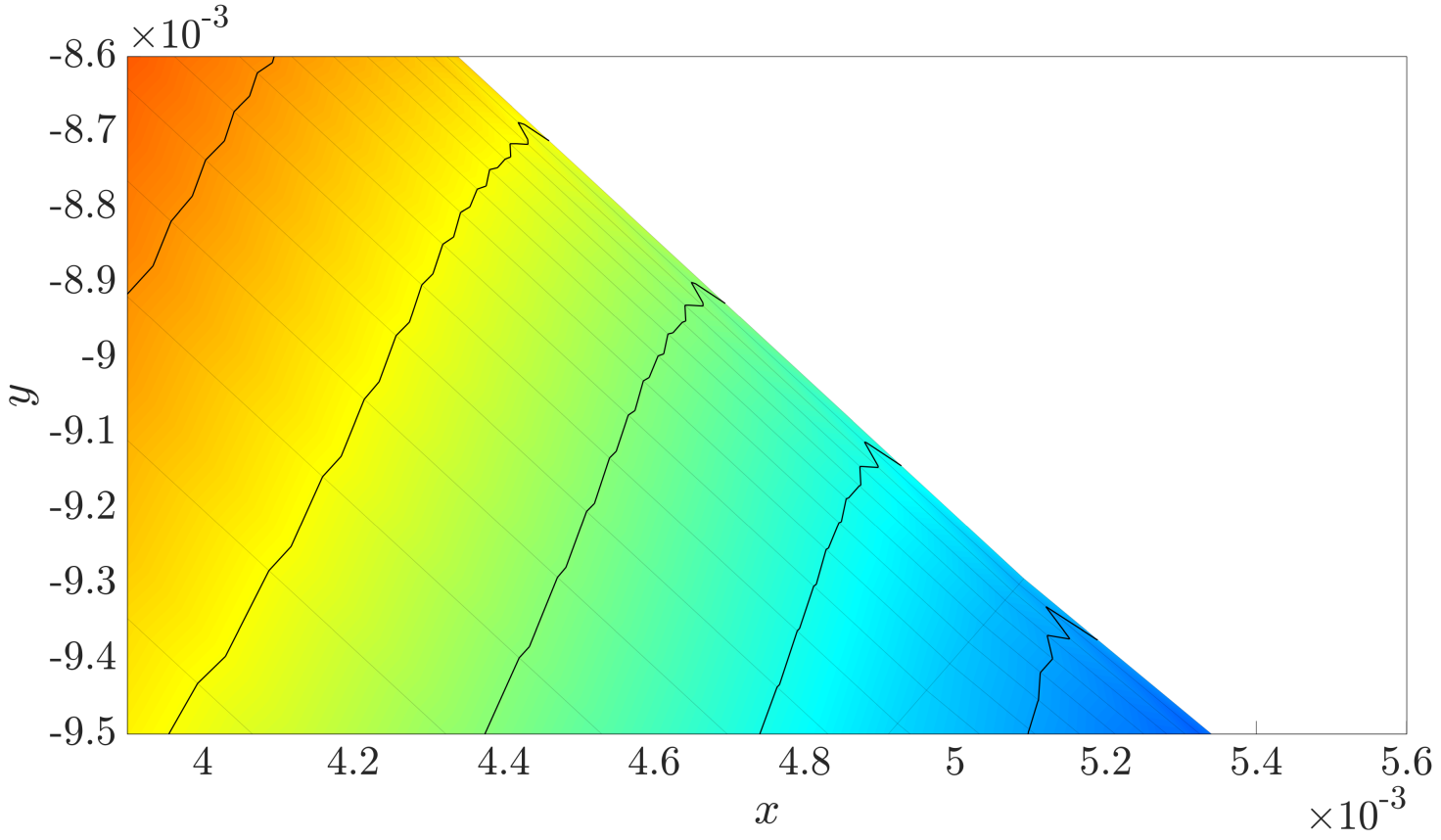}}	
	\caption{Close view of the pressure field and contour lines near the wall for $Re_\infty = 6.5 \times 10^6$, $M_{\infty} = 0.729$ and $\alpha= 2.31^\circ$ using two different solvers.}
	\label{fig:pOscillations}
\end{figure}

A second challenge to be considered when computing the spacing in the inflation layer using the pressure as key variable is that small variations in the normal direction might occur that, due to the very high stretching of the elements, can lead to extremely large derivatives. Figure~\ref{fig:pOscillations} shows a close up view of the pressure field and contour lines for two different calculations near the leading edge, using the finite difference NASA code Wind-US~\cite{cook1979aerofoil} and the finite volume FLITE system~\cite{sorensen2003b} using the same computational mesh.
Although the variation of the pressure on the area displayed is below 0.05, these small variations lead to very large derivatives when using any of the formulae presented in Section~\ref{sc:hessian}, due to the large stretching of the elements, with a height of the first element in the inflation layer of $9.9 \times 10^{-6}$ and an area of $2.4 \times 10^{-9}$.

To ensure that these small and non-desired variations of the pressure do not skew the definition of the target spacing, a pressure smoothing is proposed in this work. Given the ordering of the nodes in the inflation layer previously described, and illustrated in Figure~\ref{fig:boundaryLayerMesh}, the smoothed pressure, $\tilde{p}$ in the normal direction to the wall at node $\bx_{i,1}^w$ is assumed to be a cubic polynomial that satisfies the following four conditions:
\begin{itemize}
\item $\tilde{p}(0) = p\left(\bx_{i,1}^w\right)$,
\item $\tilde{p}'(0) = 0$,
\item $\tilde{p}\left(d_{i,\nno^{i,w}-1}\right) = p\left(\bx_{i,\nno^{i,w}-1}^w\right)$,
\item $\tilde{p}'\left(d_{i,\nno^{i,w}-1}\right) = \displaystyle \frac{ p\left(\bx_{i,\nno^{i,w}}^w\right) - p\left(\bx_{i,\nno^{i,w}-1}^w\right)}{ d_{i,\nno^{i,w}} - d_{i,\nno^{i,w}-1} }$,
\end{itemize}
where $d_{i,j} = \| \bx_{i,1}^w - \bx_{i,j}^w \|_2$.

These conditions impose the computed value of the pressure at the wall and zero normal derivative. In addition, to produce a smooth transition with the computed pressure outside the inflation layer, it imposes the value at the last node that does not belong to a triangular element and its derivative in the normal direction. The smoothing approach presented here simply aims at removing the negative influence of the undesired variations of the pressure in the spacing function computed using the Hessian. Although other, potentially more complex, smoothing approaches might be devised, for all the examples considered in this work, the cubic polynomial smoothing presented here led to satisfactory results. This strategy can be easily extended to three dimensional problems as the same type of smoothing can be defined in the normal direction to the wall.

\begin{figure}[!tb]
	\centering
	\subfigure[Key variable: $p$]			{\includegraphics[width=0.32\textwidth]{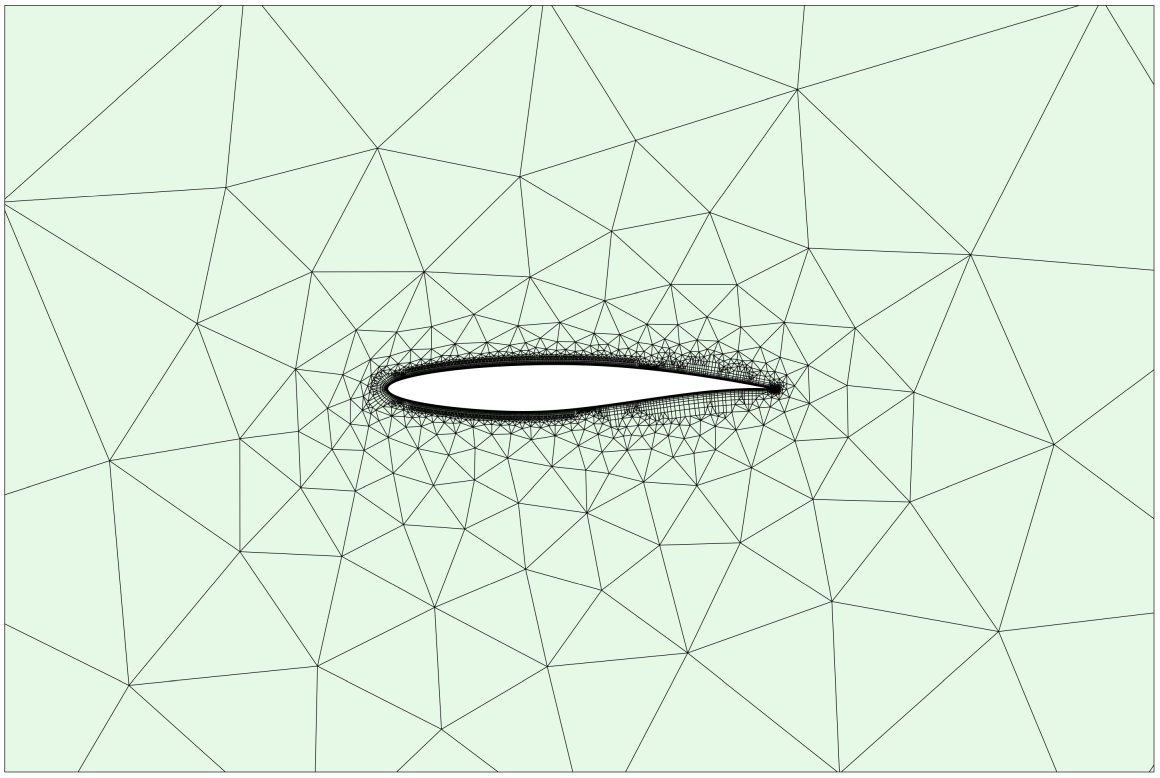}}
	\subfigure[Key variable: $\tilde{p}$]{\includegraphics[width=0.32\textwidth]{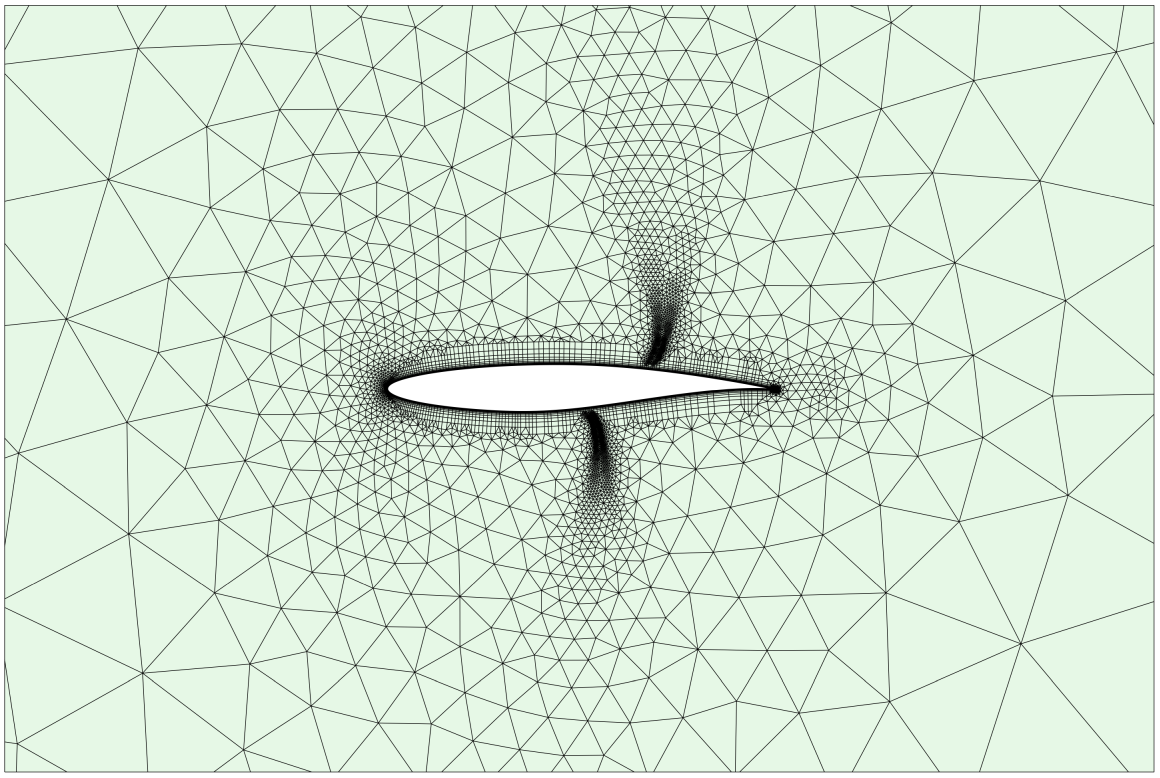}}	
	\caption{Meshes obtained with the minimum spacing provided by considering the pressure and the smoothed pressure as key variable.}
	\label{fig:meshSmoothingP}
\end{figure}

Figure~\ref{fig:meshSmoothingP} shows the meshes obtained by considering the pressure and the smoothed pressure as key variable, respectively.
It can be clearly observed that, without the proposed smoothing, the refinement is concentrated in the vicinity of the aerofoil, due to the gradients of the pressure in the normal direction. By smoothing the pressure, the variations in the normal direction do not play a relevant role and the spacing obtained provides the desired refinement in the critical zones such as leading and trailing edges and near the shocks.

\begin{Rk}
The meshes considered for compressible turbulent flows normally present a high level of stretching, not only in the inflation layer, but also downstream of the aerofoil trailing edge, as shown in Figure~\ref{fig:meshTrailing}. 
\begin{figure}[!tb]
\centering
\includegraphics[width=0.32\textwidth]{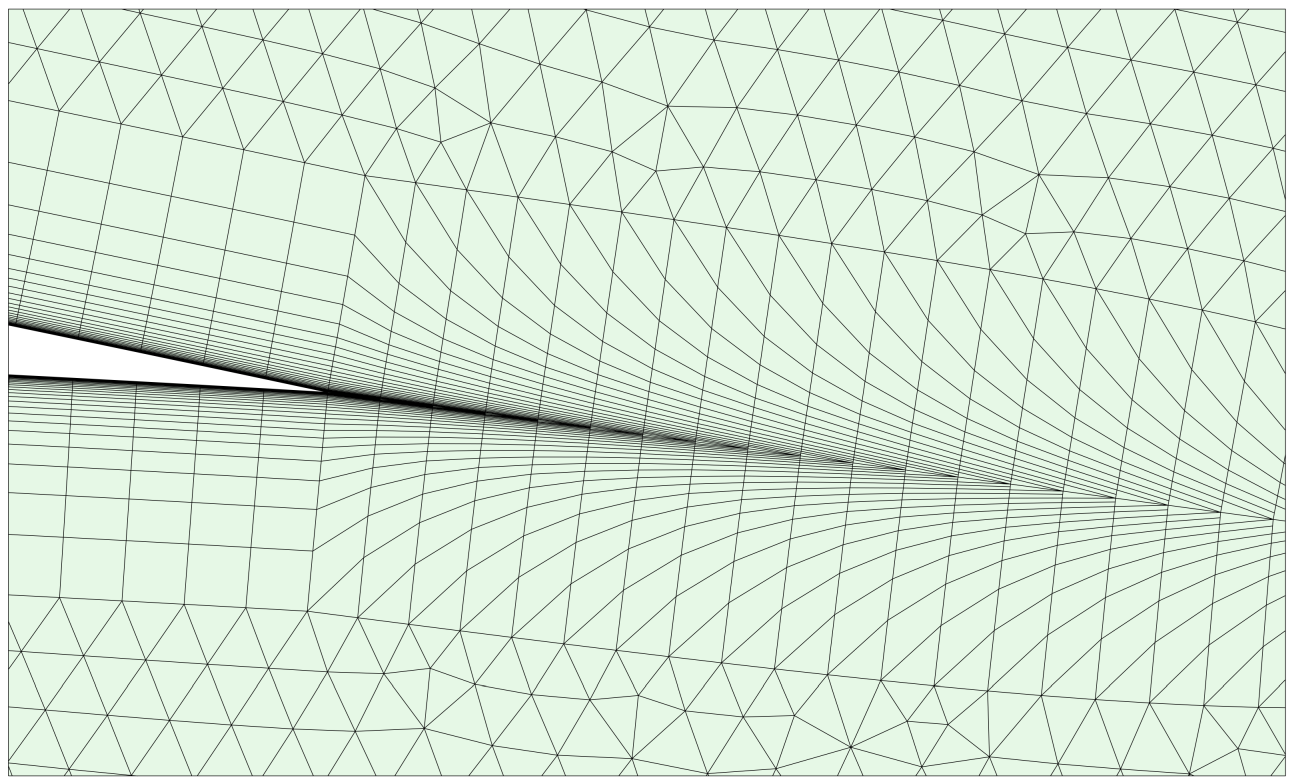}
\caption{Detail of an unstructured mesh around an aerofoil, showing the highly stretched elements downstream of the aerofoil trailing edge.}
\label{fig:meshTrailing}
\end{figure}
This is usually beneficial to capture the shear layer at low angles of attack, but it also results in very high gradients of the pressure, as in the vicinity of the wall. To avoid this region dominating the computed spacing, the smoothing of the pressure detailed in this section is also applied to the stretched elements in this region. 
\end{Rk}

\subsection{Computation of the target spacing in the background mesh} \label{sc:targetSpacingBack}

As the proposed strategy aims at training a feed-forward ANN to predict the target spacing, it is necessary to design an ANN architecture where the number of inputs and outputs is fixed. 

For problems involving a fixed geometry and variable operating conditions, the inputs will be the flow conditions (i.e., Mach number and angle of attack), whereas for problems involving variable geometries the inputs considered in this work are the control points of the NURBS curves describing the boundary. 

Considering the outputs as the spacing at each node of the computational mesh is not a viable solution for two main reasons. First, each computational mesh might have a different number of nodes. Second, even if all the computational meshes have the same number of nodes, this is expected to be a large number, because the solutions might have been computed in an over-refined mesh, leading to large ANN training times. The strategy considered here consists of defining a coarse background mesh and transferring the spacing function from the computational mesh to the background mesh. This will be done for every available training case, so that the ANN can be trained to predict the spacing at the nodes of the background mesh.

To describe the proposed approach, let us consider a background mesh $\mathcal{B}_h$ with elements $\{B_e\}_{e=1,\ldots,\nelB}$ and nodes $\{\bx_i^\texttt{B}\}_{i=1,\ldots,\nnoB}$. Similarly, let us consider a computational mesh, for one of the training cases, $\mathcal{C}_H$ with elements $\{\Omega_e\}_{e=1,\ldots,\nel}$ and nodes $\{\bx_i\}_{i=1,\ldots,\nno}$. In addition, a nodal field, corresponding to the nodal spacing, is defined in the computational mesh, namely $\{\delta_i\}_{i=1,\ldots,\nno}$. The objective is to transfer the discrete spacing $\delta$ from the computational mesh $\mathcal{C}_H$ to the background mesh $\mathcal{B}_h$.

The strategy  to transfer the spacing is adopted from~\cite{LockIMR2023} and it is briefly summarised here. For a given node of the background mesh $\bx_i^\texttt{B}$, the patch of elements containing this node is denoted by $\mathcal{P}_i^\texttt{B}$. The set of nodes of the computational mesh that belong to $\mathcal{P}_i^\texttt{B}$ is denoted by $\mathcal{X}_{\mathcal{P}_i}$. With this information, the spacing at the node $\bx_i^\texttt{B}$ of the background mesh is defined as
\begin{equation} \label{eq:conservativeIntep}
\delta_i^\texttt{B} = \min_{j \in \mathcal{X}_{\mathcal{P}_i}}\{\delta_j \}.
\end{equation}

\begin{figure}[!tb]
	\centering
	\subfigure[Reference solution]{\includegraphics[width=0.32\textwidth]{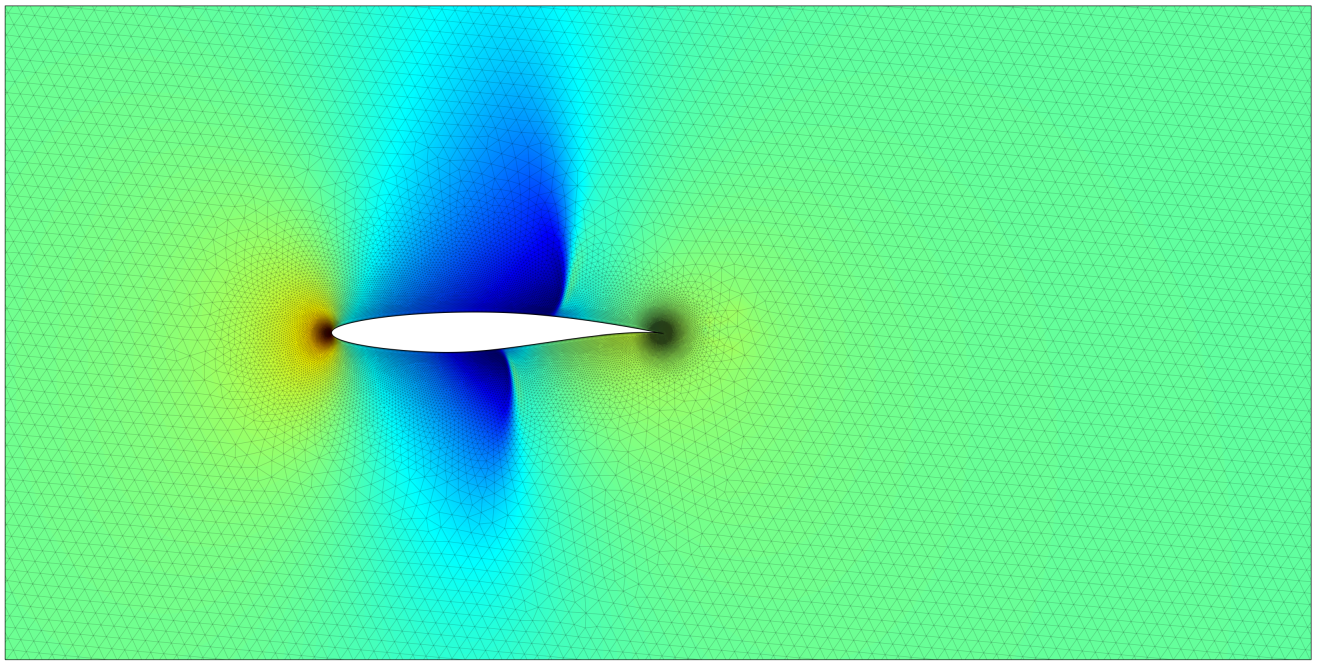}}
	\subfigure[Background mesh 1]{\includegraphics[width=0.32\textwidth]{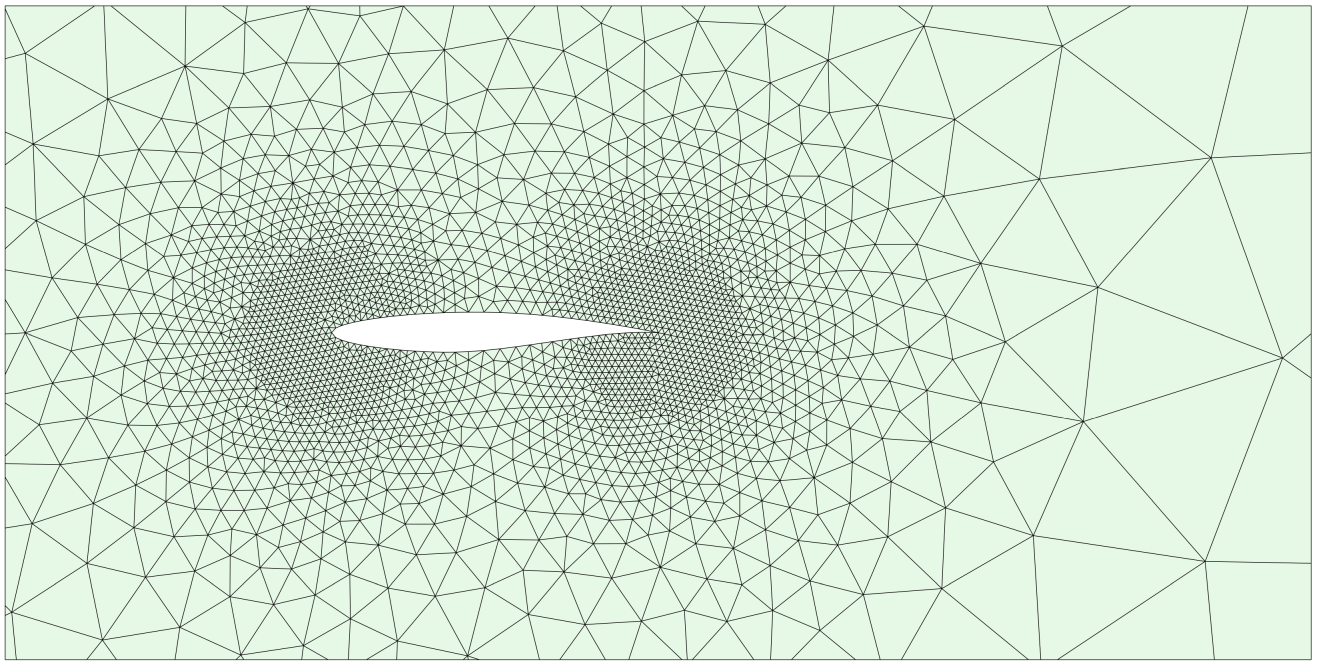}}	
	\subfigure[Background mesh 2]{\includegraphics[width=0.32\textwidth]{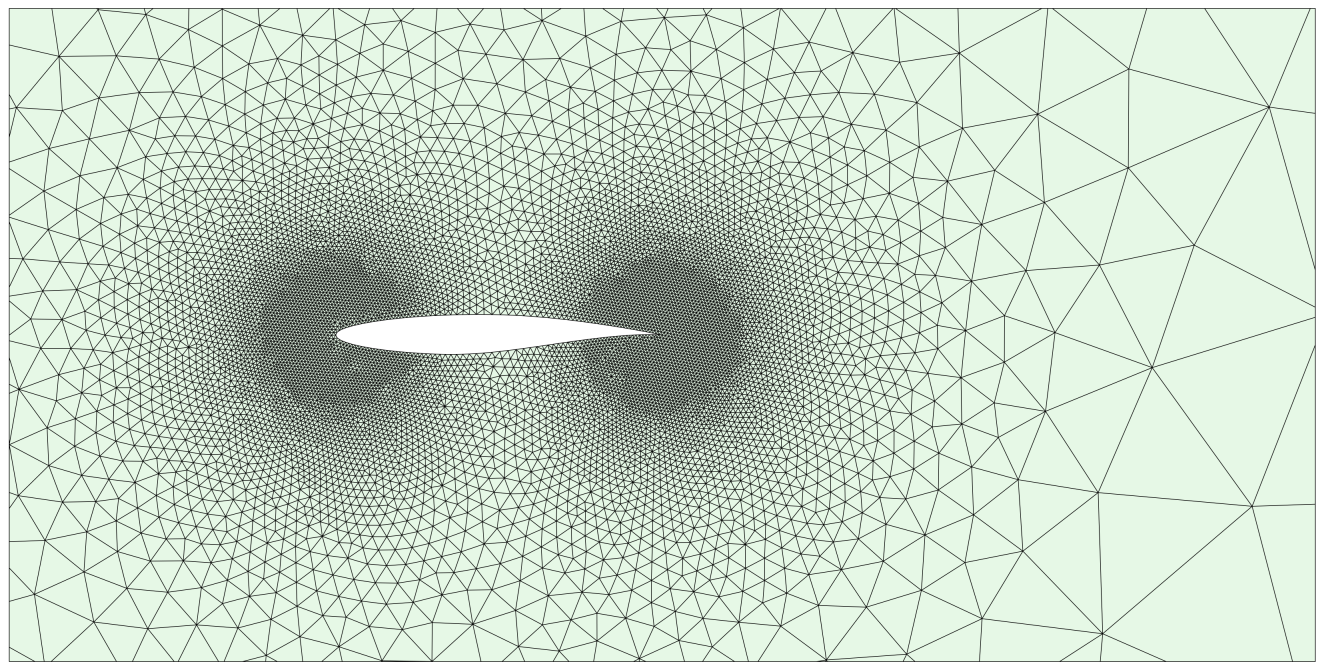}}	
	\caption{Solution on a reference mesh and two background meshes.}
	\label{fig:referenceAndBacs}
\end{figure}

To illustrate the procedure, let us consider the solution computed on a reference mesh shown in Figure~\ref{fig:referenceAndBacs}(a). Two background meshes, with different levels of refinement are utilised, as shown in Figures~\ref{fig:referenceAndBacs}(b) and ~\ref{fig:referenceAndBacs}(c).

\begin{figure}[!tb]
	\centering
	\subfigure[Reference mesh]{\includegraphics[width=0.32\textwidth]{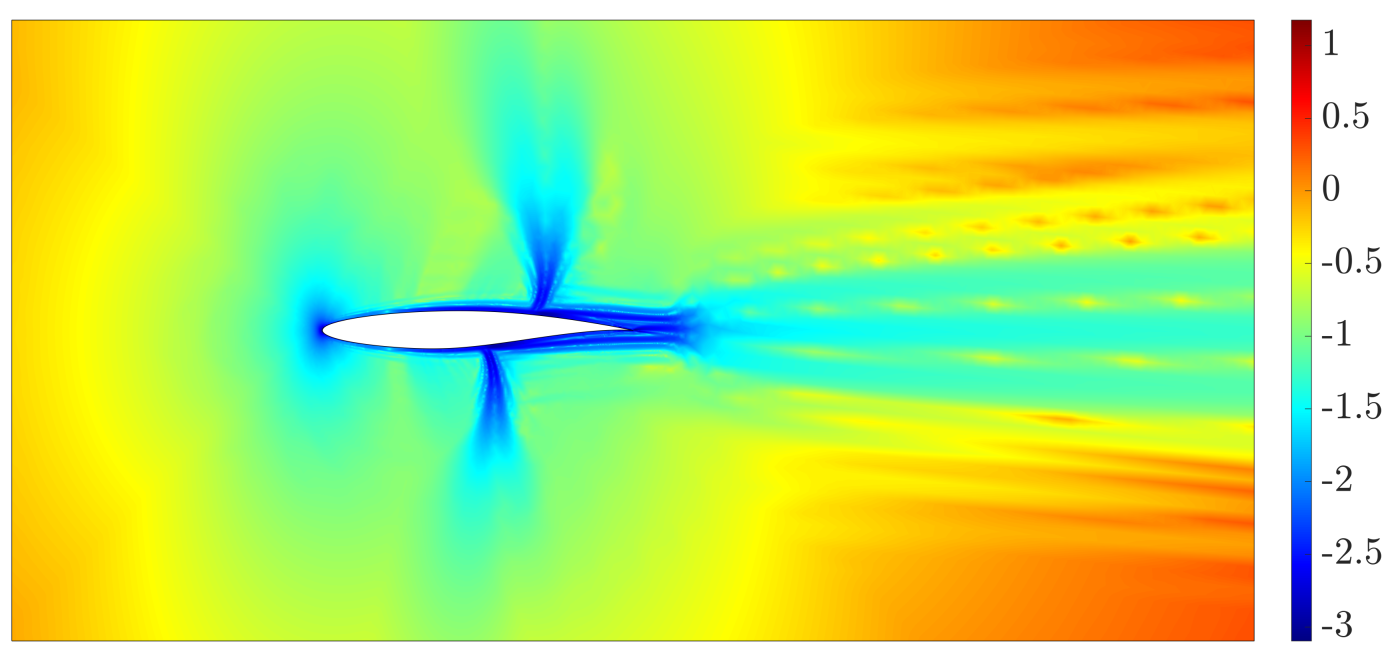}}
	\subfigure[Background mesh 1]{\includegraphics[width=0.32\textwidth]{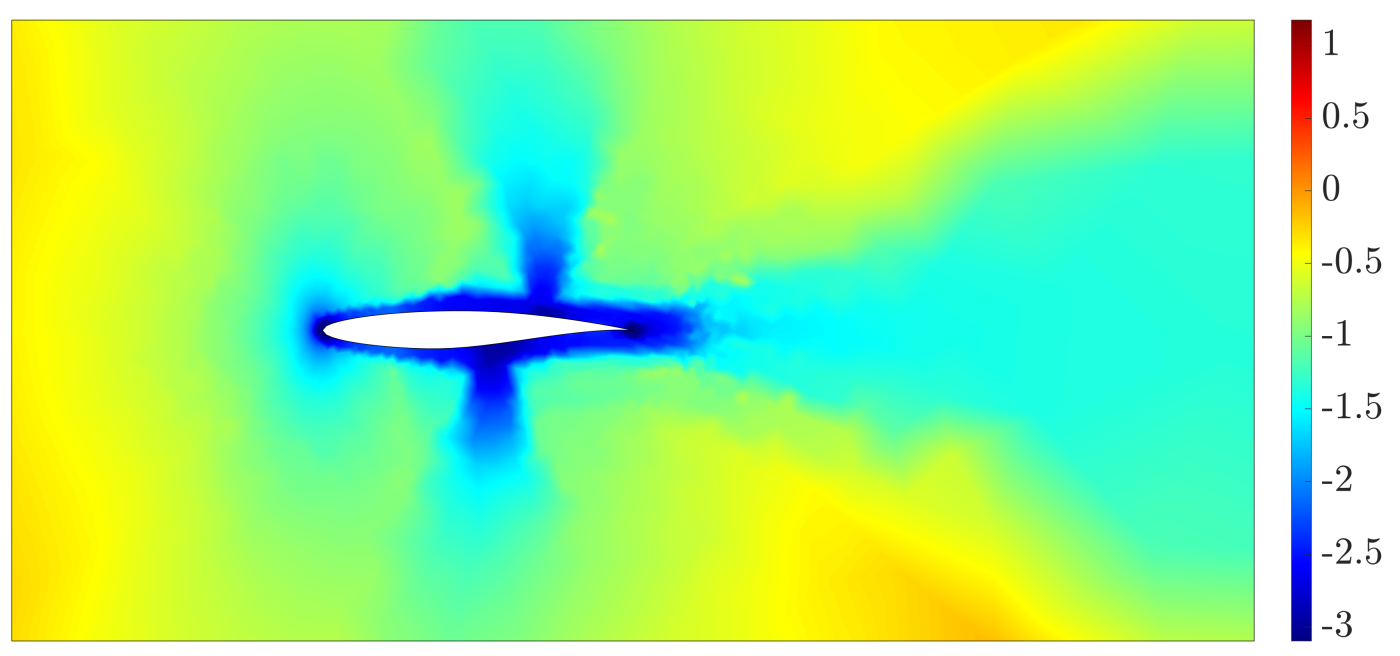}}	
	\subfigure[Background mesh 2]{\includegraphics[width=0.32\textwidth]{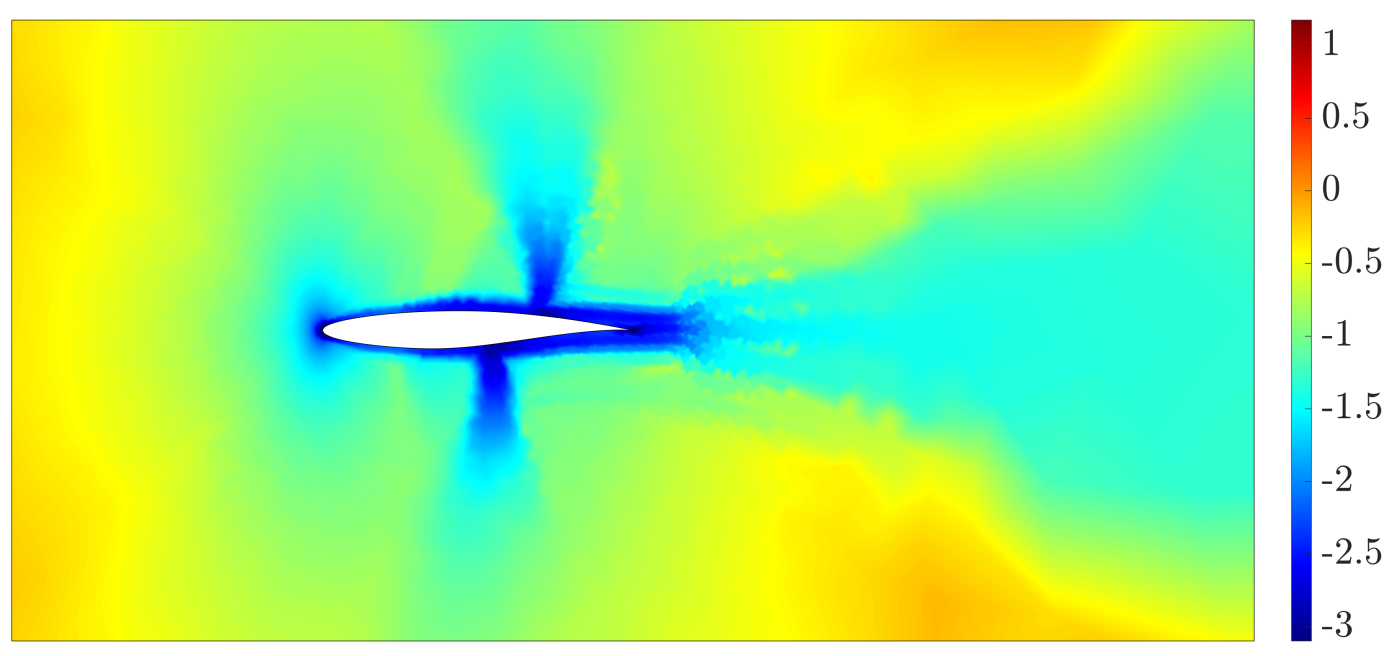}}	
	\caption{Spacing, in logarithmic scale, on the mesh where the solution is available and on the two background meshes of Figures~\ref{fig:referenceAndBacs}(b) and ~\ref{fig:referenceAndBacs}(c).}
	\label{fig:referenceAndBacsSpacing}
\end{figure}

First, the spacing is computed on the mesh where the reference solution is available, using the procedure described in the previous sections. The spacing computed using the smoothed pressure and the Mach number is shown in Figure~\ref{fig:referenceAndBacsSpacing}(a). Using the procedure described in this Section, the spacing is transferred to the two background meshes of Figures~\ref{fig:referenceAndBacs}(b) and ~\ref{fig:referenceAndBacs}(c) and depicted in Figures~\ref{fig:referenceAndBacsSpacing}(b) and ~\ref{fig:referenceAndBacsSpacing}(c), respectively.
It can be observed that transferring the spacing to a background mesh not only enables a consistent number of variables to be used in an ANN, but also introduces a desired smoothing in the discrete spacing function, when compared to the spacing function obtained in the reference mesh.

\begin{figure}[!tb]
	\centering
	\subfigure[Reference mesh]{\includegraphics[width=0.32\textwidth]{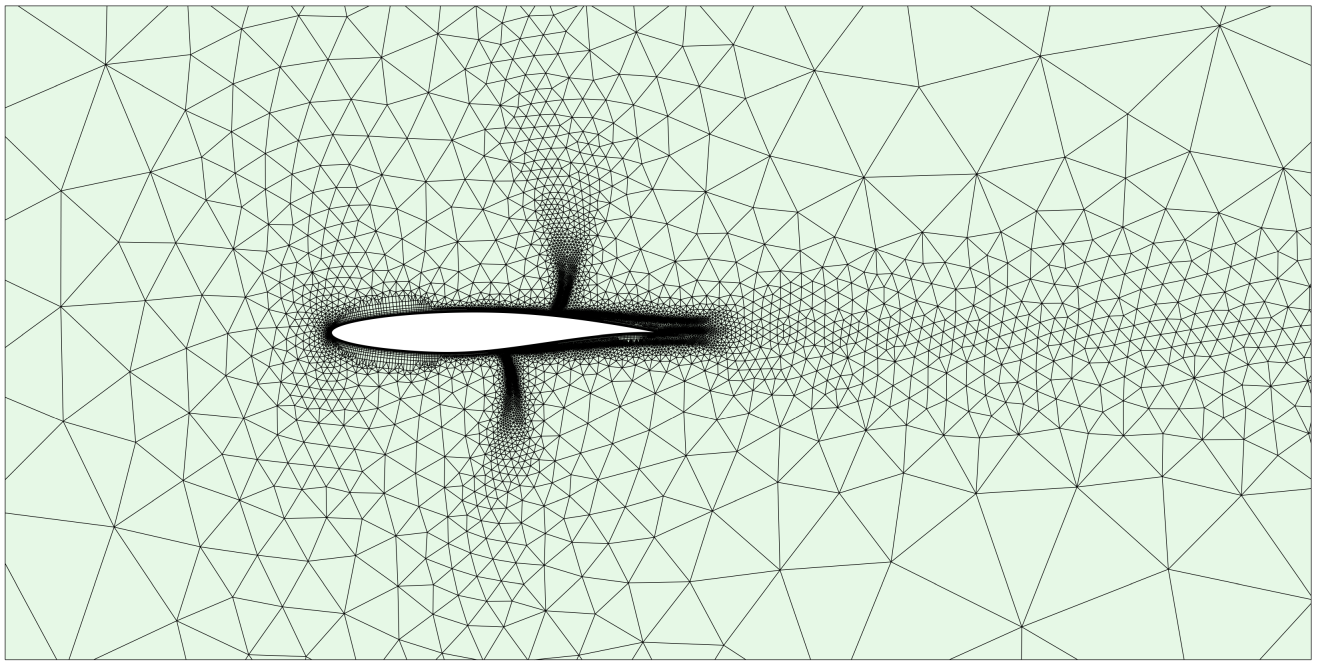}}
	\subfigure[Background mesh 1]{\includegraphics[width=0.32\textwidth]{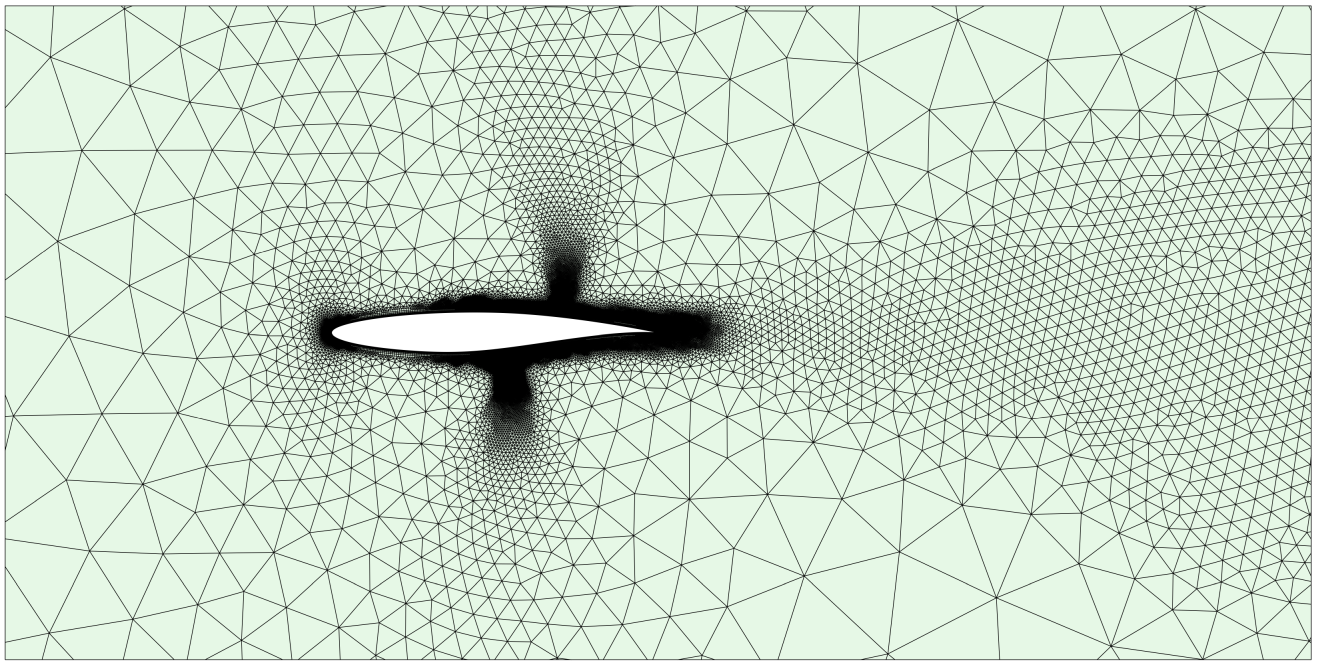}}	
	\subfigure[Background mesh 2]{\includegraphics[width=0.32\textwidth]{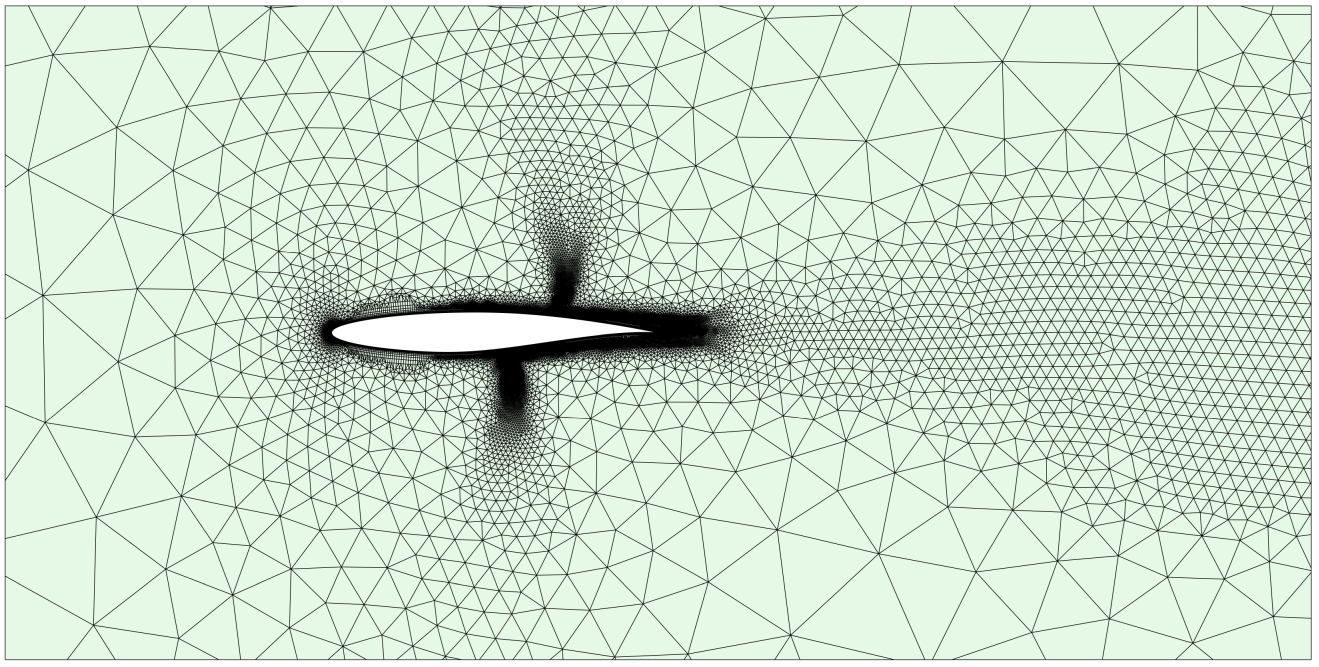}}	
	\caption{Meshes generated using the spacing function defined on a reference mesh and on the two background meshes of Figures~\ref{fig:referenceAndBacs}(b) and ~\ref{fig:referenceAndBacs}(c).}
	\label{fig:referenceAndBacsSpacingGen}
\end{figure}

Finally, utilising the spacing functions of Figure~\ref{fig:referenceAndBacsSpacing}, three meshes with the desired spacing are generated and displayed in Figure~\ref{fig:referenceAndBacsSpacingGen}.
The results show the similarity of the meshes obtained and also illustrate that for coarser background meshes the resulting generated meshes are finer. This is due to the conservative approach utilised to transfer the spacing from the reference mesh to a background mesh, given by Equation~\eqref{eq:conservativeIntep}. This conservative approach produces the desired outcome as it ensures that any local feature of the reference solution is captured by the spacing function defined in a coarser, background, mesh.

\begin{Rk}
Two special situations need to be carefully handled when using this approach. First, when considering domains with curved boundaries, it is possible to have several nodes of the computational mesh that do not belong to any element, and therefore any patch, of the background mesh. To ensure that the spacing of such nodes is not ignored, they are associated to the closest elements of the background mesh. Second, it is possible to encounter a patch associated to a node of the background mesh that does not contain any node of the computational mesh. The spacing associated to these \textit{isolated} nodes is taken as the minimum of the spacing of the nodes of the element of the computational mesh that contains the isolated node of the background mesh.
\end{Rk}

\section{Numerical results} \label{sc:examples}

Two numerical studies are considered. The first involves the prediction of near-optimal meshes for variable operating conditions, namely the free-stream Mach number, $M_\infty$, and the angle of attack, $\alpha$. The second study focuses on the prediction of near-optimal meshes for variable geometric configurations, parametrised using the control points of the NURBS that describe the geometry of an aerofoil. In all the examples, the Hessian is evaluated using the finite element recovery after splitting the quadrilateral elements into triangles, as described in Section~\ref{sc:hessian} and the spacing from a given solution is computed using the smoothed pressure and the Mach number as key variables, as described in Section~\ref{sc:BL}.

To perform the training and assess the accuracy, three different datasets are considered. The training set contains the cases used to optimise the weights and biases of the ANN. The validation set is used to avoid overfitting. Finally, the test set, which contains unseen cases is used, after training, to assess the accuracy of the predictions.

In all cases, the data required to train the ANN has been generated using the in-house vertex-centred finite volume solver FLITE~\cite{sorensen2003b}.

\subsection{Prediction of the spacing for variable operating conditions}

The first study considers a fixed geometric configuration, corresponding to the RAE2822 aerofoil, and it shows the ability to predict near-optimal meshes for different flow conditions. Numerical experiments are performed to illustrate the influence of the ANN hyperparameters and the size of the training set on the accuracy of the predictions.

\begin{figure}[!tb]
\centering
\subfigure[$M_\infty = 0.489, \alpha = 1.80^\circ$]{\includegraphics[width=0.32\textwidth]{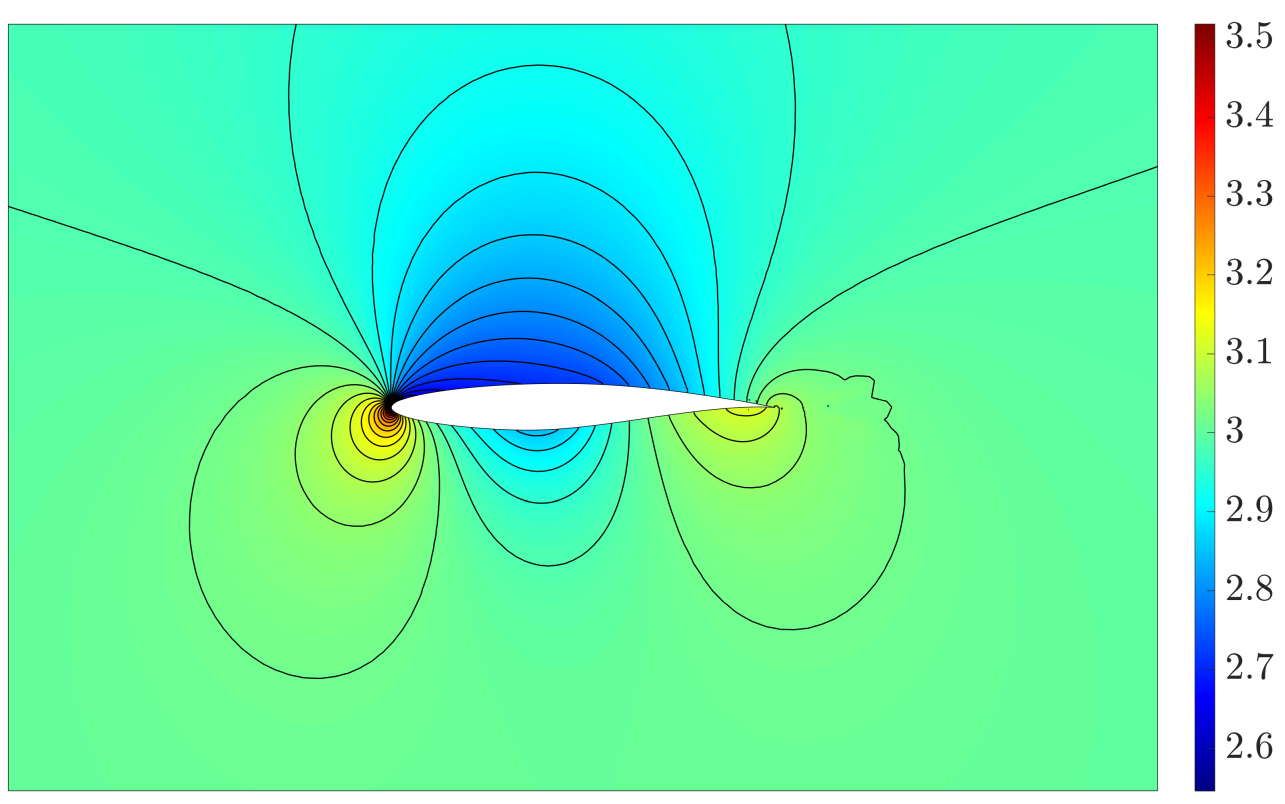}}
\subfigure[$M_\infty = 0.767, \alpha = 1.53^\circ$]{\includegraphics[width=0.32\textwidth]{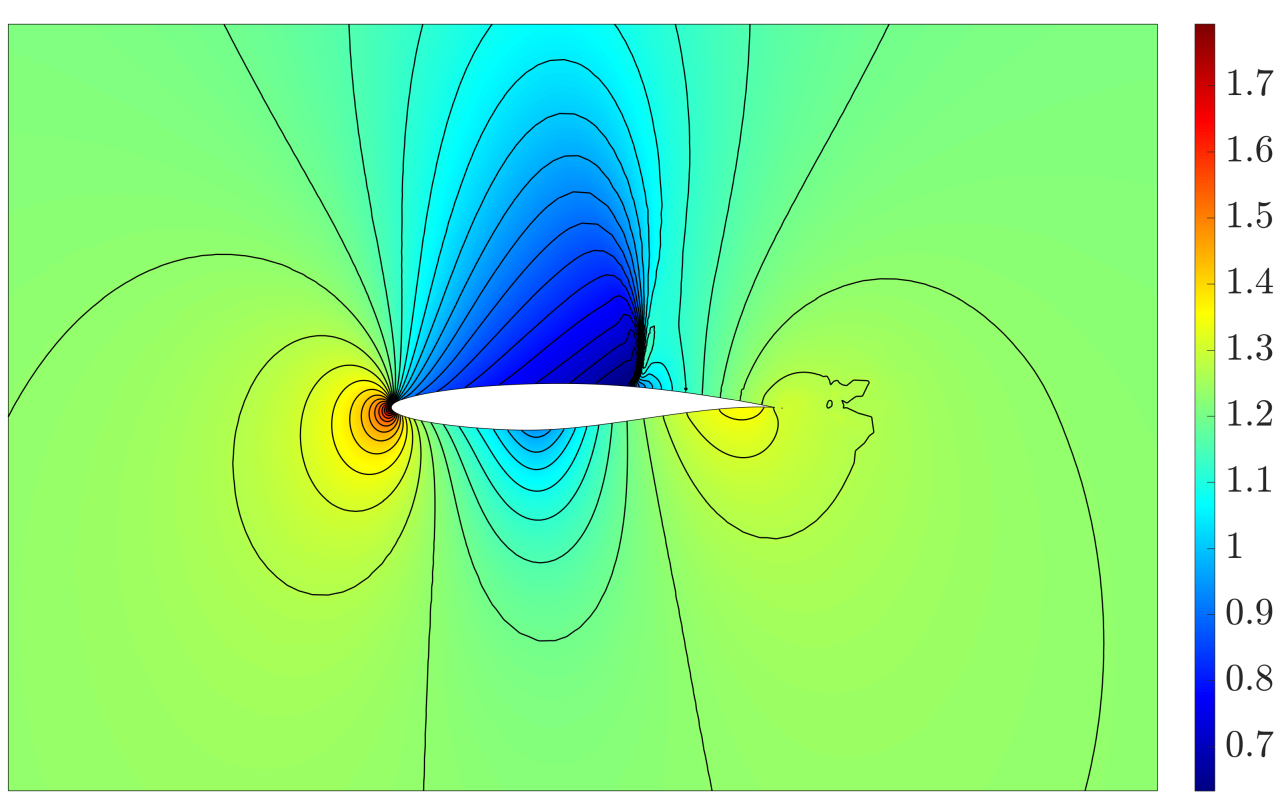}}	
\subfigure[$M_\infty = 0.792, \alpha = 0.35^\circ$]{\includegraphics[width=0.32\textwidth]{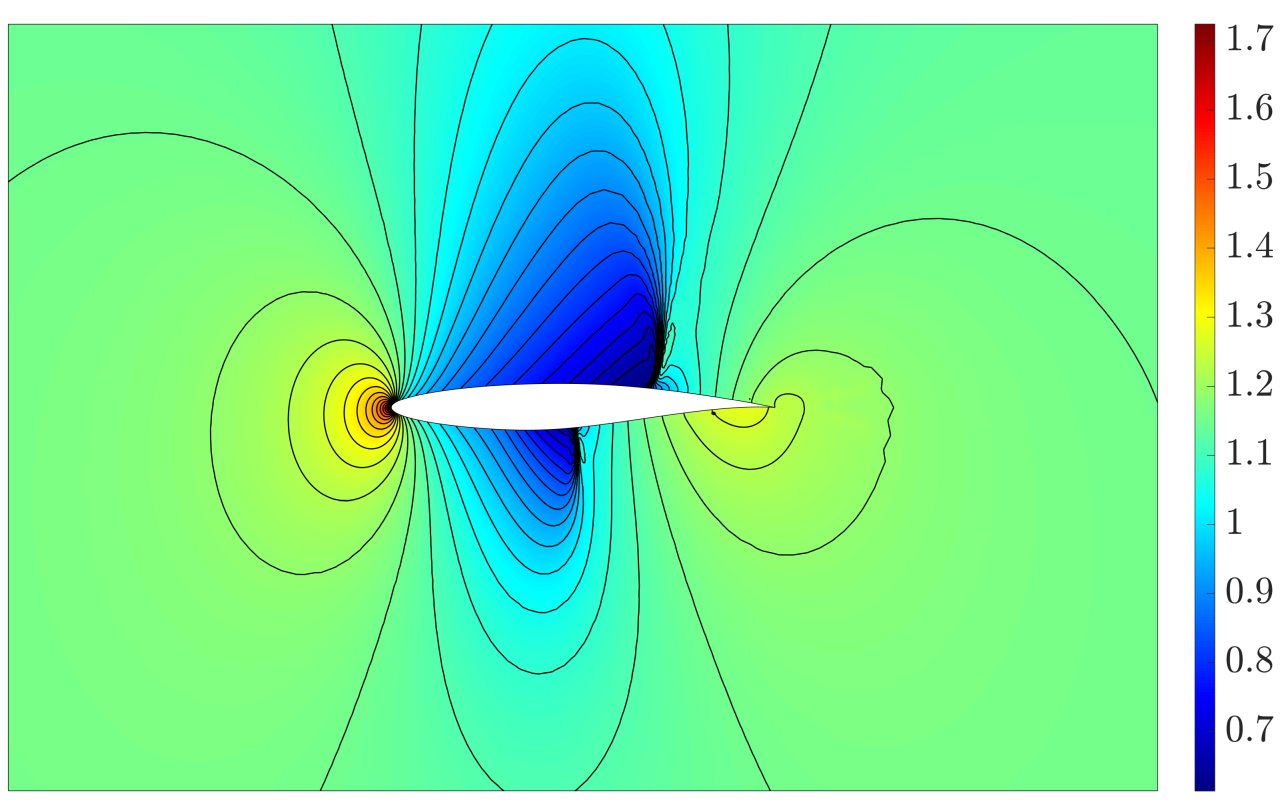}}	
\caption{Pressure field and isolines for three different flow conditions with $Re_\infty = 6.5 \times 10^6$.}
\label{fig:flowTestCases}
\end{figure}

The Reynolds number is fixed at $Re_\infty = 6.5 \times 10^6$ and the range of the flow parameters considered is $M_\infty \in [0.4,0.85]$ and $\alpha \in [0^\circ, 3^\circ]$. Therefore, all the cases involve turbulent compressible flows that span over the subsonic and transonic regimes. To illustrate the variation of the solution induced by the parameters, Figure~\ref{fig:flowTestCases} shows three solutions corresponding to three different flow conditions.
The first solution corresponds to a subsonic case, whereas the second and third solution display a transonic flow regime. The different solutions exhibit a completely different behaviour, one with no shocks, another with a strong shock on the upper surface of the aerofoil and another with one shock on the upper and one shock on the lower surface of the aerofoil. Clearly, the ideal mesh for each case is completely different, requiring refinement in different zones.

The data required to train the ANN has been generated using a fixed hybrid unstructured mesh with 59,397 nodes and 100,094 elements, with 18,042 elements being quadrilaterals in the inflation layer. The inflation layer has an initial element height of $5 \times 10^{-6}$ and a variable growing factor, namely 1.1 for the first three layers, 1.2 for the next 17 layers and 1.45 for the last 14 layers. In addition, point sources were introduced in the leading and trailing edges with a spacing of $8 \times 10^{-4}$. A constant background spacing of $4 \times 10^{-2}$ has been specified in a rectangular region of dimension $[5c, 3c]$, where $c$ is the chord length of the aerofoil. The rationale for designing the mesh follows the current industrial practice~\cite{MichalTetrahedron} that consists of generating an over-refined mesh to conduct parametric studies in order to avoid the time consuming process of generating a different mesh for each simulation, which, apart from being difficult, requires a significant level of human intervention~\cite{lock2023meshing}.

Four training sets with $N_{tr}= 20, 40, 80, 160$ training cases have been generated using the quasi-random Halton sequencing~\cite{halton1964algorithm, HaltonDescription}. To avoid over-fitting during the training process, four validation sets with $N_{val}=5,10,20,40$ cases are also generated using Halton sequencing. Finally, to assess the accuracy of the predictions a fixed set of $N_{tst} = 80$ test cases is considered, also generated using the Halton sequencing. To ensure that no extrapolation is performed, the range of the parameters is slightly reduced in the validation and tests sets, namely 3\% and 6\% reduction, on both ends, for the validation and test sets, respectively.

\begin{Rk}
When data is already available, as expected in an industrial environment, the use of cases previously executed by engineers is expected to provide a better coverage of the design space. This is because when sampling the design space with quasi-random Halton sequencing, no attention is paid to the fact that the region of the design space leading to subsonic cases requires less sampling than the area that corresponds to transonic cases. Therefore, the results presented here are expected to be conservative and even better predictions are expected when the available data has been generated by an expert engineer.
\end{Rk}

\begin{figure}[!tb]
	\centering
	\subfigure[Coarse background mesh]{\includegraphics[width=0.32\textwidth]{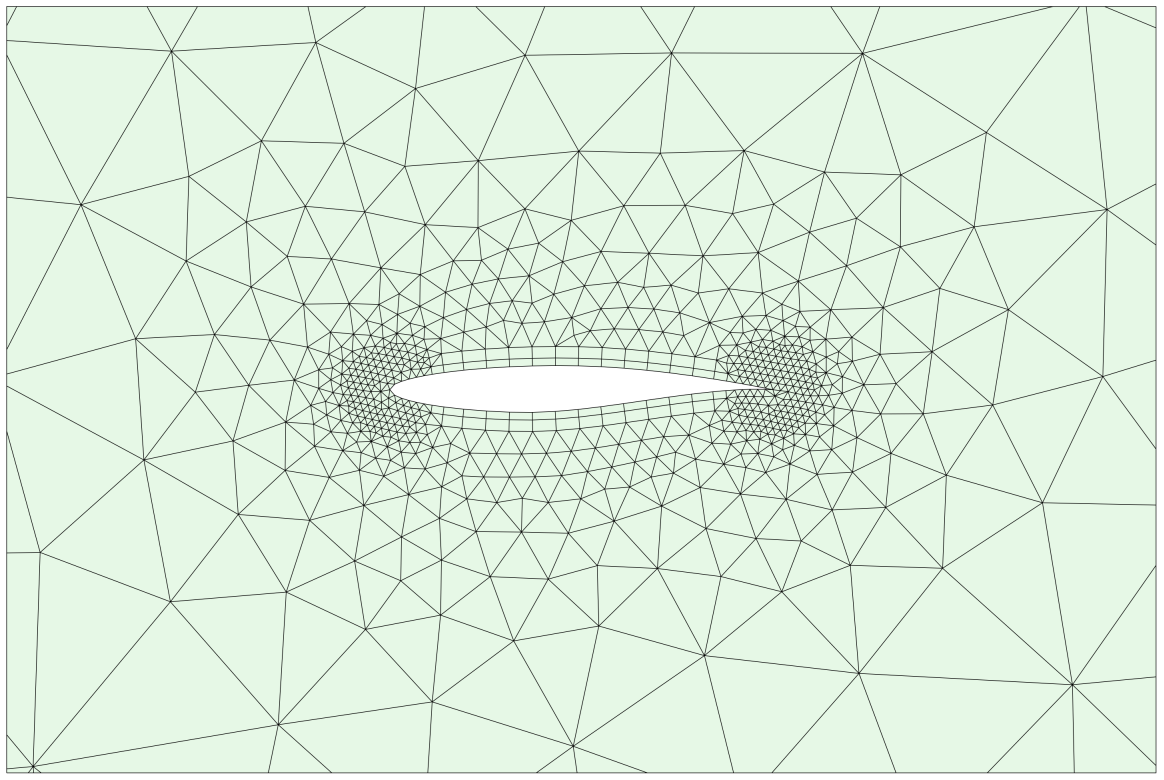}}
	\subfigure[Fine background mesh]{\includegraphics[width=0.32\textwidth]{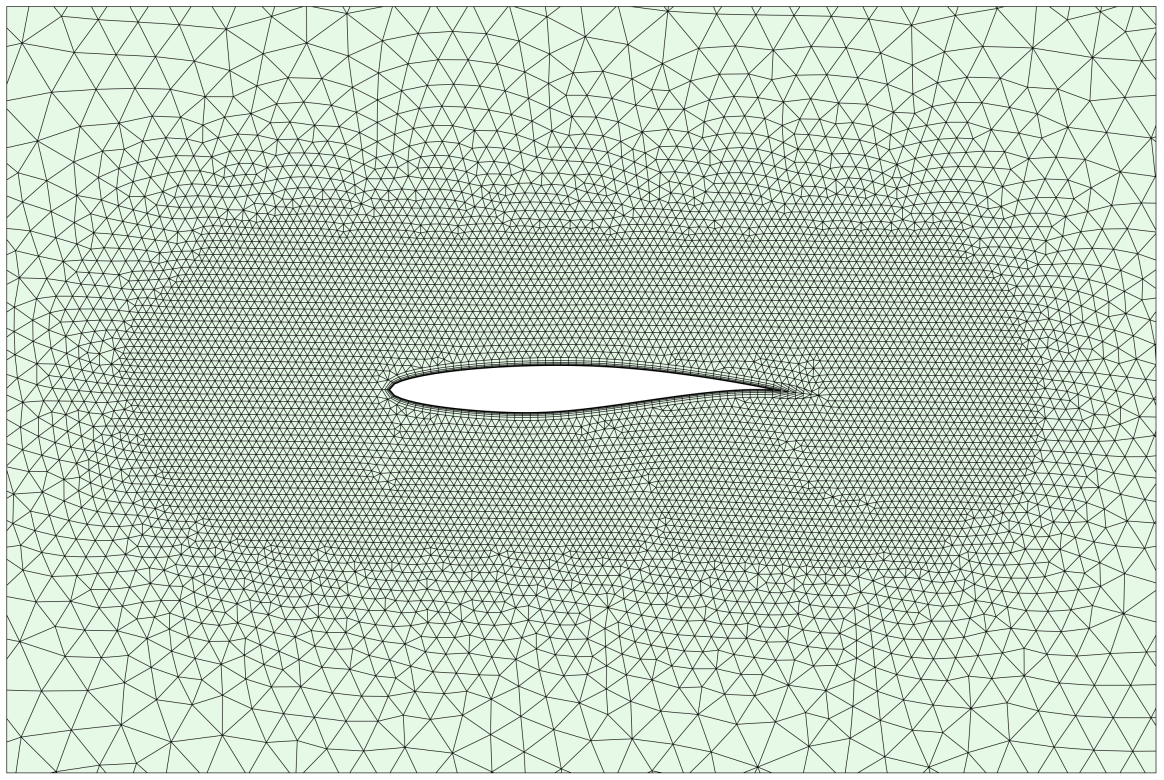}}
	\caption{Two background meshes used to transfer the spacing from the computational mesh.}
	\label{fig:flowBacMeshes}
\end{figure}

To assess the influence of the background mesh, two different background meshes are considered in this study. These are depicted in Figure~\ref{fig:flowBacMeshes}.
The first background mesh has 976 nodes, 1,780 triangular elements and 52 quadrilateral elements in the inflation layer. Local refinement has been defined by introducing two points sources, in the leading and trailing edges, and a line source connecting the leading and trailing edge. The second background mesh has 9,139 nodes, 18,076 triangular elements and 642 quadrilateral elements in the inflation layer. Local refinement has been defined by introducing a line source connecting the leading and trailing edge. It is worth noting that the anisotropic refinement near the wall in the background meshes is not designed to capture the flow physics, but to capture the expected large variation of the spacing near the aerofoil without refining isotropically.

To study the effect of the ANN hyperparameters and to compare the accuracy of the spacing prediction with the two different background meshes, a hyperparameter tuning is first performed using a simple grid search. The accuracy of the predictions is measured using the coefficient of determination, or R$^2$, defined as
\begin{equation*}
\text{R}^2 = 1 -  \frac{\displaystyle\sum_{k=1}^{N_{tr}} \sum_{i=1}^M ( y^k_i - h_i^k) ^2}{\displaystyle \sum_{k=1}^{N_{tr}} \sum_{i=1}^M ( y^k_i - \bar{y}) ^2} .
\end{equation*}
where $y^k_i$ are the target values, $h_i^k$ the predicted values and $\bar{y}$ is the mean of the target values.

\begin{figure}[!tb] 
\centering
\subfigure[Coarse background mesh]{\includegraphics[width=0.45\textwidth]{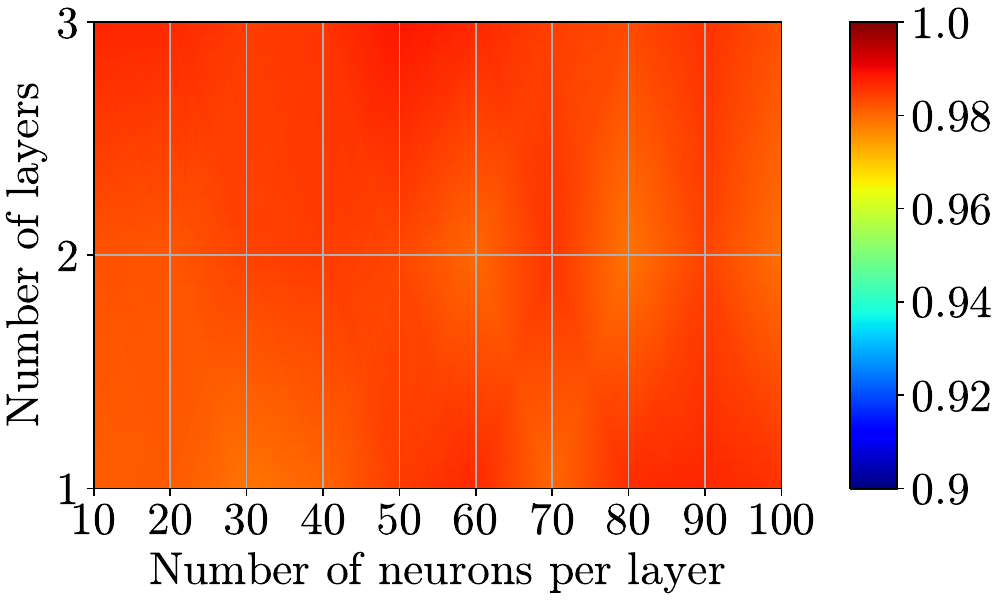}}
\subfigure[Fine background mesh]{\includegraphics[width=0.45\textwidth]{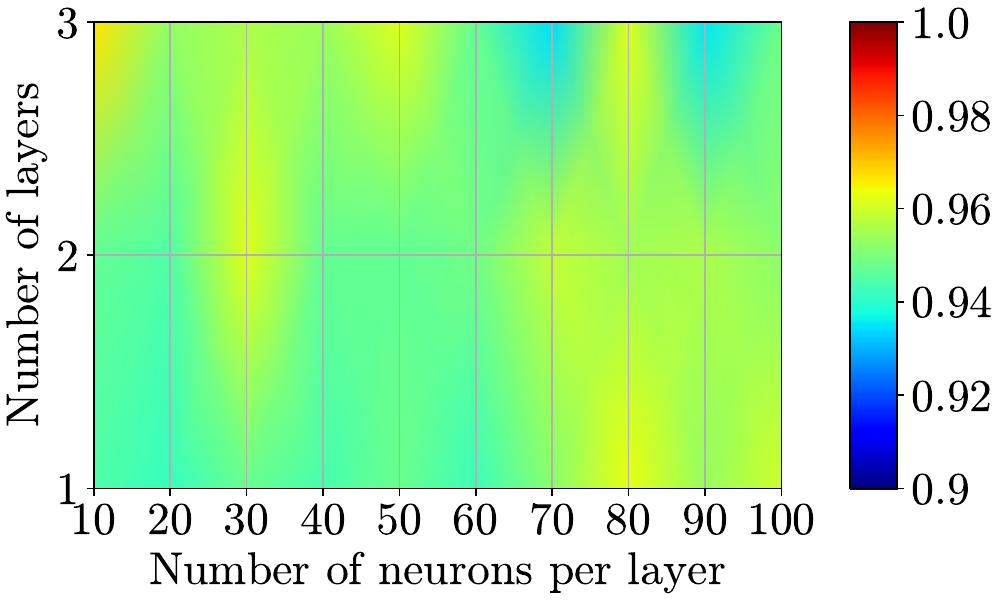}}
\caption{R$^{2}$ for the two different background meshes as a function of the number of layers and number of neurons in each layer with $N_{Tr} = 40$ training cases.}
\label{fig:flowHyperparameters} 
\end{figure}

Figure~\ref{fig:flowHyperparameters} shows the variation of the R$^{2}$ as a function of the number of layers and the number of neurons in each layer when using $N_{Tr} = 40$ training cases. 
The hyperparameter study shows that there is very little influence of the ANN architecture on the R$^{2}$, which means that a shallow ANN with only one hidden layer and 10 neurons is enough to produce accurate predictions. 

\begin{figure}[!tb]
\centering
\includegraphics[width=0.5\textwidth]{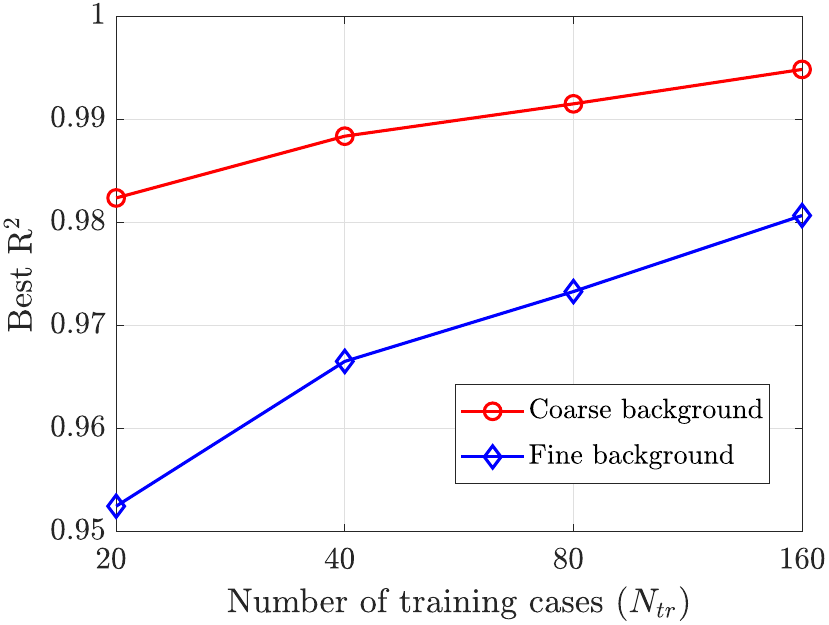}	
\caption{Best R$^2$ for the as a function of the number of training cases for the two different background meshes of Figure~\ref{fig:flowBacMeshes}.}
\label{fig:flowR2coarseFine}
\end{figure}

Using the best ANN architecture for each number of training cases, and for the two different background meshes considered, Figure~\ref{fig:flowR2coarseFine} shows the best R$^2$ as a function of the number of training cases. 
The results show that, even for a very low number of training cases, i.e., $N_{tr}=20$, the minimum R$^2$ is above 0.95 for both background meshes, with a value as high as 0.98 for the coarse background mesh. In addition, it can be clearly observed that a better R$^2$ can be obtained if the coarse background mesh is used. In fact, it is worth noting that to achieve a value of R$^2$ of around 0.98, only 20 training cases are required when the coarse background mesh is used, compared to 160 training cases when the fine background mesh is employed. This behaviour is mainly due to the increased number of outputs in the ANN when the fine background mesh is employed. In this example, the number of outputs in the ANN is 976 when using the coarse background mesh and 9,139, almost ten times higher, when using the fine background mesh. This obviously induces a higher training time for the ANN with a fine background mesh. On average, the training time for the ANN that employs the coarse background mesh is five minutes, whereas the ANN that employs the fine background mesh requires, on average, seven minutes for training.
As discussed in Section~\ref{sc:targetSpacingBack}, and illustrated in Figures~\ref{fig:referenceAndBacsSpacing} and \ref{fig:referenceAndBacsSpacingGen}, using a finer background mesh is expected to predict near-optimal meshes with less elements and a more localised refinement in the areas of interest, however this has to be balanced with the increased training time that induces using a finer background mesh.

\begin{figure}[!tb]
	\centering
	\subfigure[$M_\infty = 0.489, \alpha = 1.80^\circ$]{\includegraphics[width=0.32\textwidth]{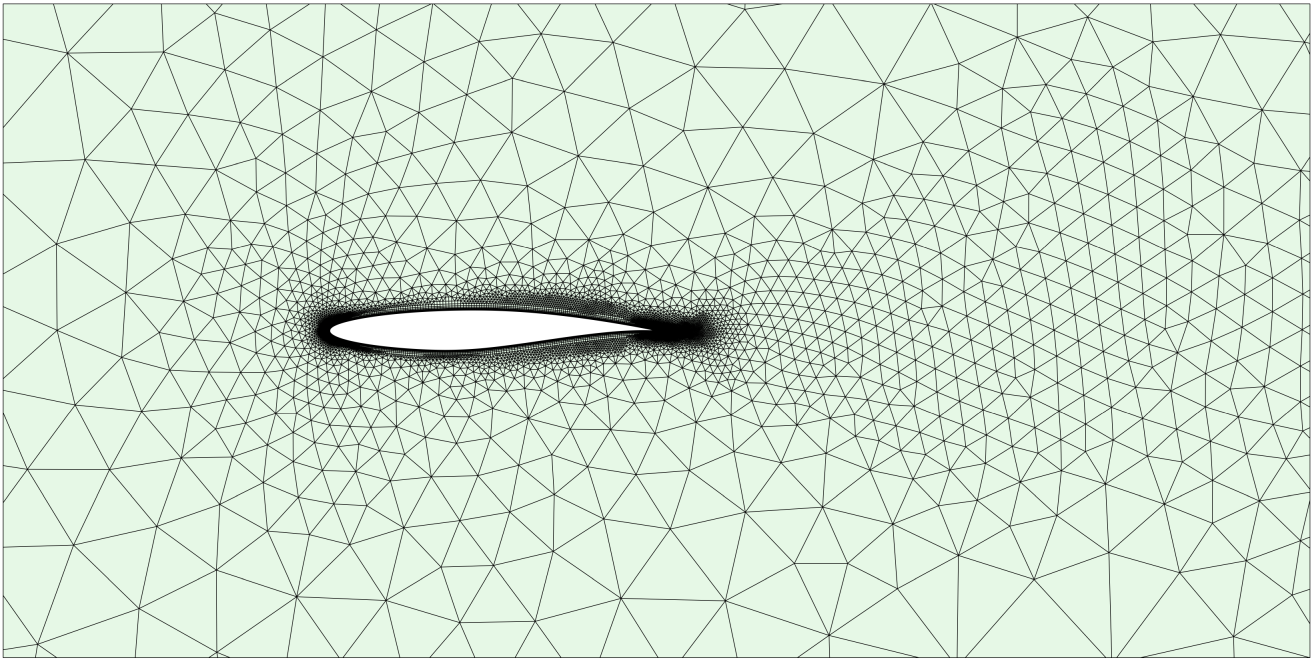}}
	\subfigure[$M_\infty = 0.767, \alpha = 1.53^\circ$]{\includegraphics[width=0.32\textwidth]{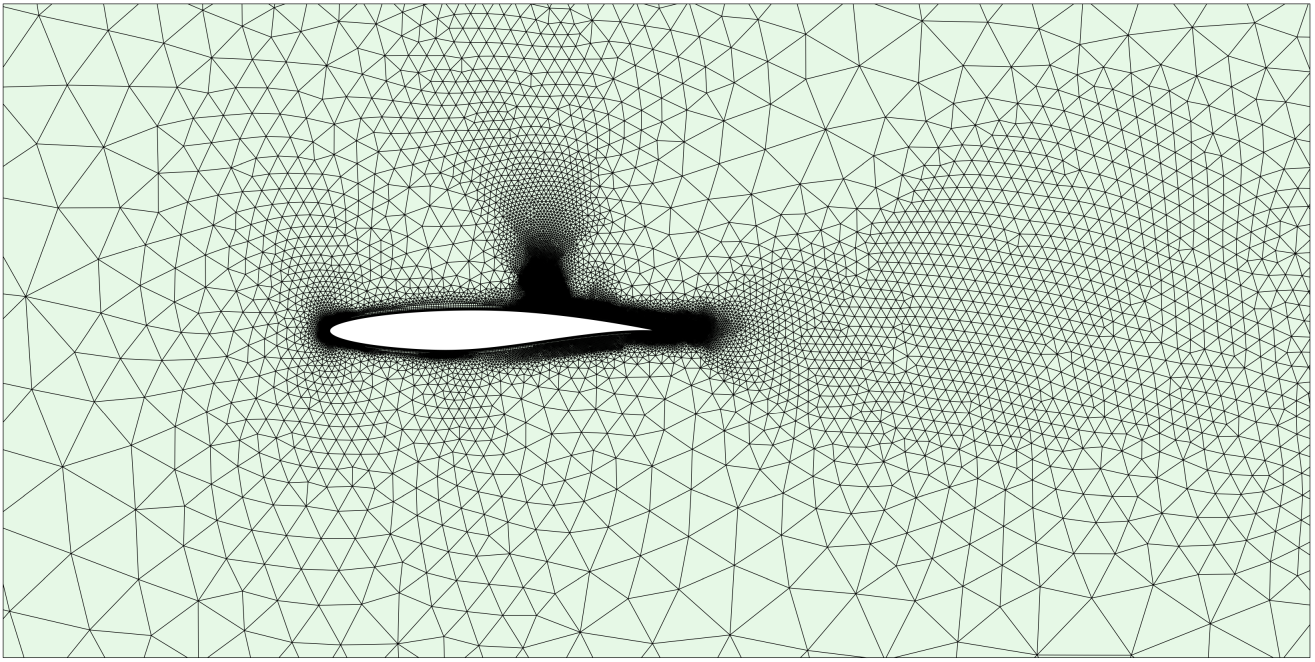}}	
	\subfigure[$M_\infty = 0.792, \alpha = 0.35^\circ$]{\includegraphics[width=0.32\textwidth]{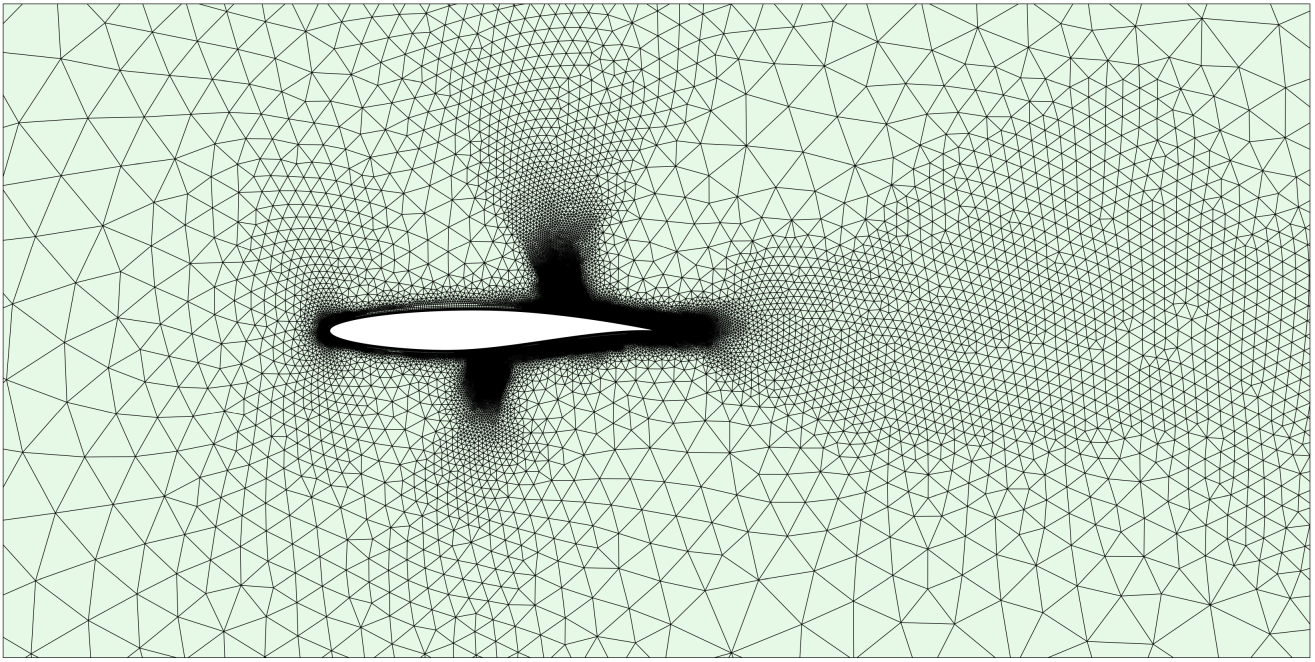}}	
	\subfigure[$M_\infty = 0.489, \alpha = 1.80^\circ$]{\includegraphics[width=0.32\textwidth]{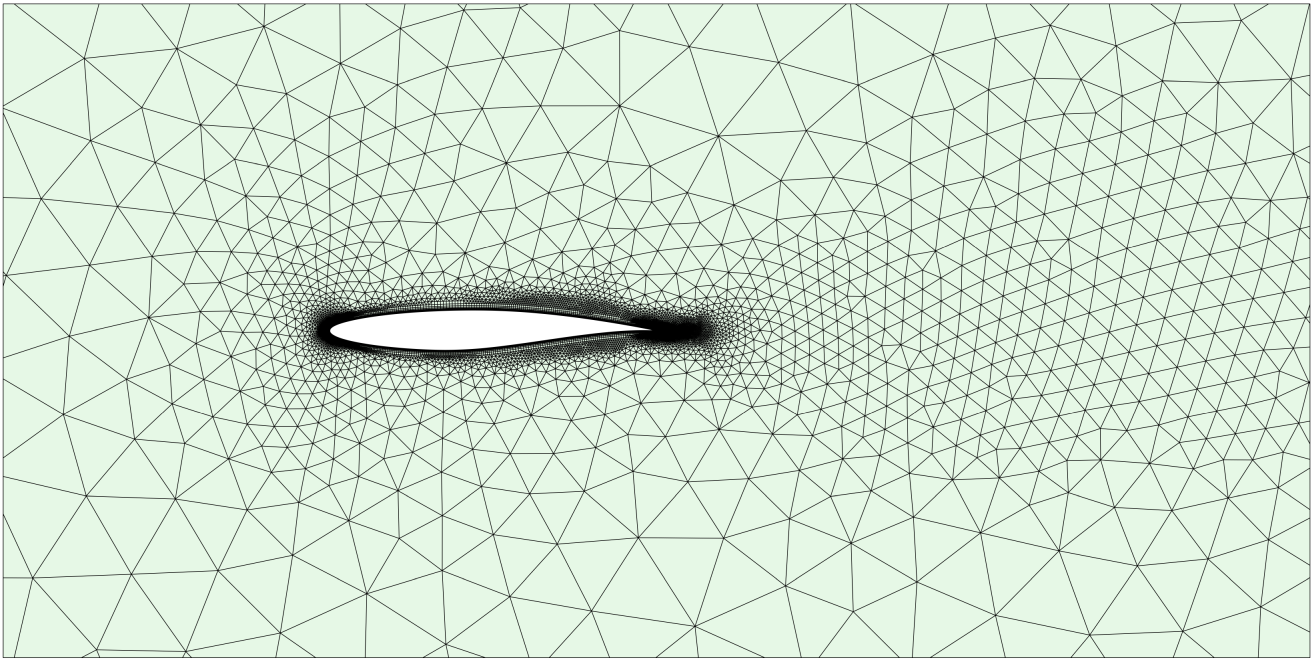}}
	\subfigure[$M_\infty = 0.767, \alpha = 1.53^\circ$]{\includegraphics[width=0.32\textwidth]{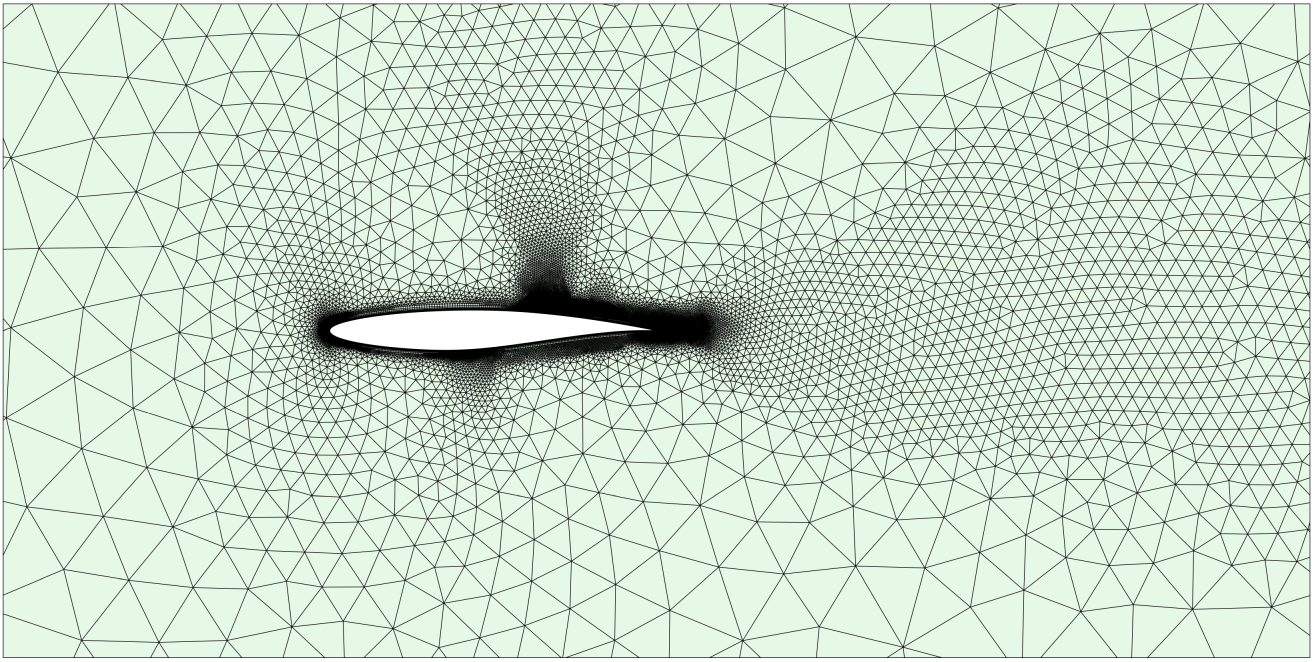}}	
	\subfigure[$M_\infty = 0.792, \alpha = 0.35^\circ$]{\includegraphics[width=0.32\textwidth]{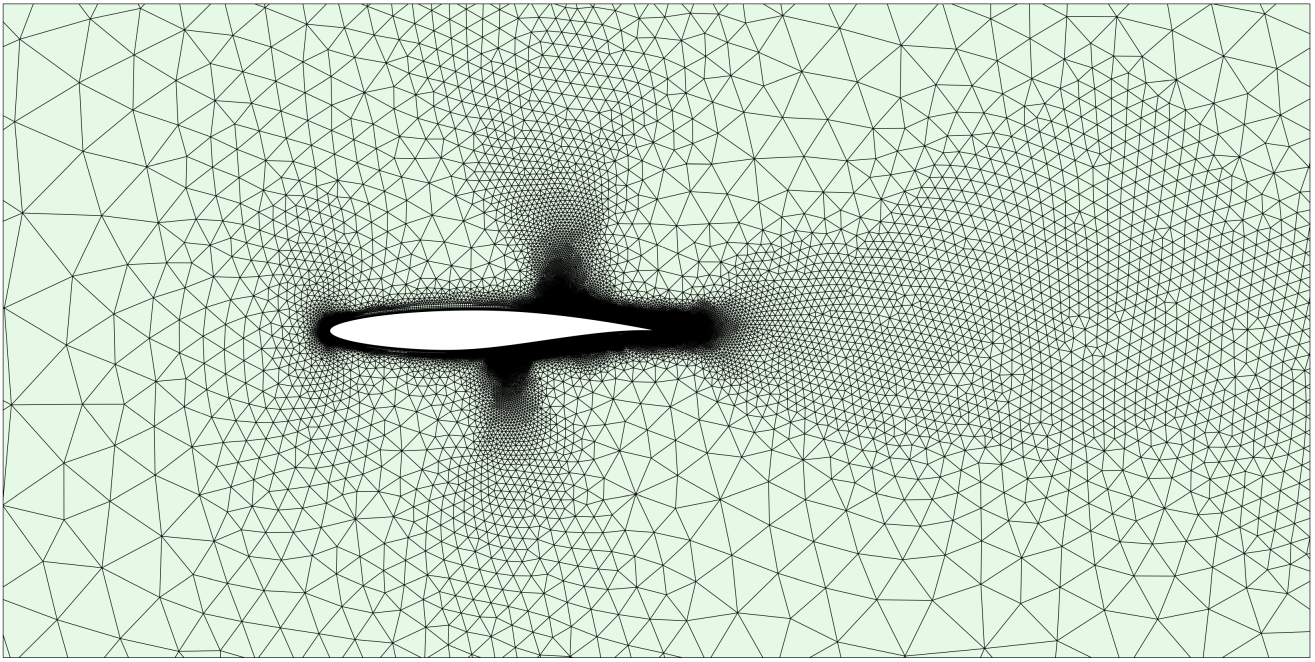}}	
	\caption{Target (top) and predicted (bottom) meshes for three unseen cases during training, corresponding to solutions of Figure~\ref{fig:flowTestCases} employing the coarse background mesh of Figure~\ref{fig:flowBacMeshes}(a).}
	\label{fig:flowMeshPredictionBac1}
\end{figure}

To visually assess the accuracy of the predictions, Figure~\ref{fig:flowMeshPredictionBac1} shows the target and predicted meshes for three unseen cases during training, corresponding to solutions of Figure~\ref{fig:flowTestCases}. The coarse background mesh of Figure~\ref{fig:flowBacMeshes}(a) has been used to define the target and predicted discrete spacing function and the ANN has been trained using 40 training cases.  
The results clearly show the ability of the trained ANN to predict the spacing with localised refinement in the areas of interest. For the subsonic case, only a local refinement near the aerofoil is predicted, whereas for the transonic cases an extra local refinement is produced in the vicinity of the shocks.

\begin{figure}[!tb]
	\centering
	\subfigure[$M_\infty = 0.489, \alpha = 1.80^\circ$]{\includegraphics[width=0.32\textwidth]{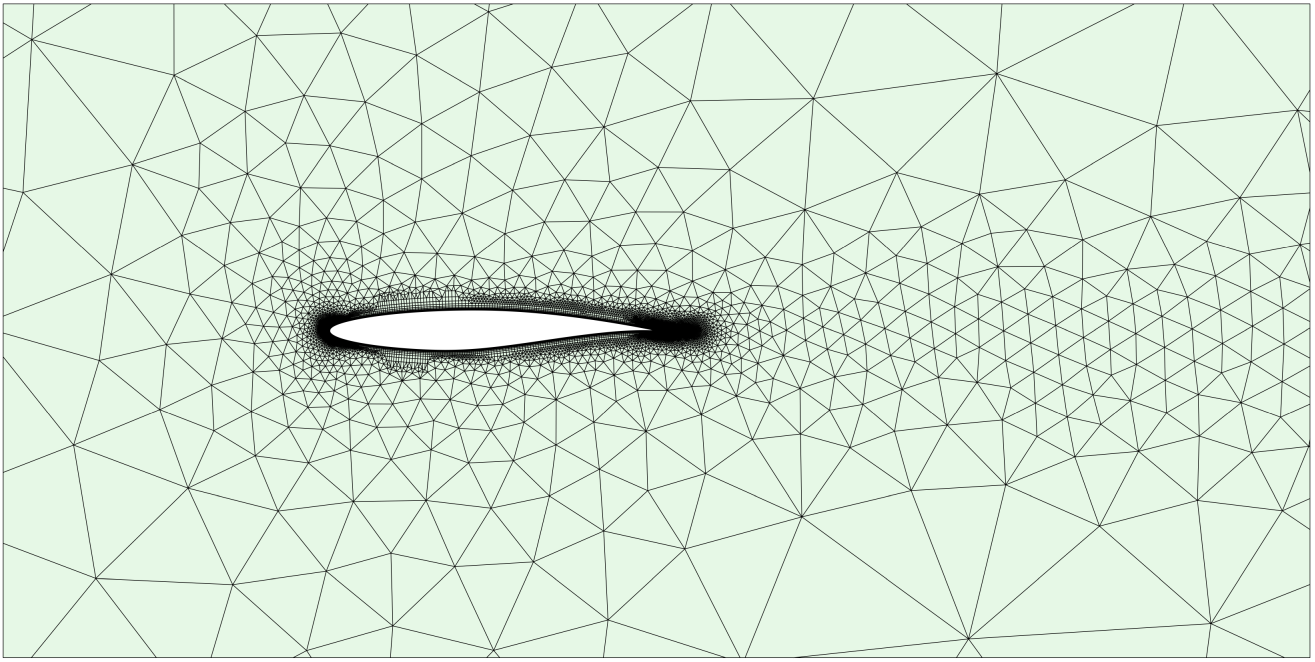}}
	\subfigure[$M_\infty = 0.767, \alpha = 1.53^\circ$]{\includegraphics[width=0.32\textwidth]{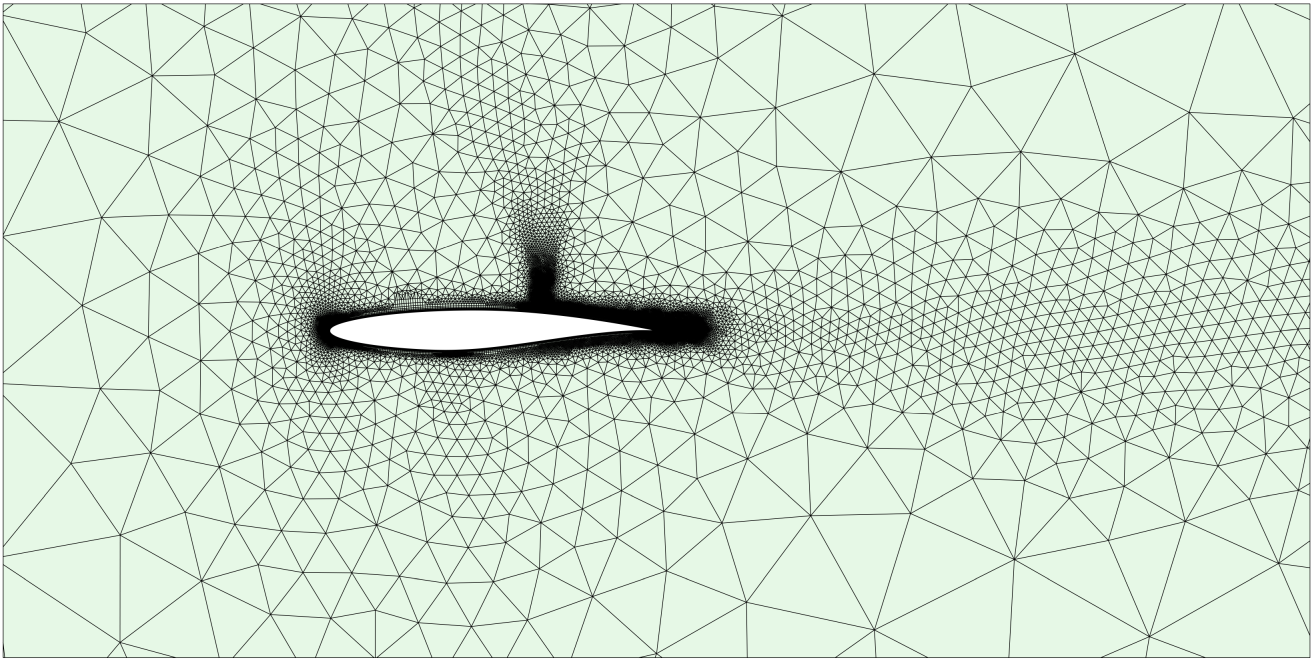}}	
	\subfigure[$M_\infty = 0.792, \alpha = 0.35^\circ$]{\includegraphics[width=0.32\textwidth]{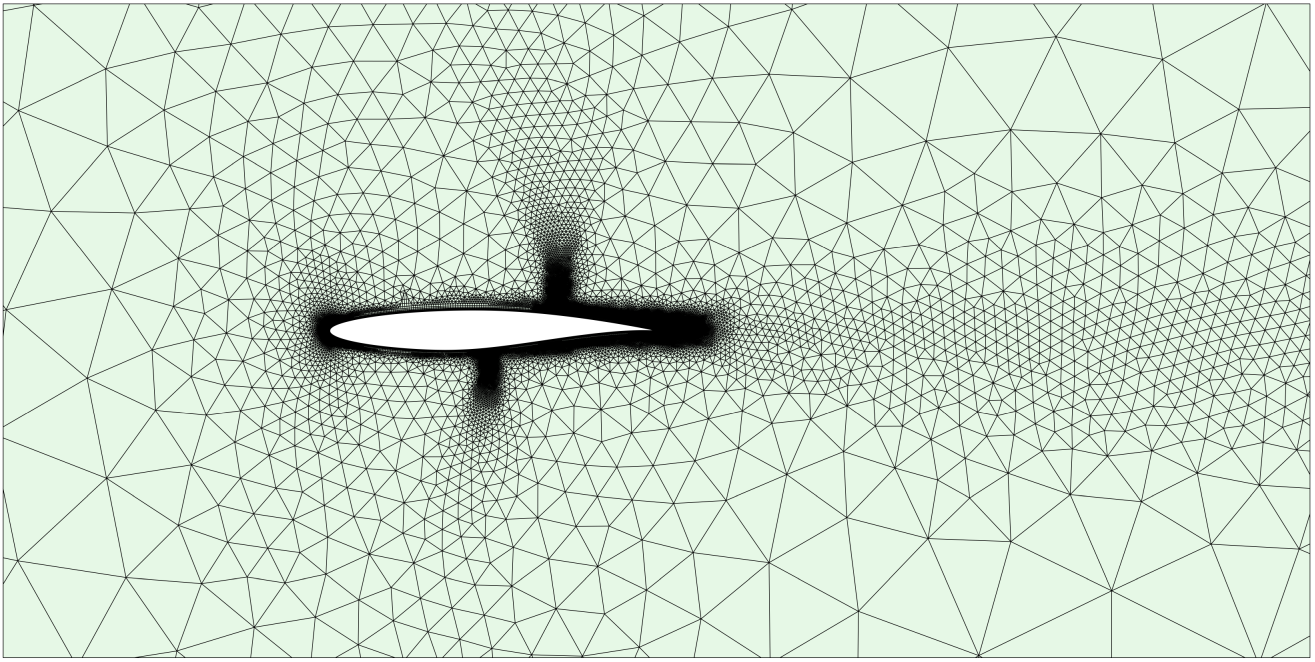}}	
	\subfigure[$M_\infty = 0.489, \alpha = 1.80^\circ$]{\includegraphics[width=0.32\textwidth]{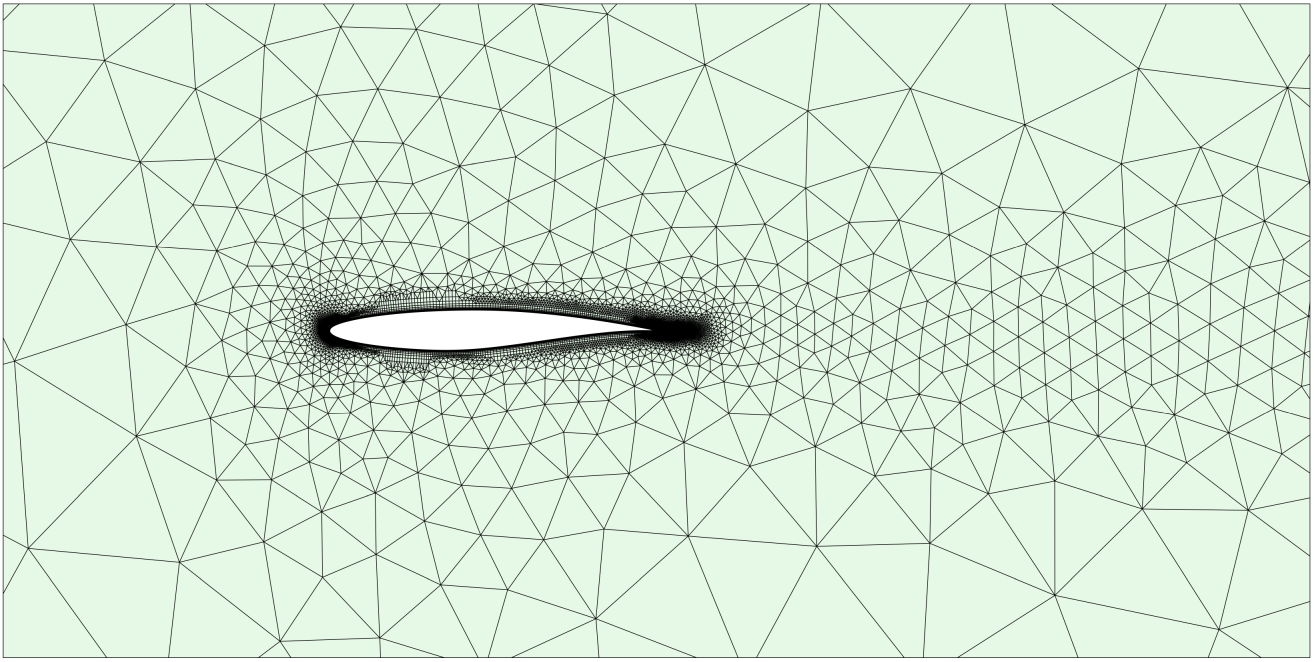}}
	\subfigure[$M_\infty = 0.767, \alpha = 1.53^\circ$]{\includegraphics[width=0.32\textwidth]{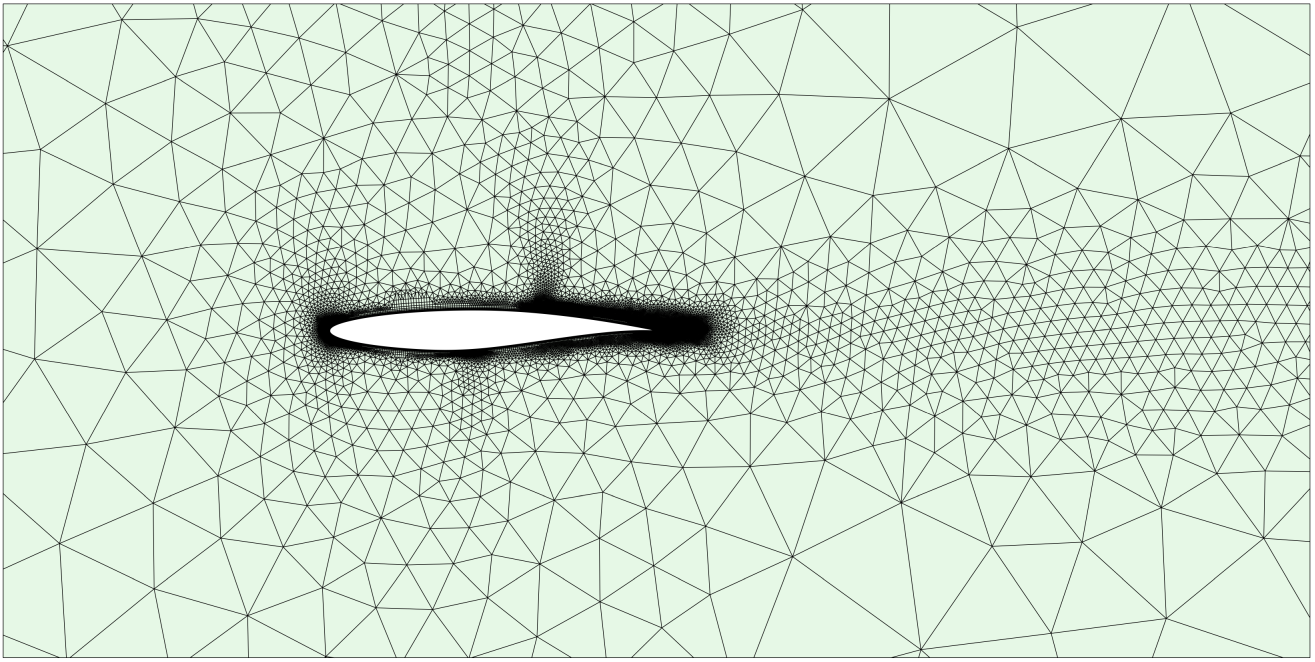}}	
	\subfigure[$M_\infty = 0.792, \alpha = 0.35^\circ$]{\includegraphics[width=0.32\textwidth]{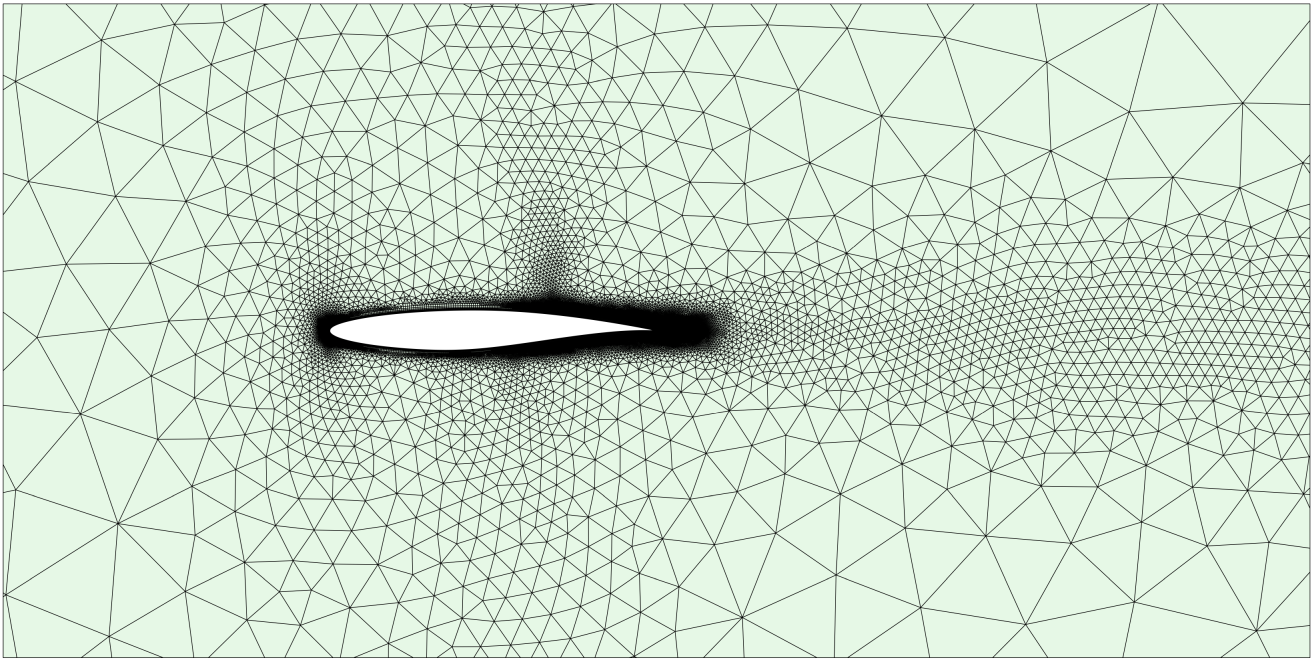}}	
	\caption{Target (top) and predicted (bottom) meshes for three unseen cases during training, corresponding to solutions of Figure~\ref{fig:flowTestCases} employing the fine background mesh of Figure~\ref{fig:flowBacMeshes}(b).}
	\label{fig:flowMeshPredictionBac2}
\end{figure}

Figure~\ref{fig:flowMeshPredictionBac2} shows the target and predicted meshes for the same three unseen cases, but using the finer background mesh of Figure~\ref{fig:flowBacMeshes}(b) and 40 training cases.
Similar to the previous example, it can be seen that the predictions produce local refinement in the regions of interest. However, it can be observed that the predictions seem to be more accurate when the coarse mesh is considered. This is mainly due to the fact that both ANNs have been trained with only 40 training cases. However, the number of outputs when using the fine background mesh is substantially higher, so to achieve a similar accuracy in the predictions, a larger number of training cases would be required.

The target mesh depends on the background mesh used, because the target spacing function is defined as a nodal field in the background mesh. As mentioned previously, this comparison clearly illustrates the compromise that a user needs to accept. To achieve a lower number of elements in the predicted meshes with the refinement localised on the areas of interest, a finer background mesh is preferred. However, this induces a more time consuming training process and requires more training cases to produce accurate predictions. In contrast, a coarser background mesh produces good predictions with a  limited number of training cases and the training times are low. However, the predictions produce meshes that contain a larger number of elements.

\begin{figure}[!tb]
	\centering
	\includegraphics[width=0.75\textwidth]{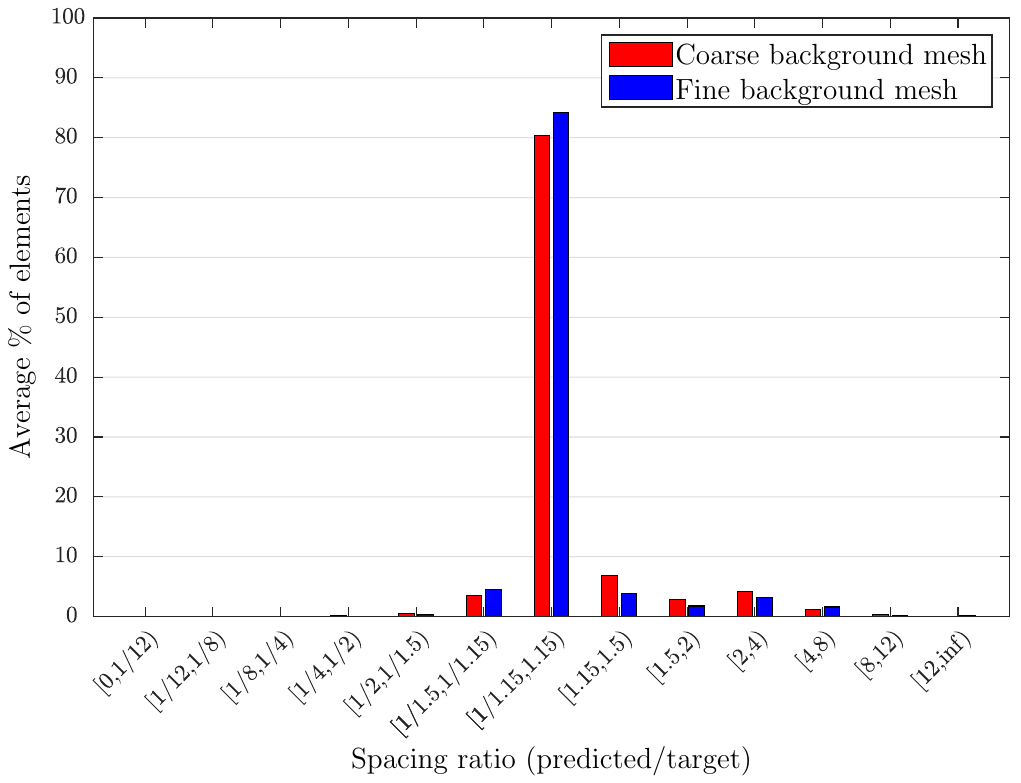}	
	\caption{Histogram of the ratio between predicted and target spacing for the example with varying flow conditions and for the two different background meshes of Figure~\ref{fig:flowBacMeshes}.}
	\label{fig:flowHistogram}
\end{figure}

To further assess the accuracy of the ANN predictions, the predicted spacing function is compared to the target spacing function by computing the ratio between the target and predicted spacing at the centroid of each element, and for all the test cases. The results are displayed as a histogram in Figure~\ref{fig:flowHistogram} for the two ANN that employ a different background mesh.
The results show again a very similar accuracy for both background meshes. It can be observed that more than 80\% of the elements have a predicted spacing between  1/1.15 and 1.15, which is considered accurate enough to produce a mesh capable of capturing all the relevant flow features appearing in the solution. For a value of the ratio higher than 1.15, the prediction induces a larger spacing than desired, resulting in under-refined areas. Similarly, a value of the ratio below 1/1.15 the prediction induces an over-refinement of the mesh. While the histograms show a slight better performance of the predictions obtained by the ANN that uses the finer background mesh, this is because the number of elements in the finer mesh is higher. When averaging the element size to produce the histograms, the very localised areas where the predicted size is not visually matching the target spacing, as shown in Figure~\ref{fig:flowMeshPredictionBac2}, do not contribute significantly to the value shown in the histogram. 

Finally, the suitability of the predicted meshes to perform simulations is studied. To this end, the three test cases of Figure~\ref{fig:flowTestCases}, unseen during training of the ANN, are considered. The predicted near-optimal meshes are utilised to perform a simulation to compare the relevant quantities of interest against a reference solution. Table~\ref{tb:flowAeroPrediction} reports the lift ($C_L$) and drag ($C_D$) coefficients obtained from a reference simulation and the ones obtained after performing a simulation with the predicted meshes using the coarse background mesh and 40 training cases.
\begin{table}[!tb]
	\centering
	\begin{tabular}{|l||c|c||c|c||c|c|}
		\hline%
		& \multicolumn{2}{c||}{$M_\infty = 0.489$} &
		\multicolumn{2}{c||}{$M_\infty = 0.767$} &
		\multicolumn{2}{c|}{$M_\infty = 0.792$} \\
		& \multicolumn{2}{c||}{$\alpha = 1.80^\circ$} &
		\multicolumn{2}{c||}{$\alpha = 1.53^\circ$} &
		\multicolumn{2}{c|}{$\alpha = 0.35^\circ$} \\
		\hline 				
		& Target & Prediction & Target & Prediction & Target & Prediction \\
		\hline
		$C_L$ & 0.448  & 0.448  & 0.545   & 0.529  & 0.323  &  0.294   \\
		\hline
		$C_D$ & 0.0069 & 0.0069 & 0.0169  & 0.0164 & 0.0177 &  0.0169  \\
		\hline
	\end{tabular}
	\caption{Comparison of the aerodynamic quantities of interest computed with the target and predicted meshes for three different flow conditions.}
	\label{tb:flowAeroPrediction}
\end{table}

The results show that for some configurations, corresponding to smooth flows, the predicted meshes lead to aerodynamic quantities of interest that perfectly match the reference results. When more complicated, transonic, cases are considered the predicted meshes lead to less accurate results as expected. From a practical point of view, the results show that the trained ANN produces good predictions of the regions that need localised refinement, even with only 40 training cases. These meshes can be considered as a very good starting point to perform an automatic adaptivity process, ensuring that no features of the solution are missed in the first computation.

\begin{figure}[!tb]
	\centering
	\subfigure[$M_\infty = 0.489, \alpha = 1.80^\circ$]{\includegraphics[width=0.32\textwidth]{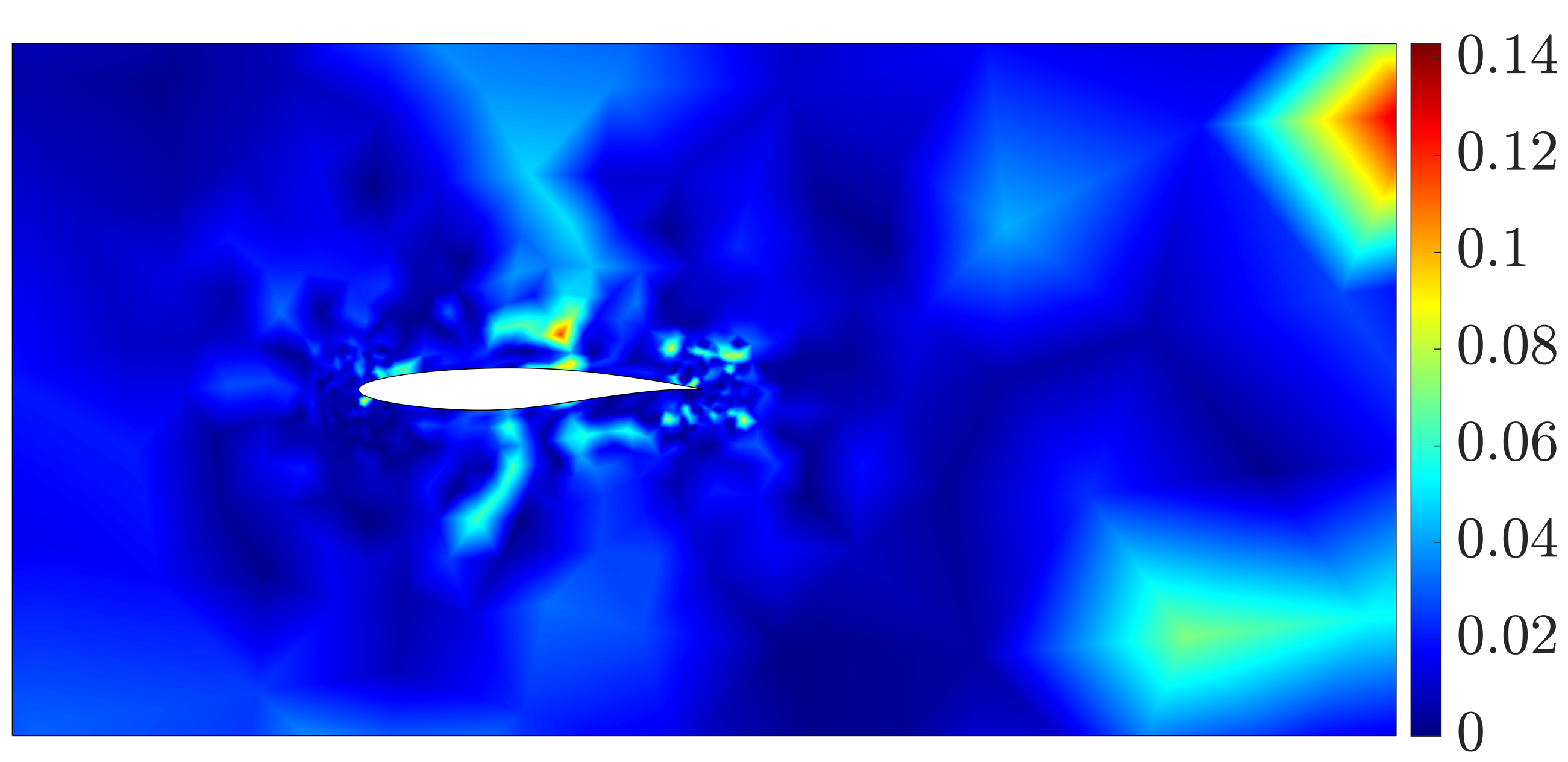}}
	\subfigure[$M_\infty = 0.767, \alpha = 1.53^\circ$]{\includegraphics[width=0.32\textwidth]{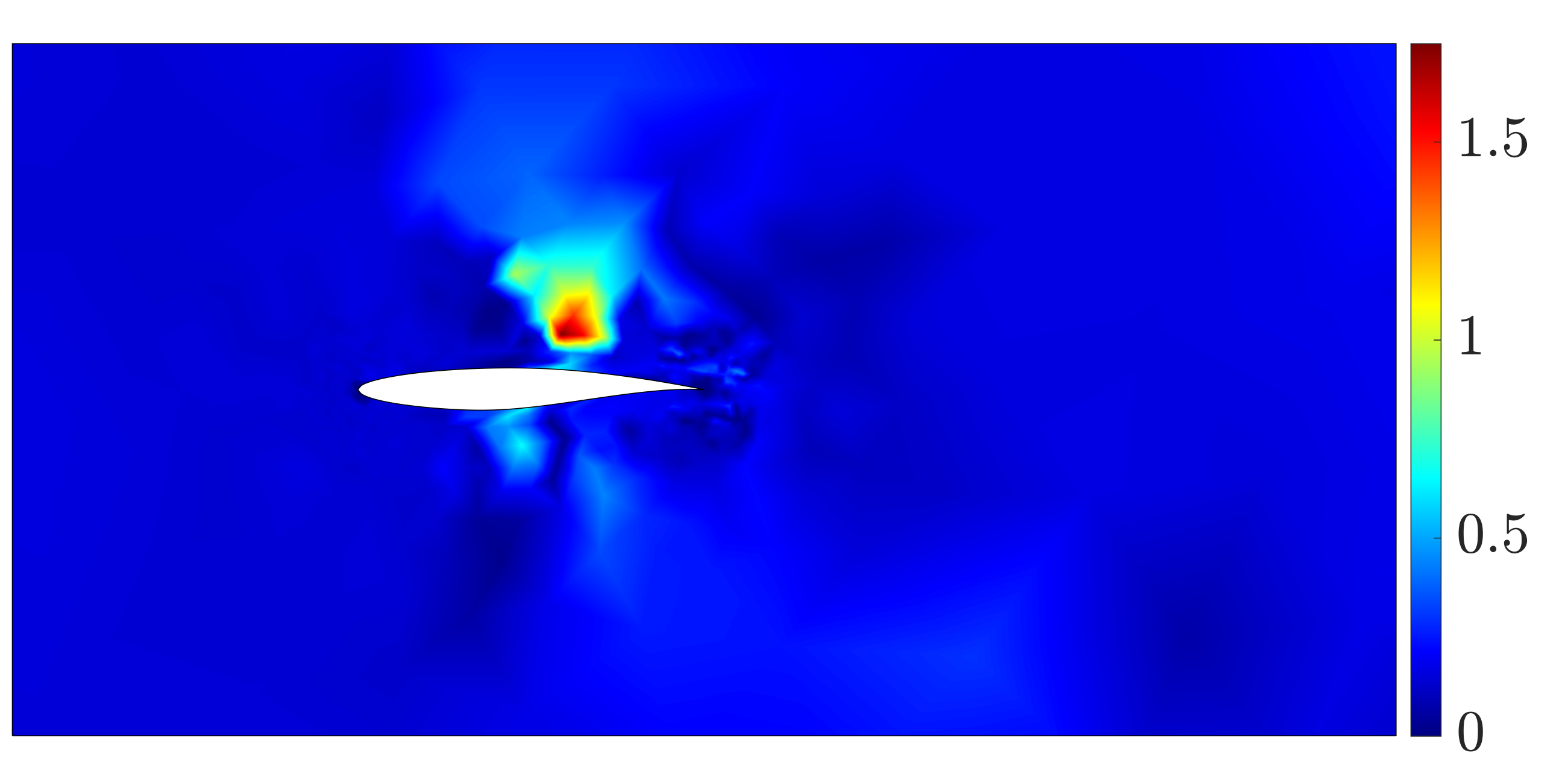}}	
	\subfigure[$M_\infty = 0.792, \alpha = 0.35^\circ$]{\includegraphics[width=0.32\textwidth]{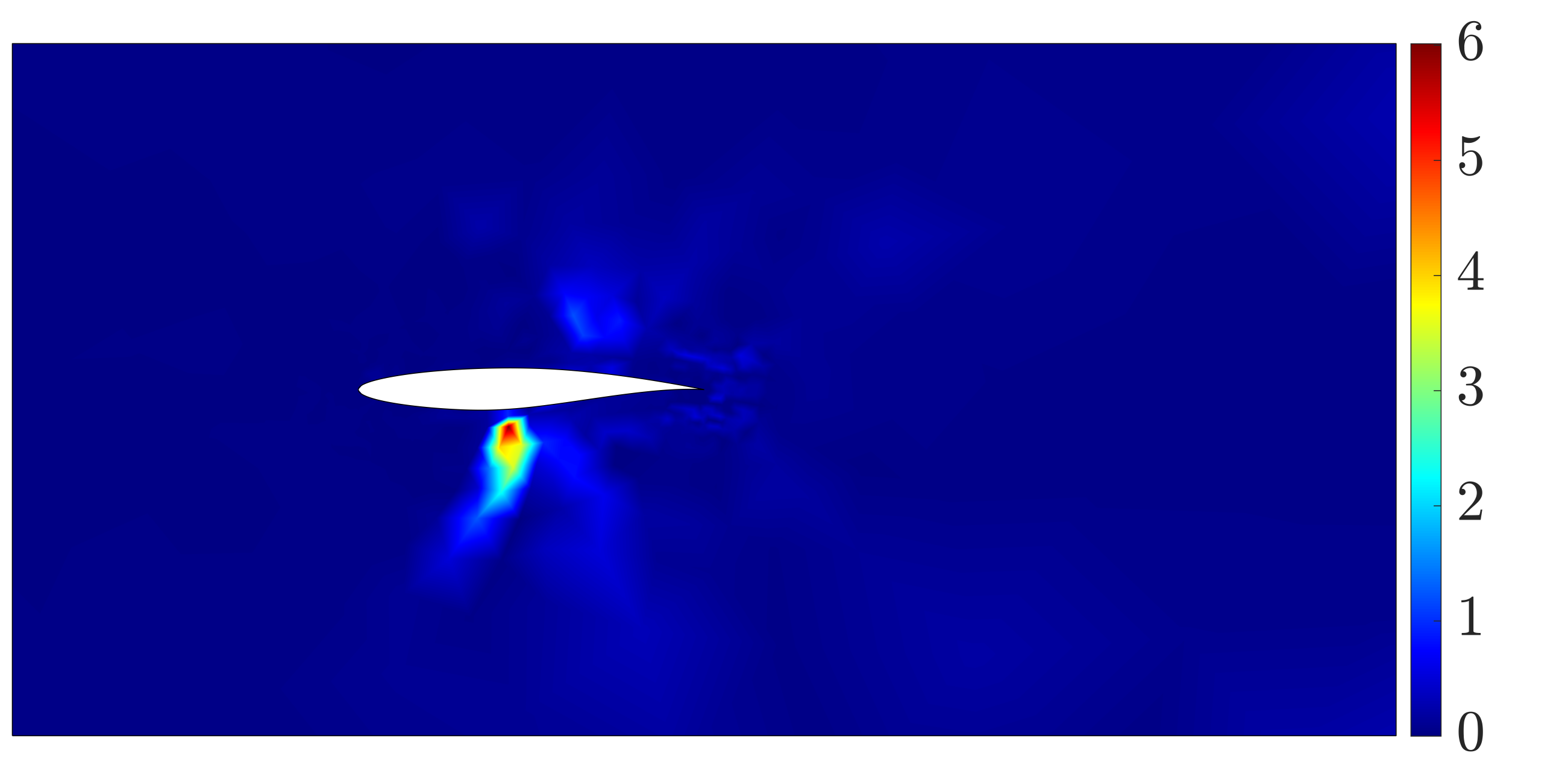}}	
	\caption{Percentage error of the predicted spacing for three unseen cases during training, corresponding to solutions of  Figure~\ref{fig:flowTestCases} employing the coarse background mesh of Figure~\ref{fig:flowBacMeshes}(a).}
	\label{fig:flowErrSpacinBac1}
\end{figure}

To further analyse the accuracy of the predictions, Figure~\ref{fig:flowErrSpacinBac1} shows the percentage error map of the predicted spacing for the three unseen cases during training, when the ANN is trained using the coarse background mesh of Figure~\ref{fig:flowBacMeshes}(a).
The results show that for the subsonic case, the error of the predicted nodal spacing is below 0.2\%. For the transonic case with one shock on the upper surface of the aerofoil, the maximum error is near the shock region and is around 1.7\%. Similarly, for the transonic case with two shocks on the upper and lower surfaces of the aerofoil the maximum error is near the shock region and is around 6\%. This behaviour is mainly attributed to the fact that in the training dataset all cases present a refinement near the aerofoil as required by the subsonic case, but only a subset of these cases present one shock and a smaller subset present two shocks.

\begin{figure}[!tb]
	\centering
	\subfigure[$M_\infty = 0.489, \alpha = 1.80^\circ$]{\includegraphics[width=0.32\textwidth]{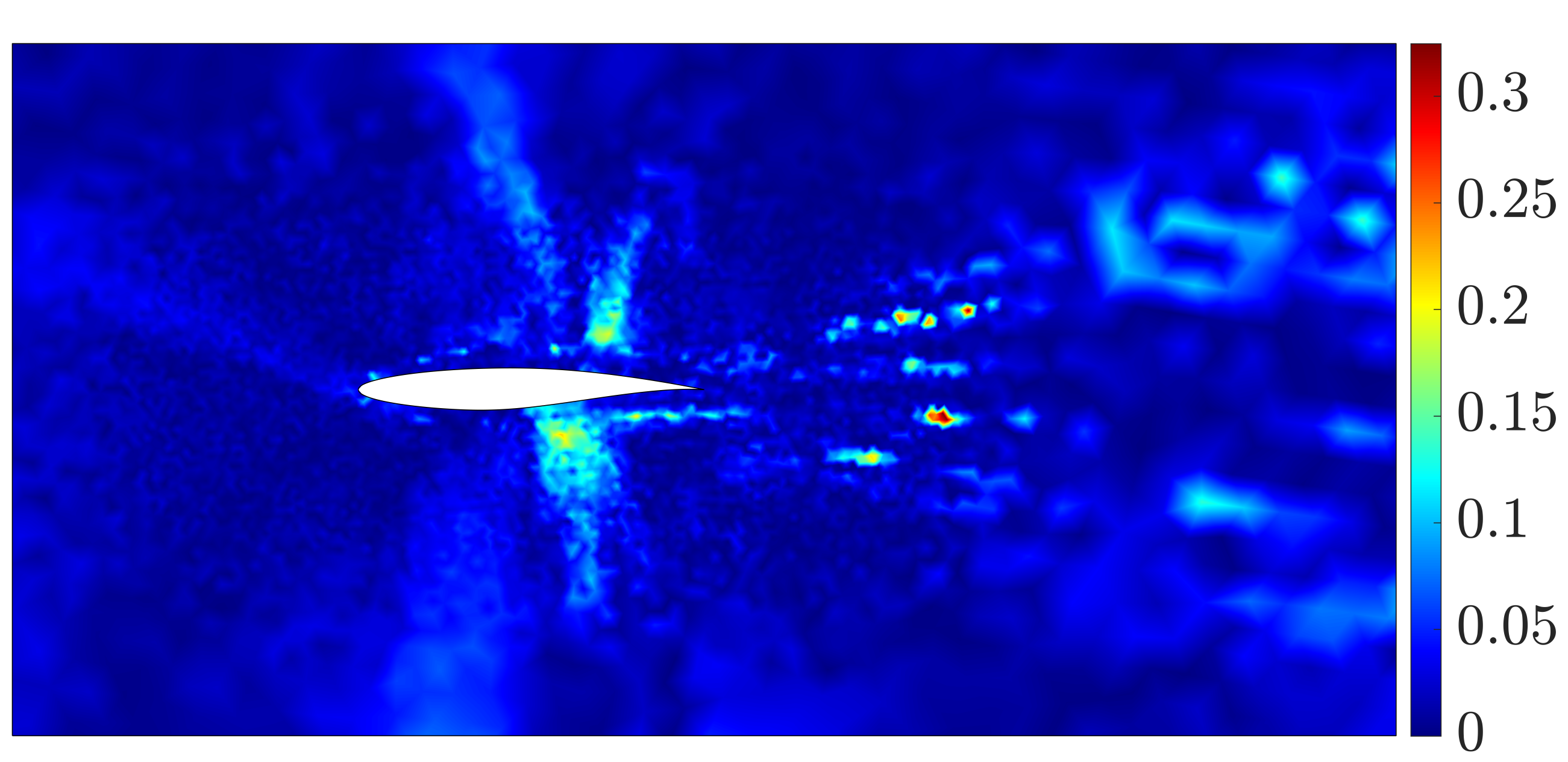}}
	\subfigure[$M_\infty = 0.767, \alpha = 1.53^\circ$]{\includegraphics[width=0.32\textwidth]{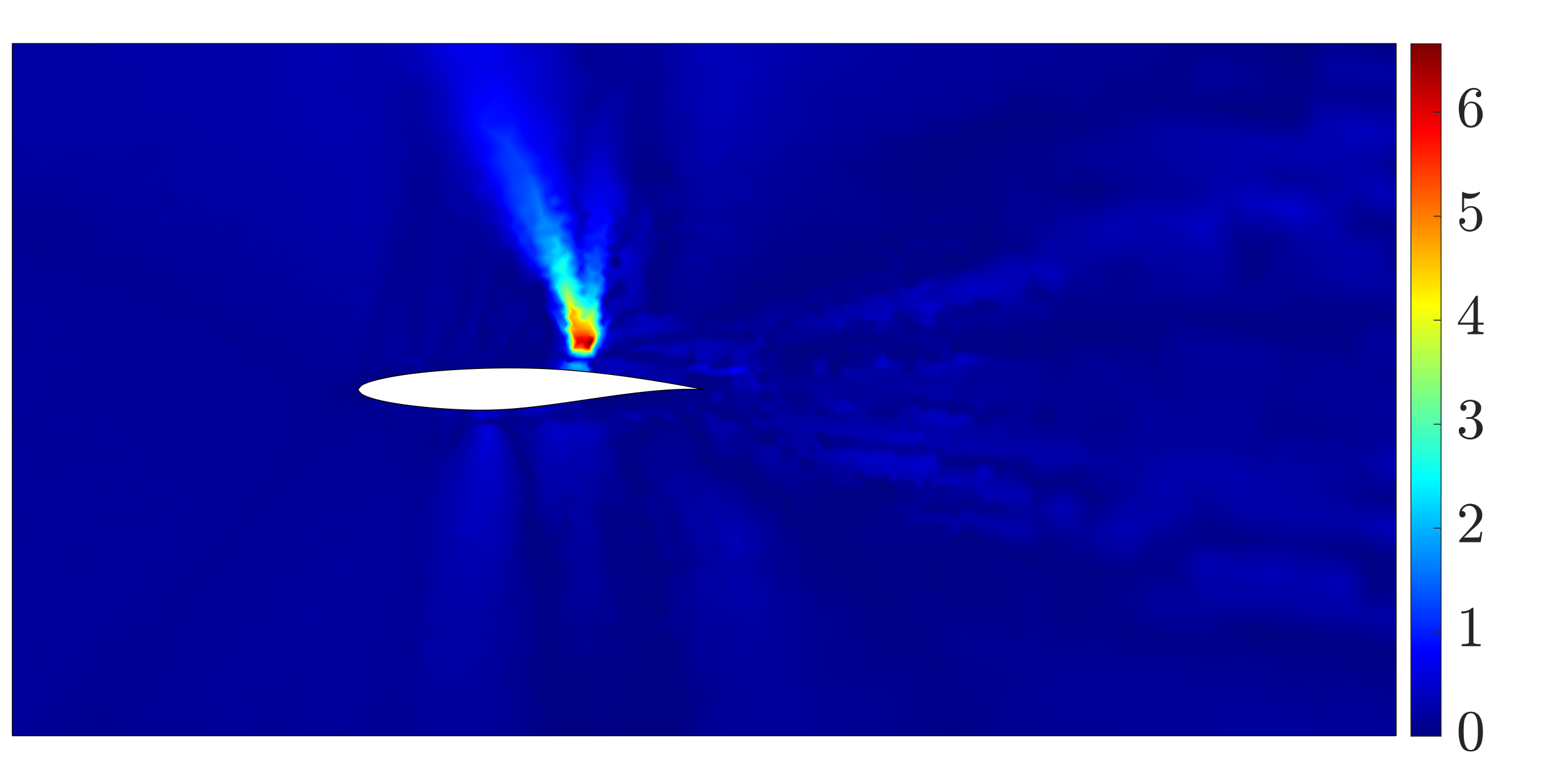}}	
	\subfigure[$M_\infty = 0.792, \alpha = 0.35^\circ$]{\includegraphics[width=0.32\textwidth]{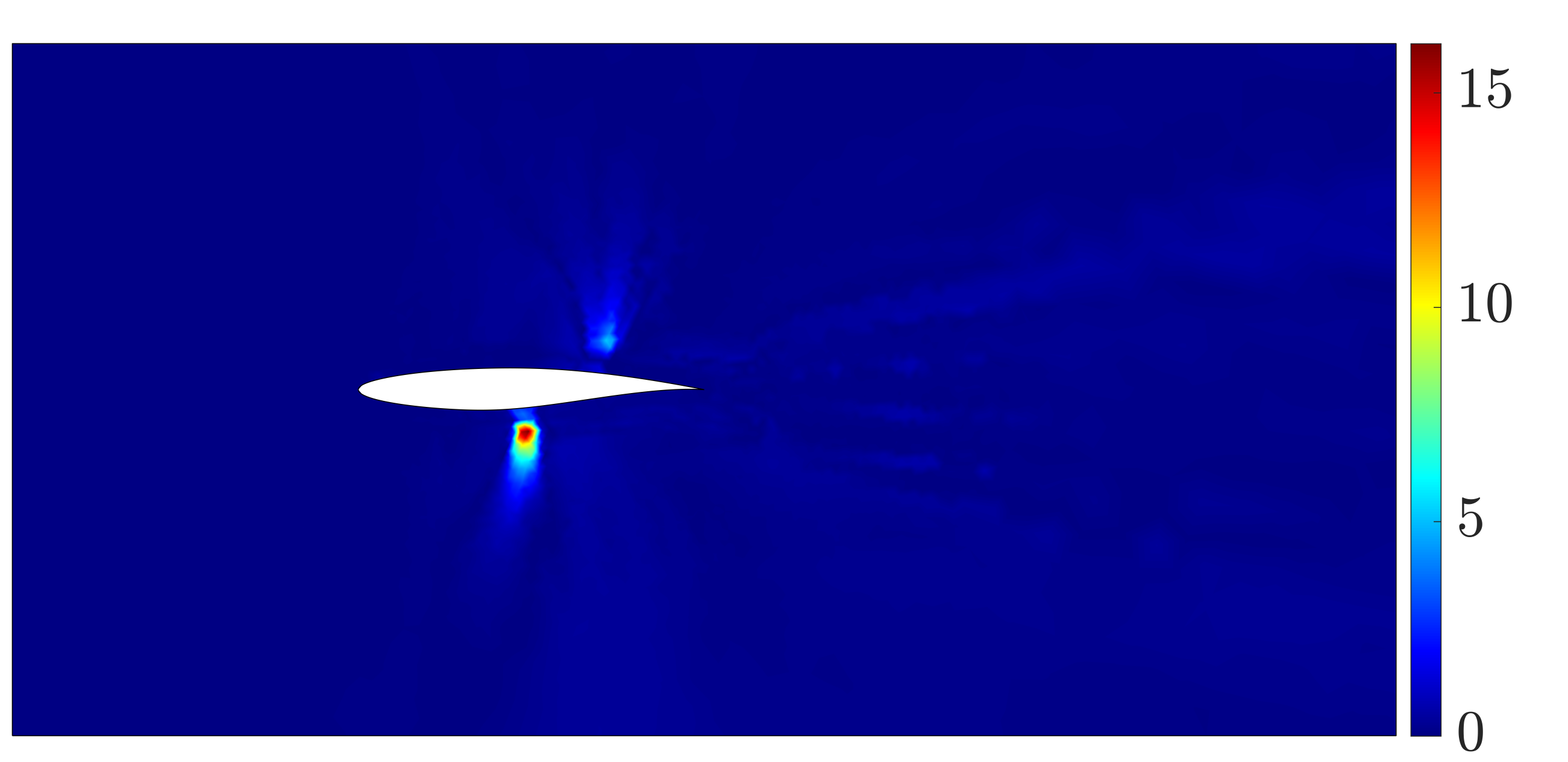}}	
	\caption{Percentage error of the predicted spacing for three unseen cases during training, corresponding to solutions of  Figure~\ref{fig:flowTestCases} employing the coarse background mesh of Figure~\ref{fig:flowBacMeshes}(b).}
	\label{fig:flowErrSpacinBac2}
\end{figure}

Finally, to compare the performance of the predictions when using a finer background mesh, Figure~\ref{fig:flowErrSpacinBac2} shows the percentage error map of the predicted spacing for the three unseen cases during training, when the ANN is trained using the coarse background mesh of Figure~\ref{fig:flowBacMeshes}(b). 
The maximum error of the predictions is again localised in the shock regions when transonic cases are considered. In addition, in all three unseen cases the maximum error is higher when using the finer background mesh and more localised. This behaviour is attributed to the fact that when a finer backgrond mesh is considered, the target spacing at a node varies more abruptly in the parametric space and larger training datasets would be required to obtain accurate predictions.

\subsection{Prediction of the spacing for variable geometries}

The next example considers the prediction of near-optimal meshes for variable geometric configurations and fixed turbulent transonic flow conditions, corresponding to $Re_\infty = 6.5 \times 10^{6}$, $M_{\infty} = 0.730$,  and $\alpha= 2.79^\circ$. The geometry of the aerofoil is parametrised using two NURBS curves, for the top and bottom surface of the aerofoil, respectively. Each NURBS curve has eight control points and the original configuration, also referred to here as the \textit{undeformed configuration}, corresponds to the control points detailed in Table~\ref{tb:nurbsCP}.
\begin{table}[!tb]
	\centering
	\begin{tabular}{|c|c|c|c|c|c|c|c|c|}
		\hline 				
		$i$ & 1 & 2 & 3 & 4 & 5 & 6 & 7 & 8 \\
		\hline
		$X_i^+$ & 0.000 & 0.000 &  0.105 &  0.391 &  0.594 &  0.799 & 0.933 & 1.000 \\
		\hline
		$Y_i^+$ & 0.000 & 0.024 &  0.059 &  0.076 &  0.059 &  0.034 & 0.012 & 0.000 \\
		\hline
		$X_i^-$ & 0.000 & 0.000 &  0.105 &  0.391 &  0.594 &  0.799 & 0.933 & 1.000 \\
		\hline
		$Y_i^-$ & 0.000 & -0.024 & -0.059 & -0.084 & -0.046 & -0.010 & 0.000 & 0.000 \\
		\hline
	\end{tabular}
	\caption{Position of control points for the NURBS curves defining the aerofoil in its original configuration. The superscripts $^+$ and $^-$ indicate points on the upper and lower parts of the aerofoil, respectively.}
	\label{tb:nurbsCP}
\end{table}

The geometric parameters are the relative position of the control points with respect to the original or undeformed configuration. Table~\ref{tb:nurbsCPvariation} details the range of variation of the position of the control points considered in this example. 
\begin{table}[!tb]
\centering
\begin{tabular}{|c|c|c|c|c|c|c|}
	\hline 				
	$i$ & 2 & 3 & 4 & 5 & 6 & 7 \\
	\hline
	$\delta X_i^+$ & $4.3 \ttimes 10^{-2}$ &  $4.3 \ttimes 10^{-2}$ &  $3.1 \ttimes 10^{-2}$ & $ 2.5 \ttimes 10^{-2}$ &  $1.5 \ttimes 10^{-2}$ &  $5.3 \ttimes 10^{-3}$  \\
	\hline
	$\delta Y_i^+$ & $2.2 \ttimes  10^{-2}$ &  $2.2 \ttimes 10^{-2}$ &  $1.5 \ttimes 10^{-2}$ & $ 1.2 \ttimes 10^{-2}$ &  $7.4 \ttimes 10^{-3}$ &  $2.7 \ttimes 10^{-3}$  \\
	\hline
	$\delta X_i^-$ & -- &  $4.3 \ttimes 10^{-2}$ &  $4.5 \ttimes 10^{-2}$ & $ 3.9 \ttimes 10^{-5}$ &  $3.2 \ttimes 10^{-2}$ &  $1.9 \ttimes 10^{-2}$  \\
	\hline
	$\delta Y_i^-$ & -- &  $2.2 \ttimes 10^{-2}$ &  $2.3 \ttimes 10^{-2}$ & $ 1.9 \ttimes 10^{-5}$ &  $1.6 \ttimes 10^{-2}$ &  $9.6 \ttimes 10^{-3}$  \\
	\hline
\end{tabular}
\caption{Variation of the position of control points for the NURBS curves defining the aerofoil. The superscripts $^+$ and $^-$ indicate the upper and lower parts of the aerofoil, respectively.}
\label{tb:nurbsCPvariation}
\end{table}
The control points corresponding to leading and trailing edges are fixed and, as pointed out in Remark~\ref{rk:nurbs}, the second control point in the bottom curve is not varied freely to ensure $\mathcal{G}^1$ continuity of the resulting aerofoil geometry. The position of this control point is defined as $\bm{B}_2^- = \bm{B}_1^- + \vartheta (\bm{B}_1^- - \bm{B}_2^+)$ where $\vartheta$ is a geometric parameter that varies between 0.5 and 1.5. The weights of all the control points are taken as one and not altered during the deformation. Therefore, the total number of geometric parameters is 23.

\begin{figure}[!tb]
	\centering
	\subfigure[Geometry 1]{\includegraphics[width=0.32\textwidth]{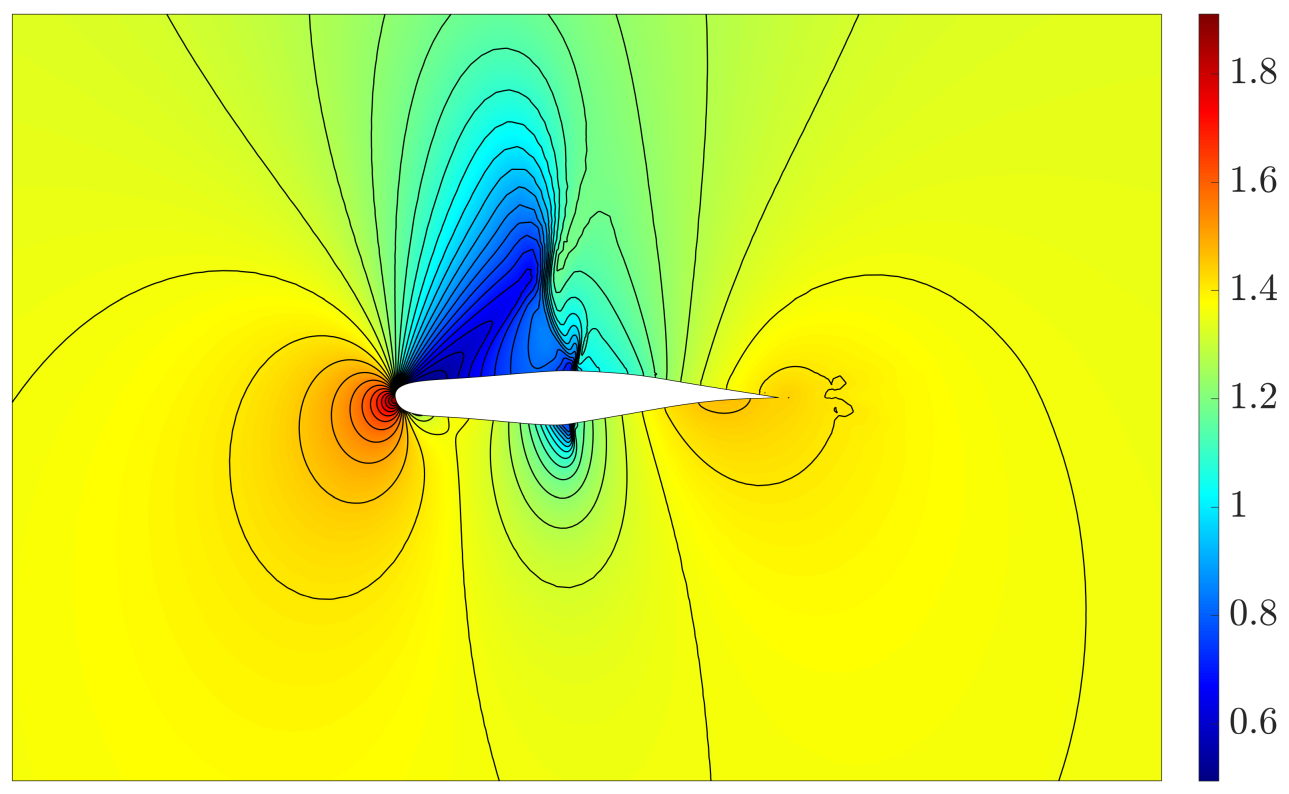}}
	\subfigure[Geometry 2]{\includegraphics[width=0.32\textwidth]{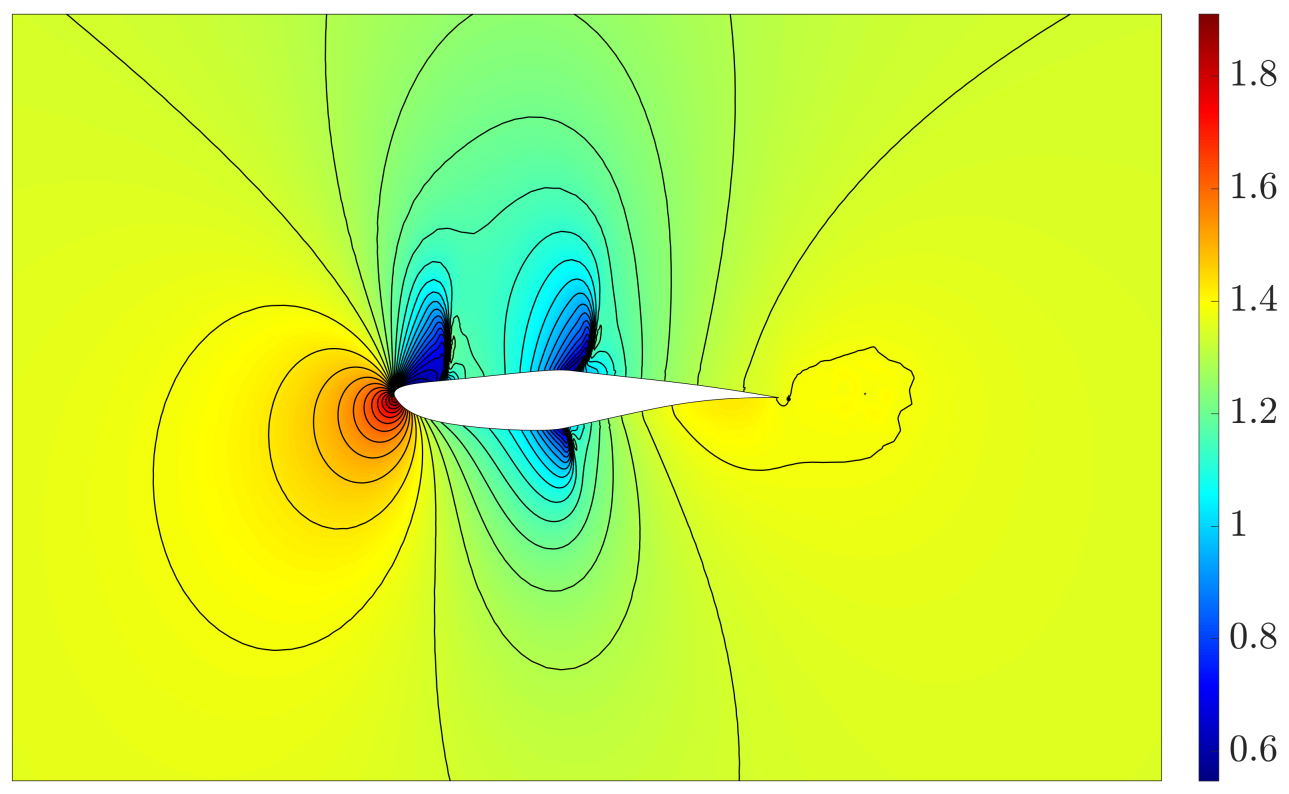}}	
	\subfigure[Geometry 3]{\includegraphics[width=0.32\textwidth]{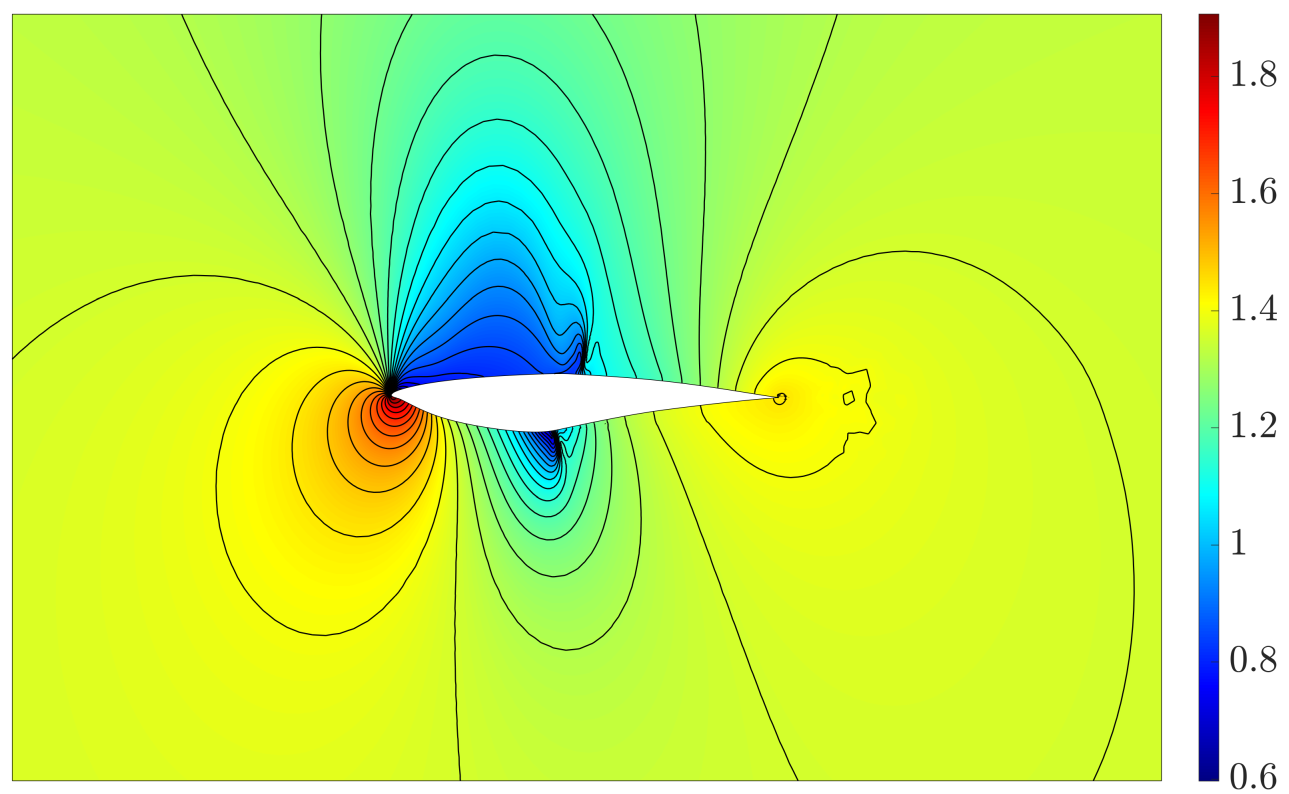}}	
	\caption{Pressure field and isolines for three different geometric configurations with $Re_\infty = 6.5 \times 10^{6}$, $M_{\infty} = 0.730$,  and $\alpha= 2.79^\circ$.}
	\label{fig:geoTestCases}
\end{figure}

To illustrate the variation of the solution induced by the variation of the geometric parameters, Figure~\ref{fig:geoTestCases} shows three solutions corresponding to three different geometric configurations.

The position of the control points for the three geometric configurations of Figure~\ref{fig:geoTestCases} are detailed in Tables~\ref{tb:nurbsCPGeo1}, \ref{tb:nurbsCPGeo2} and \ref{tb:nurbsCPGeo3}, respectively.
\begin{table}[!tb]
	\centering
	\begin{tabular}{|c|c|c|c|c|c|c|c|c|}
		\hline 				
		$i$ & 1 & 2 & 3 & 4 & 5 & 6 & 7 & 8 \\
		\hline
		$X_i^+$ & 0.000 & -0.007 &  0.139 &  0.412 &  0.598 &  0.794 &  0.931 & 1.000 \\
		\hline
		$Y_i^+$ & 0.000 &  0.043 &  0.046 &  0.071 &  0.067 &  0.031 &  0.011 & 1.000 \\ 
		\hline
		$X_i^-$ & 0.000 &  0.008 &  0.130 &  0.406 &  0.595 &  0.822 &  0.924 & 1.000 \\ 
		\hline
		$Y_i^-$ & 0.000 & -0.048 & -0.046 & -0.077 & -0.046 & -0.002 & -0.004 & 1.000 \\ 
		\hline
	\end{tabular}
	\caption{Position of control points for the NURBS curves defining the aerofoil of Figure~\ref{fig:geoTestCases}(a).}
	\label{tb:nurbsCPGeo1}
\end{table}
\begin{table}[!tb]
	\centering
	\begin{tabular}{|c|c|c|c|c|c|c|c|c|}
		\hline 				
		$i$ & 1 & 2 & 3 & 4 & 5 & 6 & 7 & 8 \\
		\hline
		$X_i^+$ & 0.000 & -0.019 &  0.123 &  0.413 &  0.583 & 0.801 &  0.934 & 1.000 \\ 
		\hline
		$Y_i^+$ & 0.000 &  0.034 &  0.042 &  0.076 &  0.056 & 0.034 &  0.011 & 1.000 \\ 
		\hline
		$X_i^-$ & 0.000 &  0.021 &  0.108 &  0.393 &  0.595 & 0.809 &  0.948 & 1.000 \\ 
		\hline
		$Y_i^-$ & 0.000 & -0.037 & -0.074 & -0.091 & -0.046 & 0.002 & -0.000 & 1.000 \\  
		\hline
	\end{tabular}
	\caption{Position of control points for the NURBS curves defining the aerofoil of Figure~\ref{fig:geoTestCases}(b).}
	\label{tb:nurbsCPGeo2}
\end{table}
\begin{table}[!tb]
	\centering
	\begin{tabular}{|c|c|c|c|c|c|c|c|c|}
		\hline 				
		$i$ & 1 & 2 & 3 & 4 & 5 & 6 & 7 & 8 \\
		\hline
		$X_i^+$ & 0.000 & -0.035 &  0.075 &  0.412 &  0.594 &  0.795 &  0.928 & 1.000 \\ 
		\hline
		$Y_i^+$ & 0.000 &  0.009 &  0.047 &  0.065 &  0.054 &  0.035 &  0.010 & 1.000 \\ 
		\hline
		$X_i^-$ & 0.000 &  0.022 &  0.093 &  0.362 &  0.595 &  0.813 &  0.926 & 1.000 \\ 
		\hline
		$Y_i^-$ & 0.000 & -0.006 & -0.059 & -0.098 & -0.046 & -0.017 & -0.006 & 1.000 \\ 
		\hline
	\end{tabular}
	\caption{Position of control points for the NURBS curves defining the aerofoil of Figure~\ref{fig:geoTestCases}(c).}
	\label{tb:nurbsCPGeo3}
\end{table}

The plots in Figure~\ref{fig:geoTestCases} show the increased complexity of the pressure fields that result from varying the geometry of the aerofoil, compared to the previous example with varying operating conditions. The first example shows two shocks in the upper and lower surface of the aerofoil respectively. The second geometry induces a more complicated flow pattern with two shocks in the upper surface of the aerofoil and also a relatively strong shock in the lower surface. Finally, the third geometry involves smoother gradients but still with shocks in the upper and lower surfacert of the aerofoil. It can be observed that predicting near-optimal meshes in this example is even more challenging. Not only is the number of parameters substantially higher, but the nature of the parameters induces more extreme variations in the flow features that require a completely different mesh, with refinement in different regions for each case.

\begin{figure}[!tb]
	\centering
	\subfigure[Geometry 1]{\includegraphics[width=0.32\textwidth]{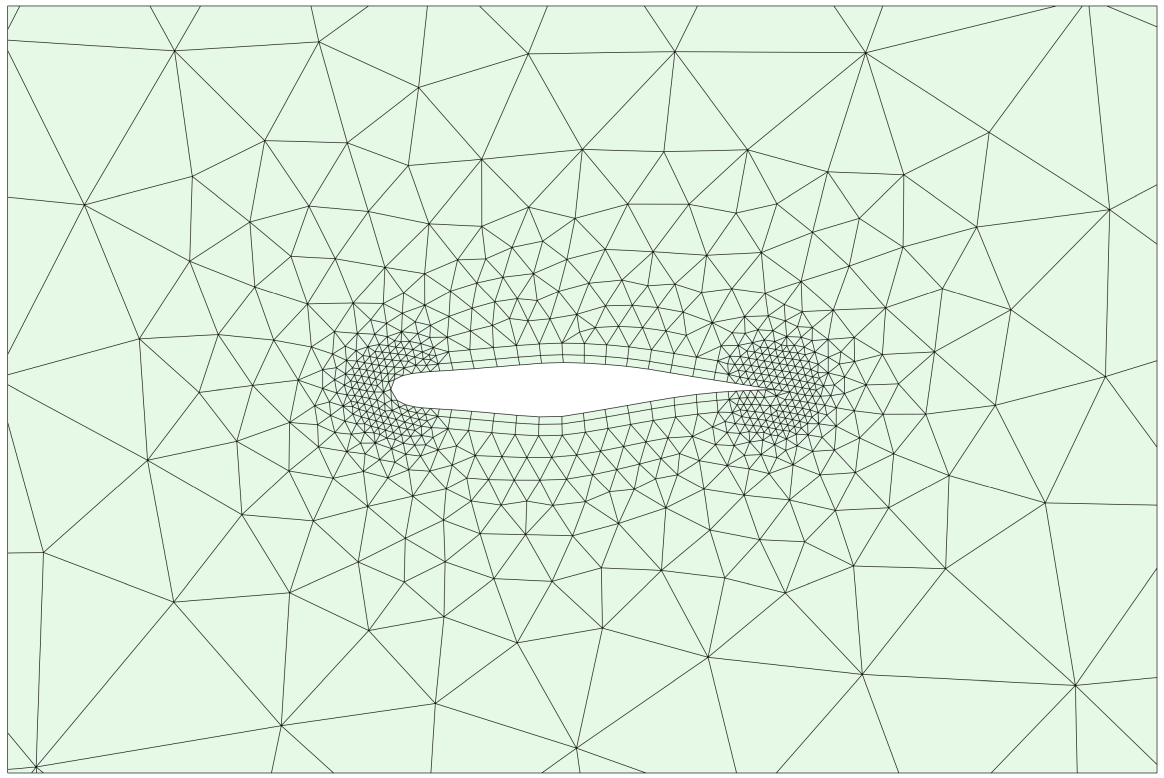}}
	\subfigure[Geometry 2]{\includegraphics[width=0.32\textwidth]{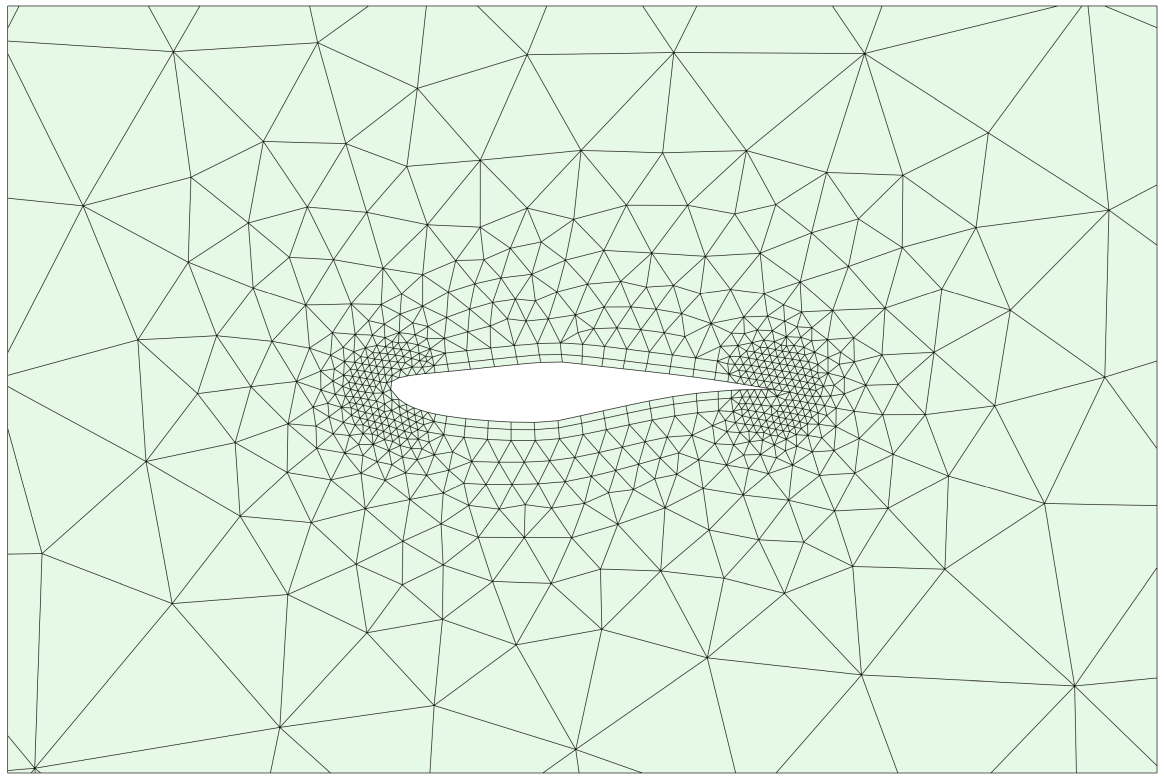}}	
	\subfigure[Geometry 3]{\includegraphics[width=0.32\textwidth]{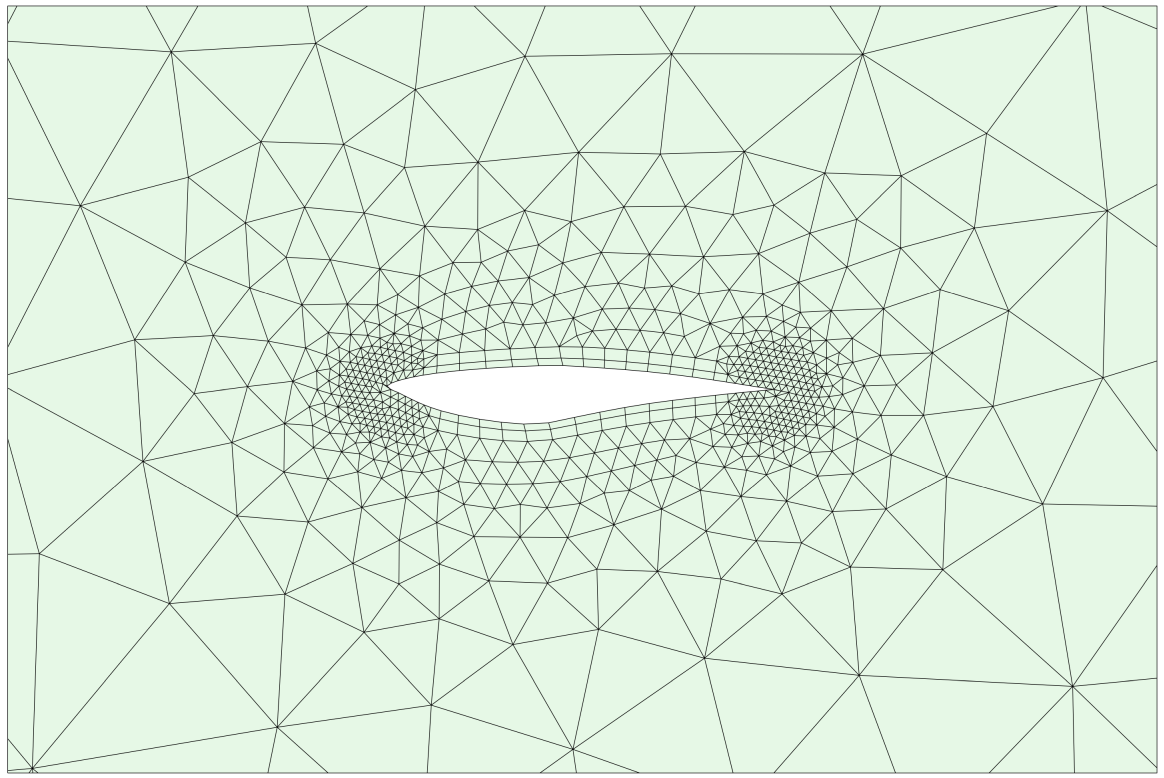}}	
	\caption{Background meshes obtained using an elastic deformation for the three geometric configurations of Figure~\ref{fig:geoTestCases}.}
	\label{fig:geoBacMeshes}
\end{figure}

To ensure that the background mesh employed for each geometric configuration contains the same number of nodes, the mesh morphing approach described in Section~\ref{sc:meshMorphing} is employed. A coarse background mesh of the undeformed configuration is generated and deformed for each unseen geometric configuration using an elasticity approach. Figure~\ref{fig:geoBacMeshes} shows three deformed configurations of the background mesh for the geometries of Figure~\ref{fig:geoTestCases}.

The background meshes have 976 nodes, 1,780 triangular elements and 52 quadrilateral elements in the inflation layer, which is the same as the coarse mesh used in the previous example. Local refinement has been defined by introducing two points sources, in the leading and trailing edges, and a line source connecting the leading and trailing edge. Again, it is worth pointing out that the anisotropic refinement near the wall in the background meshes is not designed to capture the flow physics, but to capture the expected large variation of the spacing near the aerofoil.

Training sets with $N_{tr}= 20, 40, 80, 160, 320, 640$ training cases have been generated using Halton sequencing. Given the high dimensionality of the parameter space, the so-called scrambled Halton sequencing~\cite{vandewoestyne2006good} is utilised to avoid the undesired correlation in the sampling that occurs with the traditional Halton sequencing in high dimensional problems. As in the previous example, validation sets with $N_{val}=5,10,20,40,80,160$ cases are also generated using scrambled Halton sequencing. The accuracy of the ANN predictions is assessed using a fixed set of $N_{tst} = 74$ test cases. Finally, the range of the parameters is slightly reduced in the validation and tests sets, namely 3\% and 6\% reduction, at both ends, for the validation and test sets, respectively, to avoid extrapolation.

\begin{figure}[!tb] 
	\centering
	\subfigure[$N_{tr} = 40$]{\includegraphics[width=0.45\textwidth]{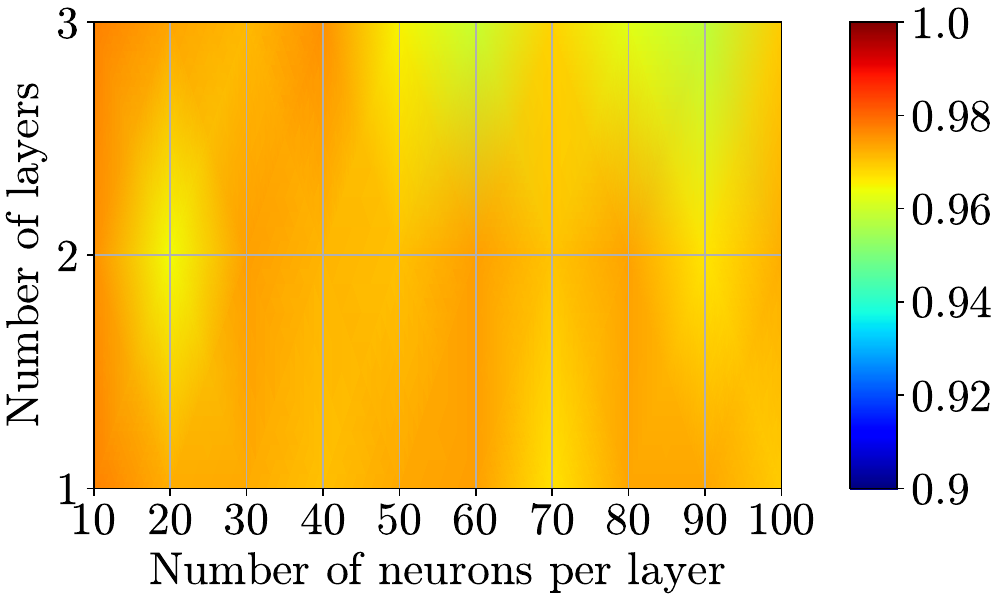}}
	\subfigure[$N_{tr} = 80$]{\includegraphics[width=0.45\textwidth]{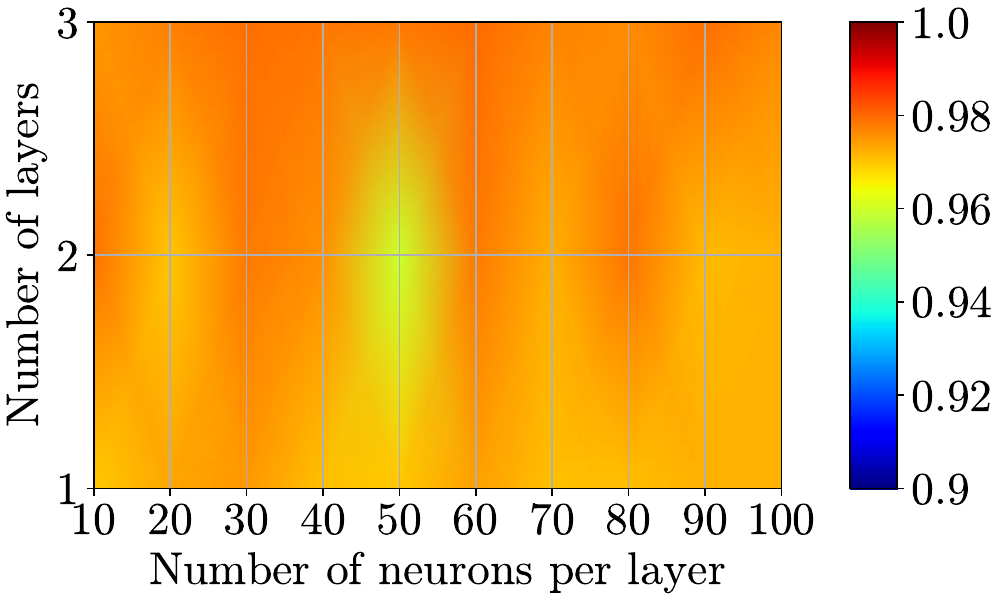}}\\
	\subfigure[$N_{tr} = 320$]{\includegraphics[width=0.45\textwidth]{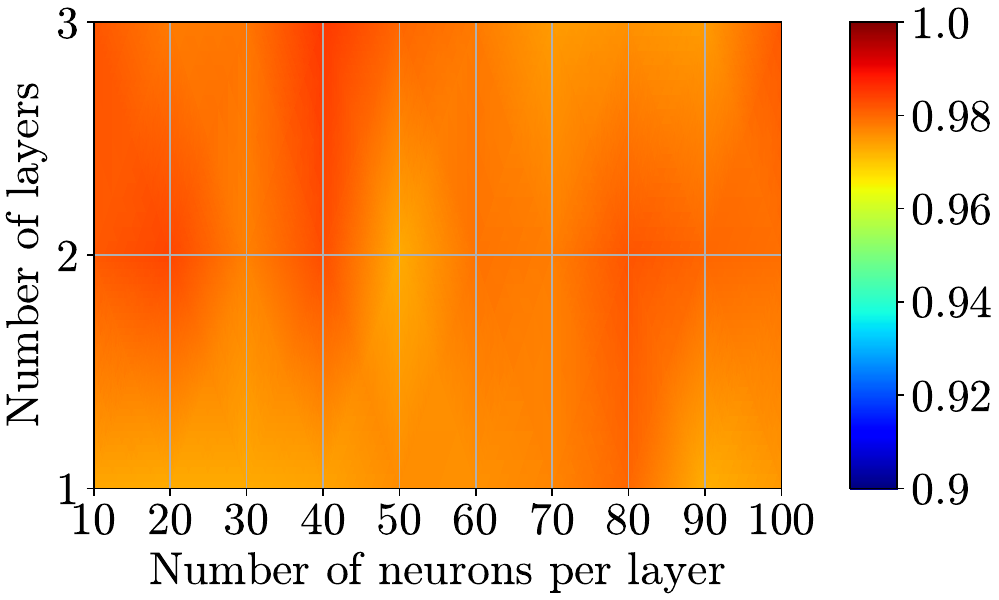}}
	\subfigure[$N_{tr} = 640$]{\includegraphics[width=0.45\textwidth]{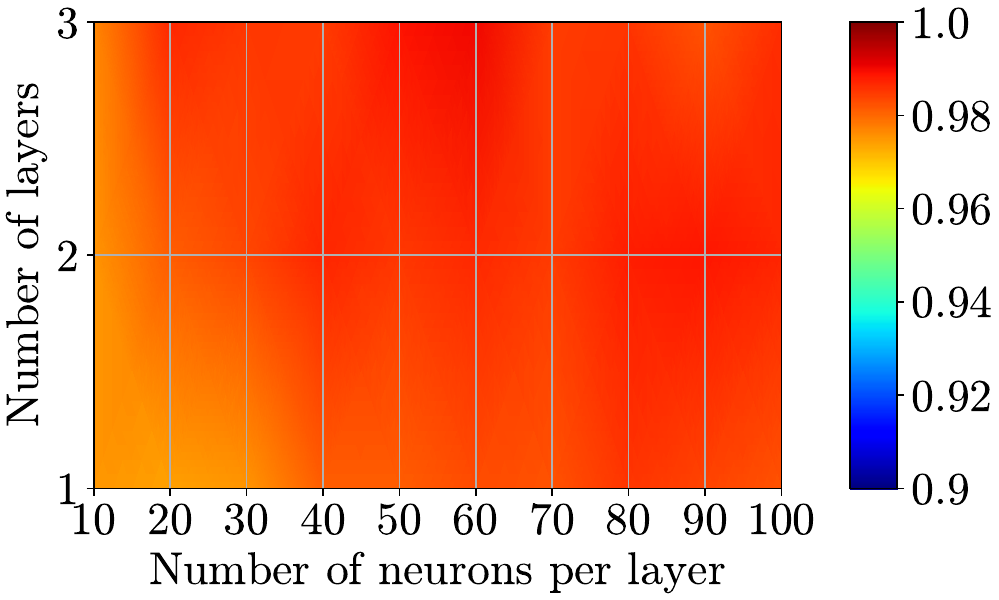}}
	\caption{R$^{2}$ for different number of training cases ($N_{tr}$) as a function of the number of layers and number of neurons in each layer.}
	\label{fig:geometryHyperparameters} 
\end{figure}

Next, the effect of the ANN hyperparameters is studied employing a simple grid search.  Figure~\ref{fig:geometryHyperparameters} shows the variation of the R$^{2}$ as a function of the number of layers and the number of neurons in each layer when using an increasing number of training cases.
The results clearly show that high values of the R$^{2}$ can be achieved with small training sets. Comparing the results with the hyperparameter study in the previous example, it is clear that larger training datasets are required to achieve values of the R$^{2}$ close to 1, given the substantial increase in the number of inputs. In this example, a training set with 640 samples provides high values of the R$^{2}$, even for shallow networks with a low number of neurons.

\begin{figure}[!tb]
	\centering
	\includegraphics[width=0.5\textwidth]{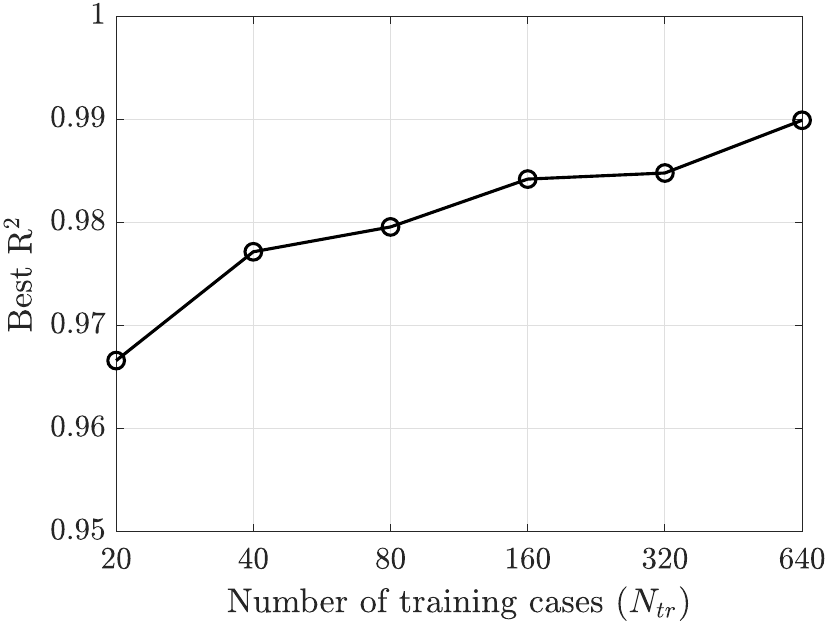}	
	\caption{Best R$^2$ for the as a function of the number of training cases for the problem with varying geometric parameters.}
	\label{fig:geometryBestR2}
\end{figure}

To better observe the influence of the number of training cases in the accuracy of the predictions, Figure~\ref{fig:geometryBestR2}	shows the best R$^2$ as a function of the number of training cases. 
The results confirm the accuracy of the predictions in this example and they also illustrate the increased difficulty of performing accurate predictions with large number of parameters. In the previous example, a value of R$^{2}$ near 0.99 was achieved with 80 training cases, whereas in this example the number of training cases required to achieve a similar R$^{2}$ is 640. In the previous example, with only two parameters, a sampling with 80 training cases corresponds to 8.9 cases per parameter, whereas in the current example a sampling with 640 cases corresponds to only 1.3 cases per parameter. This analysis already indicates the potential of the proposed ANN approach to accurately predict the spacing function for unseen geometric configurations.

\begin{figure}[!tb]
	\centering
	\subfigure[Geometry 1]{\includegraphics[width=0.32\textwidth]{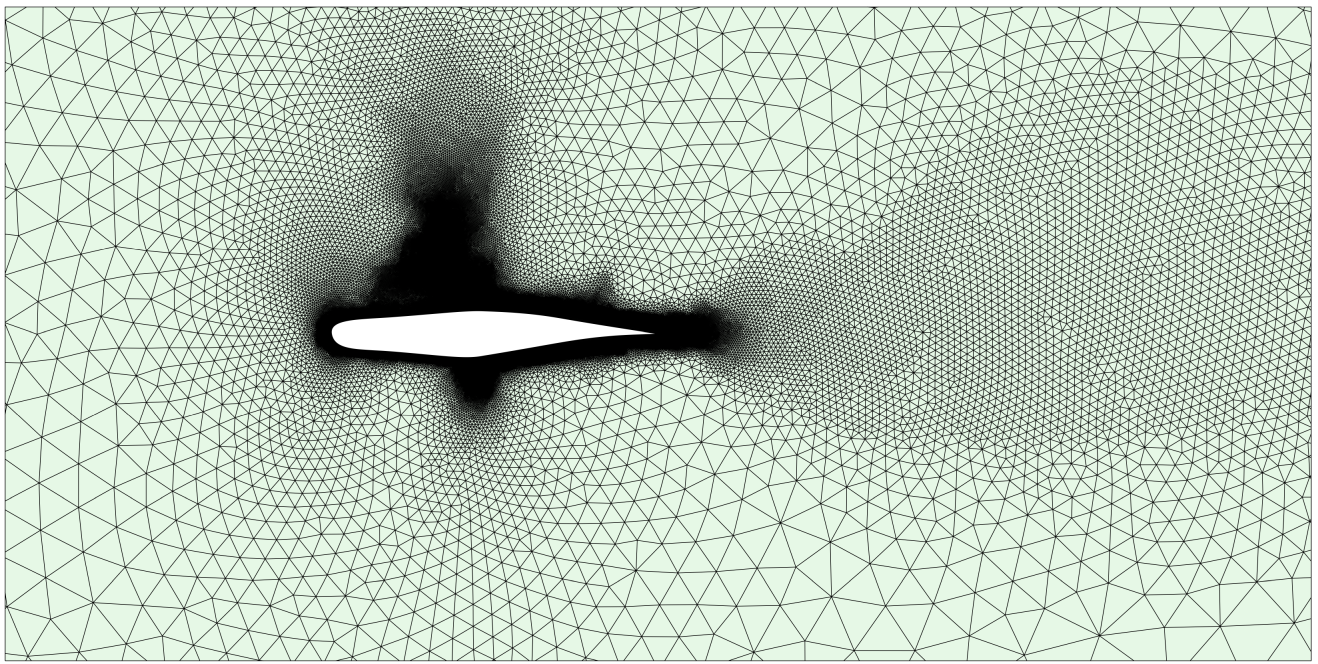}}
	\subfigure[Geometry 2]{\includegraphics[width=0.32\textwidth]{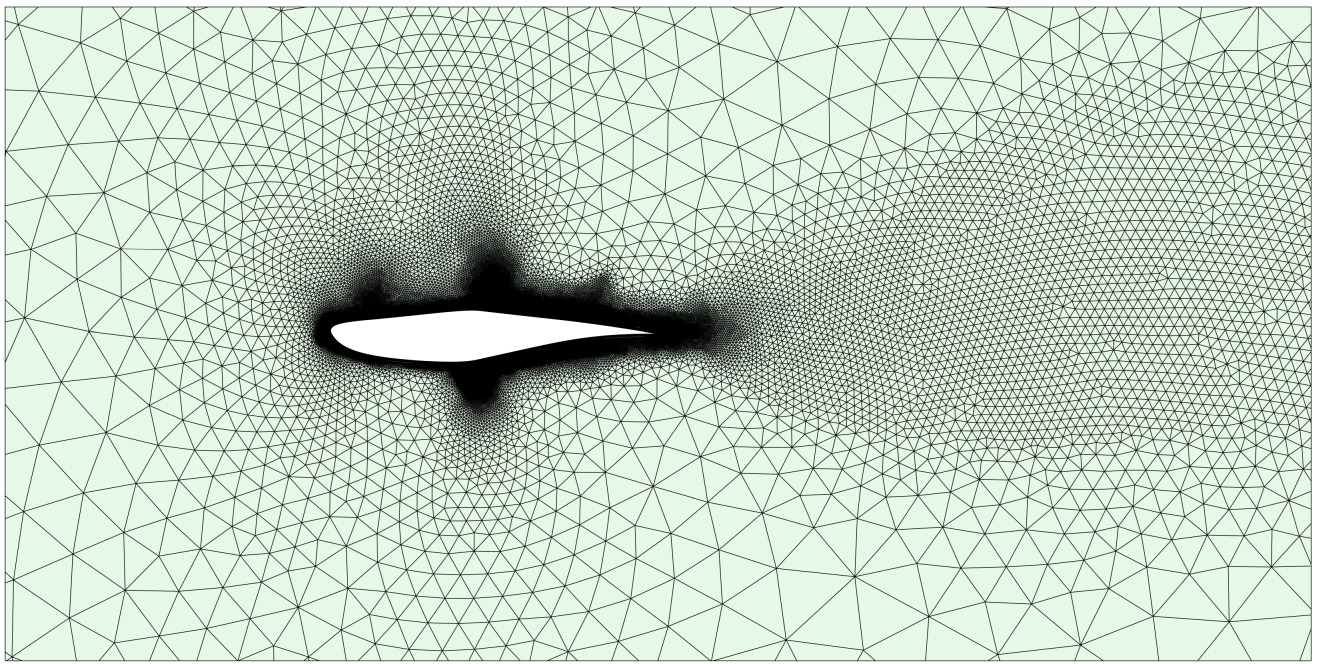}}	
	\subfigure[Geometry 3]{\includegraphics[width=0.32\textwidth]{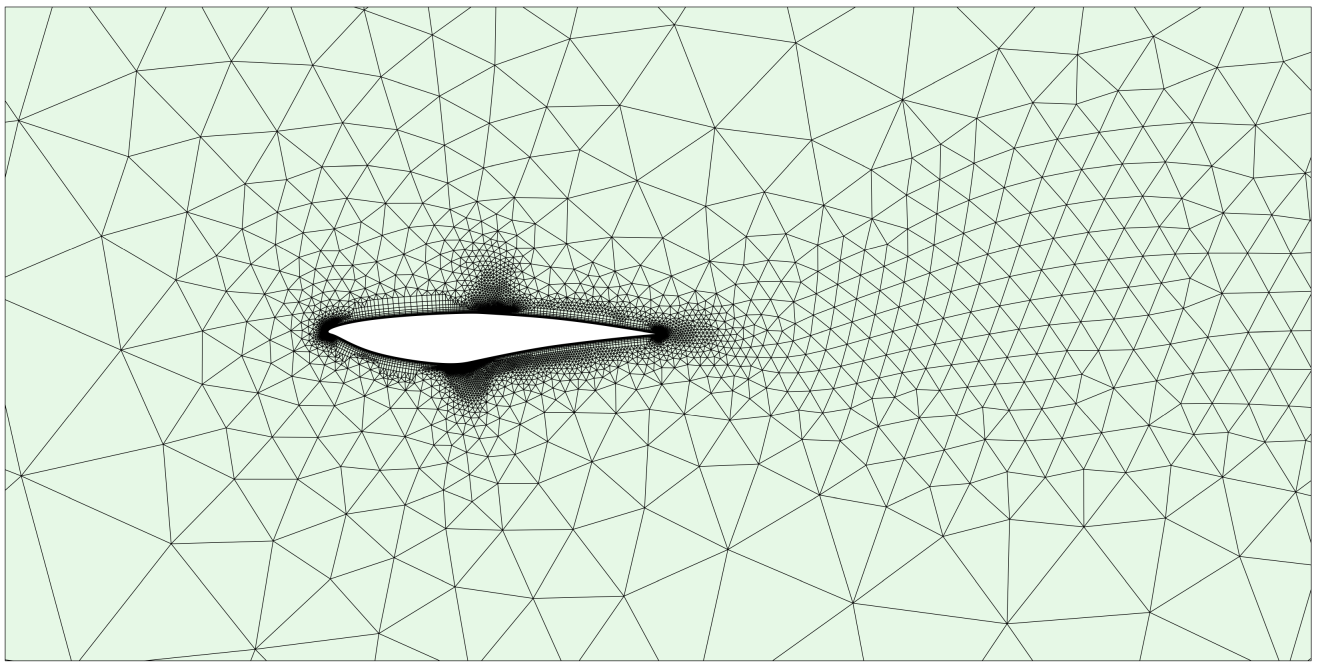}}	
	\subfigure[Geometry 1]{\includegraphics[width=0.32\textwidth]{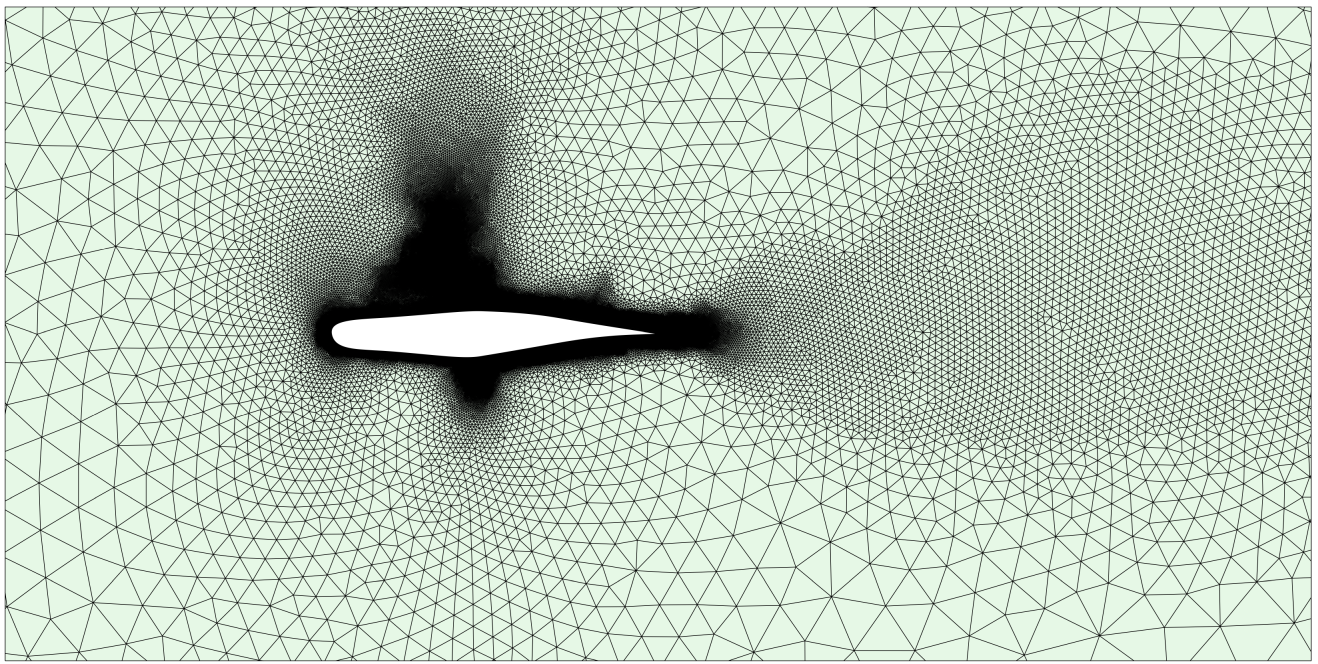}}
	\subfigure[Geometry 2]{\includegraphics[width=0.32\textwidth]{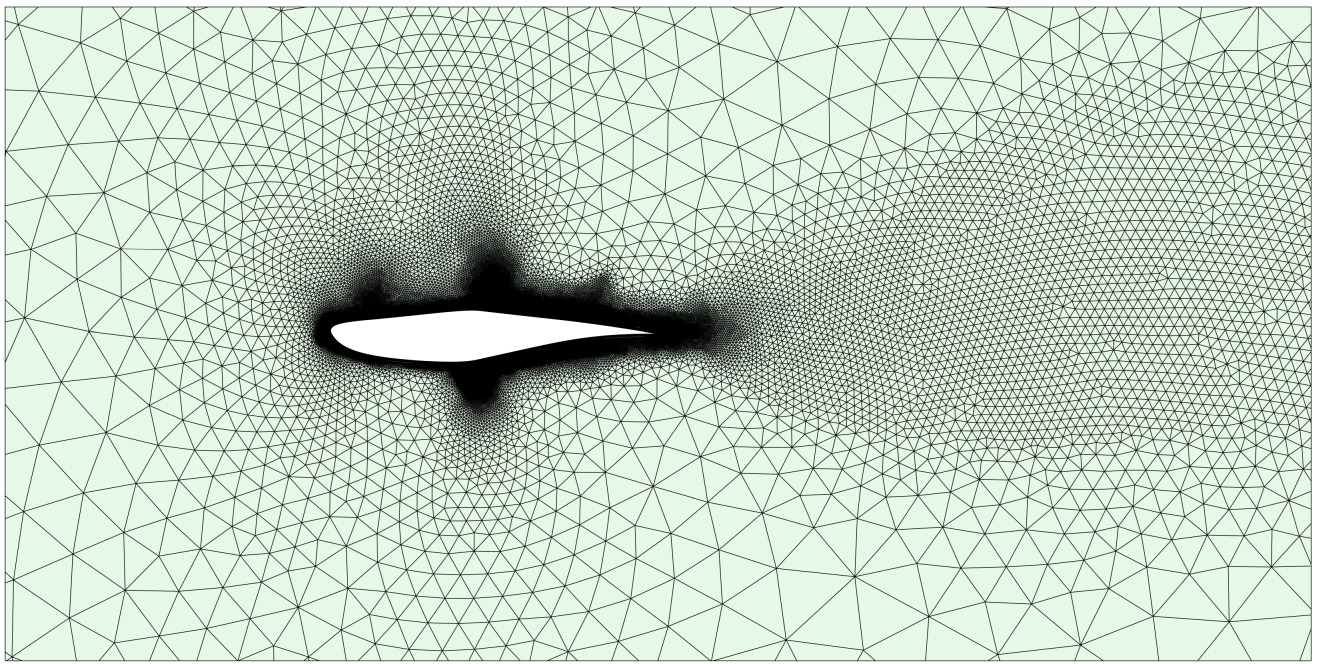}}	
	\subfigure[Geometry 3]{\includegraphics[width=0.32\textwidth]{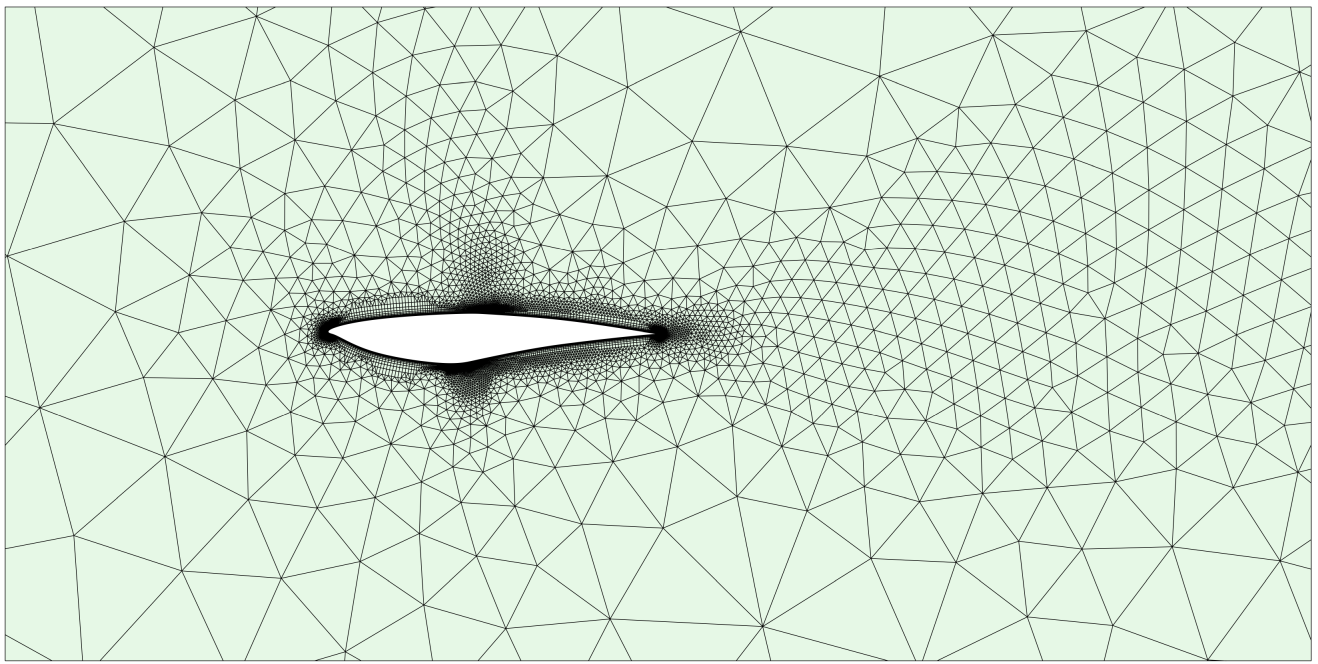}}	
	\caption{Target (top) and predicted (bottom) meshes for three geometric configurations unseen during training, corresponding to solutions of Figure~\ref{fig:geoTestCases} employing the coarse background meshes of Figure~\ref{fig:geoBacMeshes}.}
	\label{fig:geometryMeshPrediction}
\end{figure}

To illustrate the predictive capability of the trained ANNs, Figure~\ref{fig:geometryMeshPrediction} shows the target and predicted meshes for three unseen cases during training, corresponding to solutions of Figure~\ref{fig:geoTestCases}.

The predictions are obtained with the ANN trained using 640 training cases. The results show that for three geometric configurations with very different target meshes, the predicted meshes successfully provide the required local refinement near the regions where the solutions present a large gradient. 

\begin{figure}[!tb]
	\centering
	\includegraphics[width=0.75\textwidth]{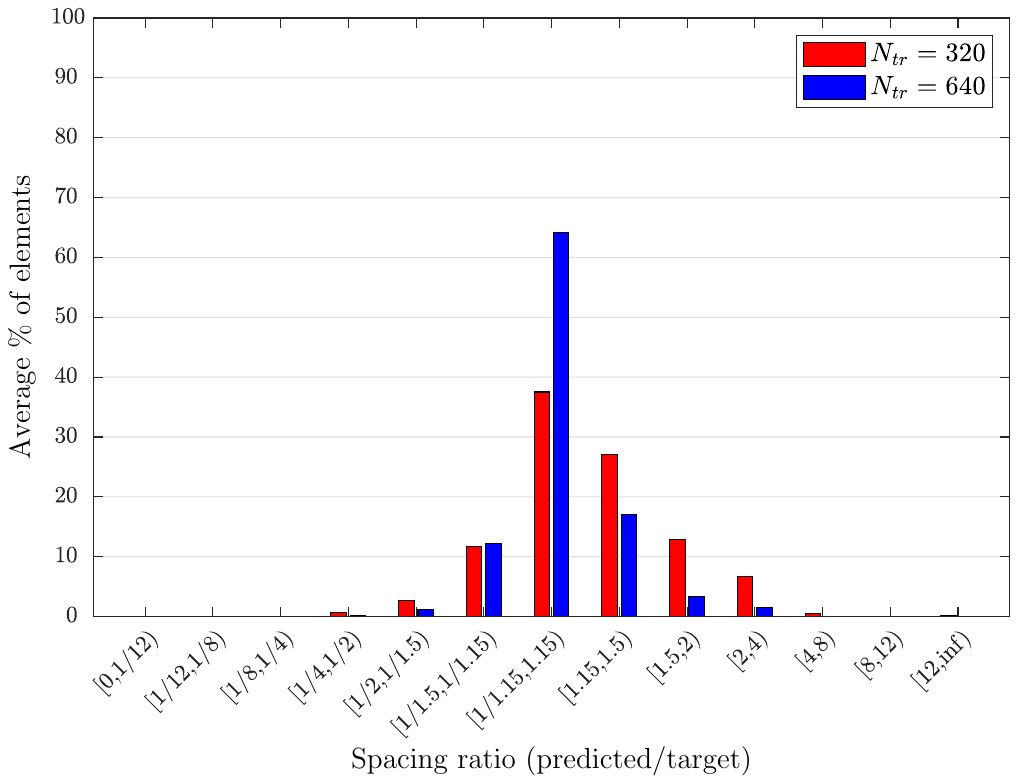}	
	\caption{Histogram of the ratio between predicted and target spacing for the example with varying geometric parameters and for two ANNs trained with a different number of training cases.}
	\label{fig:geometryHistogram}
\end{figure}

As in the previous example, to further assess the accuracy of the ANN predictions, the predicted spacing function is compared to the target spacing function by computing the ratio between the target and predicted spacing at the centroid of each element, and for all the test cases. The results are displayed as a histogram in Figure~\ref{fig:geometryHistogram} for the two different training sets with 320 and 640 cases.
The results show the ability of the trained ANNs to accurately predict the spacing for the large majority of the elements. Using an ANN trained with 320 cases, 76\% of the elements have a predicted spacing between  1/1.15 and 1.15, which is considered accurate enough to produce a mesh capable of capturing all the relevant flow features appearing in the solution. Increasing the number of training cases to 640 leads to a significant benefit, as it produces meshes where almost 70\% of the elements have a predicted spacing between  1/1.15 and 1.15. In addition, more than 90\% of the elements have a predicted spacing between  1/1.5 and 1.5.

To conclude, the suitability of the predicted meshes to perform simulations is analysed. For the three test cases of Figure~\ref{fig:geoTestCases}, unseen during training of the ANN, the predicted near-optimal meshes are utilised to perform a simulation to compare the relevant quantities of interest against a reference solution. Table~\ref{tb:geoAeroPrediction} reports the lift ($C_L$) and drag ($C_D$) coefficients obtained from a reference simulation and the ones obtained after performing a simulation with the predicted meshes using 640 training cases.
\begin{table}[!tb]
	\centering
	\begin{tabular}{|l||c|c||c|c||c|c|}
		\hline%
		& \multicolumn{2}{c||}{Geometry 1} &
		\multicolumn{2}{c||}{Geometry 2} &
		\multicolumn{2}{c|}{Geometry 3} \\
		\hline 				
		& Target & Prediction & Target & Prediction & Target & Prediction \\
		\hline
		$C_L$ & 0.689   & 0.683   &  0.354   & 0.343  & 0.566  & 0.539    \\
		\hline
		$C_D$ & 0.0212  & 0.0202  &  0.0347  & 0.0349  & 0.0186 & 0.0219   \\
		\hline
	\end{tabular}
	\caption{Comparison of the aerodynamic quantities of interest computed with the target and predicted meshes for three different geometric configurations.}
	\label{tb:geoAeroPrediction}
\end{table}

This study confirms the ability of the ANN to accurately predict the required spacing for unseen geometric configurations. For some configurations, the predicted meshes lead to aerodynamic quantities of interest that closely match the reference results, whereas for geometric configurations that lead to more complex flow features a larger discrepancy can be observed in the quantities of interest. Given the large number of geometric parameters considered in this example, more accurate predictions can obviously be obtained if the training set is increased. However, an alternative to avoid the generation of large data sets is to utilise the predicted meshes as a starting point for an adaptive process. The successful prediction of the regions where refinement is required will ensure that features are not missed in the first iteration of an adaptive process, and it is expected that this will lead to faster convergence of an adaptive process without human intervention for generating an initial mesh.

\section{Concluding remarks} \label{sc:conclusions}

A methodology to predict a near-optimal spacing function for turbulent compressible flow simulations has been presented. The strategy enables the use of available accurate solutions to learn the spacing required to capture the flow features of unseen cases.

The strategy proposed involves three main stages, namely computing the target spacing for a given solution, transferring the spacing to a (potentially morphed) background mesh and training an ANN. After training the ANN it can be used to predict the spacing for unseen geometric configuration or flow conditions. The work proposed focuses on three main challenges that appear when dealing with turbulent compressible viscous flows. First, it is necessary to consider several key variables to compute the target spacing. The combination of the pressure and Mach number is selected in this work. Second, a special treatment of stretched elements is proposed to ensure that the target spacing is a true representation of the given solution. Finally, the effect of using different numerical strategies to compute the Hessian of a key variable is discussed.

The results presented consider two problems with variable flow conditions and geometric parameters. In both cases accurate predictions can be made with limited training data and numerical studies are presented to show the influence of the ANN architecture and the size of the training dataset. The problem involving geometric parameters considers as geometric parameters the positions of the control points of the NURBS describing the boundary. Therefore, it provides a tight coupling with the CAD model.

The suitability of the meshes for performing simulations is also assessed by comparing the aerodynamic quantities of interest of reference results and the results obtained with the predicted meshes. Despite the reduced number of training cases employed, a good accuracy is obtained, indicating that the proposed strategy provides a very good initial mesh that can be further enhanced using an adaptivity procedure.

Future work will involve the use of an anisotropic metric to define the spacing. This requires the prediction of a metric per point rather than a single scalar.

\section*{Acknowledgements}

The financial support of the Engineering and Physical Sciences Research Council (Grant Number: EP/T009071/1) is gratefully acknowledged. The authors would like to thank Dr Jason Jones for the help provided in preprocessing the results of the simulation shown in Figure~\ref{fig:pOscillations}(a). 

\section*{Competing interests}

The authors have no competing interests to declare that are relevant to the content of this article.

\bibliographystyle{unsrt}
\bibliography{references}
	
\end{document}